\documentclass[useAMS,usegraphicx,usenatbib]{mn2e}
\usepackage{times}   
\usepackage{mathptm} 
\usepackage{color}
\usepackage{amsmath}

\title[MAGMA2]
{The Lagrangian hydrodynamics code MAGMA2}
\author[Rosswog ]
{S. Rosswog\thanks{E-mail: stephan.rosswog@astro.su.se}$^{1}$\\
$^1$ The Oskar Klein Centre, Department of Astronomy, AlbaNova, Stockholm University, SE-106 91 Stockholm, Sweden}

\newcommand{\Ma}{\texttt{MAGMA2}\,}
\newcommand{\ma}{\texttt{MAGMA2}}
\newcommand{\Ar}{\texttt{Arepo}\,}

\newcommand{\Pha}{\texttt{PHANTOM}\,}
\newcommand{\Ga}{\texttt{GASOLINE2}\,}
\def\bea{\begin{eqnarray}}
\def\eea{\end{eqnarray}}
\def\be{\begin{equation}}
\def\ee{\end{equation}}
\def\msun{M$_{\odot}$}
\def\Msun{M$_{\odot}$ }

\begin{document}

\date{Draft version}

\pagerange{\pageref{firstpage}--\pageref{lastpage}} \pubyear{2020}

\maketitle

\label{firstpage}

\begin{abstract}
We present the methodology and performance of the new Lagrangian hydrodynamics code \ma,
a Smoothed Particle Hydrodynamics code that benefits from a number of non-standard enhancements.
By default it uses high-order smoothing kernels and wherever gradients are needed, they are calculated 
via accurate matrix inversion techniques, but a more conventional formulation with kernel gradients
has also been implemented for comparison purposes. We also explore a matrix inversion formulation 
of SPH with a symmetrisation in the particle indices that is not frequently used. We find 
interesting advantages of this formulation in some of the tests, for example,  a substantial
reduction of surface tension effects for non-ideal particle setups and more accurate peak densities
in Sedov blast waves. 
\Ma uses artificial viscosity, but enhanced  by techniques that are commonly used in finite volume schemes 
such as reconstruction and slope limiting. While simple to implement, this approach efficiently suppresses 
particle noise, but at the same time drastically reduces dissipation in locations where it is not needed 
and actually unwanted. We demonstrate the performance of the new code in a  number of challenging  
benchmark tests including e.g. multi-dimensional vorticity creating Schulz-Rinne-type Riemann problems 
and more astrophysical  tests such as a collision between two stars to demonstrate its robustness and  
excellent conservation properties.
\end{abstract}


\section{Introduction}

A Lagrangian formulation of  hydrodynamics is a natural choice for many astrophysical problems. 
Smoothed Particle Hydrodynamics (SPH) \citep{lucy77,monaghan77} is  the most wide-spread
Lagrangian method in astrophysics. It is entirely mesh-free and the equations can  be symmetrised
in a way so that mass, energy, momentum and angular momentum are conserved by construction.
As it is derived, SPH is entirely dissipationless and therefore needs to be augmented by additional 
measures to produce appropriate amounts of entropy in shocks. This is traditionally done via artificial 
viscosity, but also Riemann solver approaches have been explored 
\citep{inutsuka02,cha03,cha10,murante11,puri14}. \\
In SPH one calculates the density via a smooth weighting of nearby particle masses
and this smooth density estimate enters the calculation of pressures that drive the motion.
The internal energy, in contrast, is evolved via a straight-forward discretion of the Lagrangian
energy conservation law and does therefore not involve a smoothing process. This potentially
different inherent smoothness of both quantities can lead to unintended "pressure blips" when setting up contact 
discontinuities. These blips  cause surface tension effects that can suppress weakly triggered fluid 
instabilities \citep{agertz07,mcnally12}. Such effects can be counterbalanced by a careful setup 
of initial conditions with consistent smoothness, by alternative 
expressions for SPH volume elements \citep{ritchie01,saitoh13,hopkins13,rosswog15b,cabezon17}
or by adding artificial conductivity terms to smooth out sharp transitions in the internal 
energy \citep{price08a,valdarnini12}.\\
These issues have spurred further developments on Lagrangian hydrodynamics, many of which
have employed techniques from finite volume Eulerian hydrodynamics. The \Ar code \citep{springel10b},
for example, tesselates space into Voronoi cells and evolves the hydrodynamic equations via
a Riemann solver-based finite volume strategy. Such finite volume approaches,
however, are not bound to Voronoi or other meshes and can actually also be applied to hydrodynamic
schemes which use particles. Several such finite  volume particle schemes have been suggested in 
the applied mathematics literature \citep{benmoussa99,vila99,hietel00,junk03}, but
they have only recently found their way into astrophysics \citep{gaburov11,hopkins15a,hubber18}
where they have delivered accurate  results.\\
Also on the SPH-side there have been several new developments. Apart from the above mentioned
volume element improvements, substantially more accurate gradient estimates have been implemented
 \citep{garcia_senz12,cabezon12a,rosswog15b}. 
The perhaps most advanced SPH scheme to date \citep{frontiere17} uses a reproducing kernel 
methodology \citep{liu95} together with a sophisticated artificial dissipation scheme. They 
find good performance in a number of benchmark tests that are generally considered difficult for SPH.
While the increasing cross-fertilisation between different methods has been very beneficial, 
the boundaries between different numerical schemes and their naming conventions have started to
blur.\\
In this paper we describe the new Lagrangian hydrodynamics code \ma,
an SPH code that benefits from many improvements compared to more
traditional SPH methods. It uses, for example, high-order kernels, calculates gradients
via matrix-inversion techniques and uses slope-limited velocity reconstructions within
an artificial viscosity approach. Many of these techniques have been scrutinised in a
both Newtonian and special-relativistic context in \cite{rosswog15b}. The artificial viscosity
approach  that we are using is oriented at the recent work by \cite{frontiere17} which
use fixed dissipation parameters, but reduce the effective viscosity by linearly reconstructing
the velocities to inter-particle midpoints. In this paper we further expand on these ideas 
and throughout this paper we keep the dissipation parameters constant. We have also
implemented a new way to steer time-dependent dissipation by monitoring local
entropy violations and thus identifying "troubled particles" that need more dissipation.
This new approach is discussed in a separate study  \citep{rosswog20b} and will not 
be used here.
The main purpose of this paper is to document \Ma as a new simulation tool and  to
demonstrate its performance in a series of benchmark tests.\\
The paper is structured as follows. In Sec.~\ref{sec:method} we describe the methodology
used in \Ma and 
Sec.~\ref{sec:tests} is dedicated to benchmark tests. We begin with smooth advection, then
various shock tests, subsequently explore Kelvin-Helmholtz and Rayleigh-Taylor instabilities 
and combined, vorticity-creating shocks. We conclude our test series with astrophysical tests
such as stellar collisions and tidal disruption events that demonstrate the versatility, robustness
and the excellent conservation properties in practical examples. Sec.~\ref{sec:summary} briefly 
summarises our study.

\section{Methodology}
\label{sec:method}
Many of the design choices in \Ma are informed by our recent work \citep{rosswog15b} 
where we have carefully explored various SPH ingredients such as the calculation of gradients, dissipation 
triggers or kernel functions. Methods that have been described in detail elsewhere are only
briefly summarised, while we focus here on those elements that are new.\\
For a general introduction to the Smooth Particle Hydrodynamics (SPH) method, we refer
to  review articles \citep{monaghan05,springel10a,price12a,rosswog15c}, a 
detailed step-by-step derivation of both Newtonian and relativistic SPH can be found 
in \cite{rosswog09b}.

\subsection{Ideal hydrodynamics}
\label{sec:ideal_hydro}
There are many different correctly symmetrised SPH discretizations of the ideal fluid equations.
We have implemented three different SPH formulations into \ma, which we briefly summarise below.
For all of them we use high-order kernels and a novel dissipation scheme, but the formulations
differ in the way gradients are calculated (two use matrix inversions, one
kernel gradients), and in the symmetrisation of the equations.
\subsubsection{Matrix inversion formulation 1 (MI1)}
\label{sec:IA1}
The  standard SPH-approach of using direct gradients of (radial) kernel functions
is a straight-forward way to ensure exact conservation\footnote{See e.g. Sec.2.4 of \cite{rosswog09b} 
for detailed discussion of conservation in SPH. "Exact conservation" means up to potential violations
due to finite accuracy due to time integration or approximations of the gravitational forces,
e.g. due to the use of a tree. These latter terms, however, are usually fully controllable, though
at some computational expense.}.
The resulting gradients, however, are of moderate accuracy only 
\citep{abel11,garcia_senz12,cabezon12a,rosswog15b,rosswog15c,cabezon17}.
Improvements of many orders of magnitude in gradient accuracy can be achieved,
see  Fig. 1 in \cite{rosswog15b} at the (low) price of inverting a 3x3-matrix (in 3D).\\
The first matrix inversion formulation MI1  is the Newtonian limit of  a special-relativistic formulation  that was 
derived and extensively tested in \cite{rosswog15b}\footnote{The corresponding formulation
was called $\mathcal{F}_3$ in the original paper.}. It reads explicitly:
\bea
\rho_a&=& \sum_b m_b W_{ab}(h_a),
\label{eq:dens_sum}\\
\frac{d\vec{v}_a}{dt}&=& - \sum_b m_b \left\{ \frac{P_a}{\rho_a^2} \vec{G}_a +  \frac{P_b}{\rho_b^2} \vec{G}_b\right\}
\label{eq:momentum_IA},\\
\left(\frac{du_a}{dt}\right)&=& \frac{P_a}{\rho_a^2} \sum_b m_b \vec{v}_{ab} \cdot \vec{G}_a
 \label{eq:energy_IA},
\eea
where $\rho, \vec{v}, u$ and $h$ denote mass density, velocity, specific internal energy and smoothing length.
The particle mass is denoted by  $m$, $P$ is the gas pressure, $\vec{v}_{ab}= \vec{v}_a - \vec{v}_b$  and 
$W$ is the chosen SPH kernel function. The gradient functions are given by
\be
\left(\vec{G}_{a}\right)^k= \sum_{d=1}^3 C^{kd}(\vec{r}_a,h_a) (\vec{r}_b - \vec{r}_a)^d W_{ab}(h_a),
\label{eq:Ga}
\ee
\be
\left(\vec{G}_{b}\right)^k= \sum_{d=1}^3 C^{kd}(\vec{r}_b,h_b) (\vec{r}_b - \vec{r}_a)^d W_{ab}(h_b),
\label{eq:Gb}
\ee
where $W_{ab}(h)= W(|\vec{r}_a-\vec{r}_b|,h)$.  The "correction matrix" $C$ accounts for 
the local particle distribution and is calculated as 
\be
\left(C^{ki} (\vec{r},h)\right)= \left( \sum_b \frac{m_b}{\rho_b} (\vec{r}_b - \vec{r})^k (\vec{r}_b - \vec{r})^i W(|\vec{r}-\vec{r}_b|,h)\right)^{-1}.
\label{eq:corr_mat}
\ee
The functions $\vec{G}_{k}$ are anti-symmetric  with respect to the exchange of the involved position vectors
and therefore allow for exact conservation in a similar way as the anti-symmetric $\nabla_a W_{ab}$ in standard SPH.
For more details we refer to \cite{rosswog15b}.
The practical conservation will be scrutinised below in a violent, off-center collision between two stars, see 
Sec.~\ref{sec:stellar_collision}, which shows that conservation with matrix inversion gradients is on par
with standard SPH.

\subsubsection{Matrix inversion formulation 2 (MI2)}
\label{sec:IA2}
The key property of the SPH-equations to achieve conservation, is to ensure the correct (anti-) symmetries  
in the particle indices, see for example Sec. 2.4 in \cite{rosswog09b} for detailed explanation. This
can be achieved in an infinite number of different ways. One can, for example, 
start from  the Lagrangian form of the Euler equations and use \citep{monaghan92}
\be
\frac{\nabla P}{\rho}= \frac{P}{\rho^\sigma} \nabla \left( \frac{1}{\rho^{1-\sigma}}\right) + 
\frac{1}{\rho^{2-\sigma}}   \nabla \left( \frac{P}{\rho^{\sigma-1}} \right),
\ee
where $\sigma$ is real parameter. This leads to a momentum equation of the form
\be
\frac{d\vec{v}_a}{dt}= - \sum_b m_b \left[ \frac{P_a}{\rho_a^\sigma \rho_b^{2-\sigma}} + 
                                                                  \frac{P_b}{\rho_a^{2-\sigma} \rho_b^{\sigma}}  \right] \nabla_a W_{ab},
\label{eq:dvdt_IA2}
\ee
and 
\be
\frac{du_a}{dt}= \frac{P_a}{\rho_a^\sigma} \sum_b m_b \frac{\vec{v}_{ab}}{\rho_b^{2-\sigma}} \nabla_a W_{ab}.
\ee 
as an energy equation. We implement these equations, as in Sec.~\ref{sec:IA1}, by replacing the kernel gradients by the functions
Eqs.~(\ref{eq:Ga}) and (\ref{eq:Gb}). In the tests shown below we use $\sigma=1$, so that final equations read
\bea
\frac{d\vec{v}_a}{dt}&=& - \sum_b m_b \left\{ \frac{P_a+P_b}{\rho_a \rho_b} \right\}\vec{G}_{ab} 
\label{eq:momentum_IA},\\
\left(\frac{du_a}{dt}\right)&=& \sum_b m_b \left\{\frac{P_a}{\rho_a \rho_b} \right\} \vec{v}_{ab} \cdot \vec{G}_{ab}
 \label{eq:energy_IA},
\eea
where $\vec{G}_{ab}= 0.5(\vec{G}_a + \vec{G}_b$) and we use the density estimate  of Eq.~(\ref{eq:dens_sum})\footnote{We had also
performed some experiments with $\sigma=0$ and $\sigma=2$, which give good, but slightly worse results in Sedov explosions
than $\sigma=1$.}. If one
replaces the matrix inversion gradient functions $\vec{G}$ by the corresponding kernel gradients one recovers
the SPH equations that have been successfully used in the Gasoline2 code \citep{wadsley17}.\\
While both of the matrix inversion formulations MI1 and MI2 yield very similar results on most benchmark tests, they 
show noticeable differences in some of them, for example in the Sedov explosion and in
the "square test", see below. Where the results are practically indistinguishable, we will only show 
one set of results, only where visible differences occur will we show plots for different formulations.  

\subsubsection{Standard kernel gradient formulation (stdGrad)}
\label{sec:stdGrad}
For comparison purposes we have also implemented a more conventional kernel-gradient formulation
\bea
\frac{d\vec{v}_a}{dt}&=& - \sum_b m_b
\left\{
\frac{P_a}{\rho_a^2} \nabla_a W_{ab}(h_a) +
\frac{P_b}{\rho_b^2} \nabla_a W_{ab}(h_b) 
\right\}
\label{eq:momentum},\\
\frac{d u_a}{dt}&=& \frac{P_a}{\rho_a^2}
\sum_b m_b \vec{v}_{ab} \cdot \nabla_a W_{ab}(h_a) \label{eq:energy},
\label{eq:std_eqs}
\eea
where the density is calculated as in our default version, see Eq.~(\ref{eq:dens_sum}).
This equation set can be derived from a Lagrangian \citep{monaghan01,springel02}, but
note that we have for simplicity  omitted the so-called "grad-h terms" that 
contain derivatives of the kernel with respect to the smoothing length. These 
terms further improve the numerical conservation, but our tests in \cite{rosswog07c} 
have shown that the violations of exact conservation are very small, even under 
conditions that are considered as "worst case" \citep{hernquist93}. This will be further confirmed in our
collision test in Sec.~\ref{sec:stellar_collision} which demonstrates that conservation is
excellent even without these correction terms.
\begin{figure}
\includegraphics[width=1.\columnwidth]{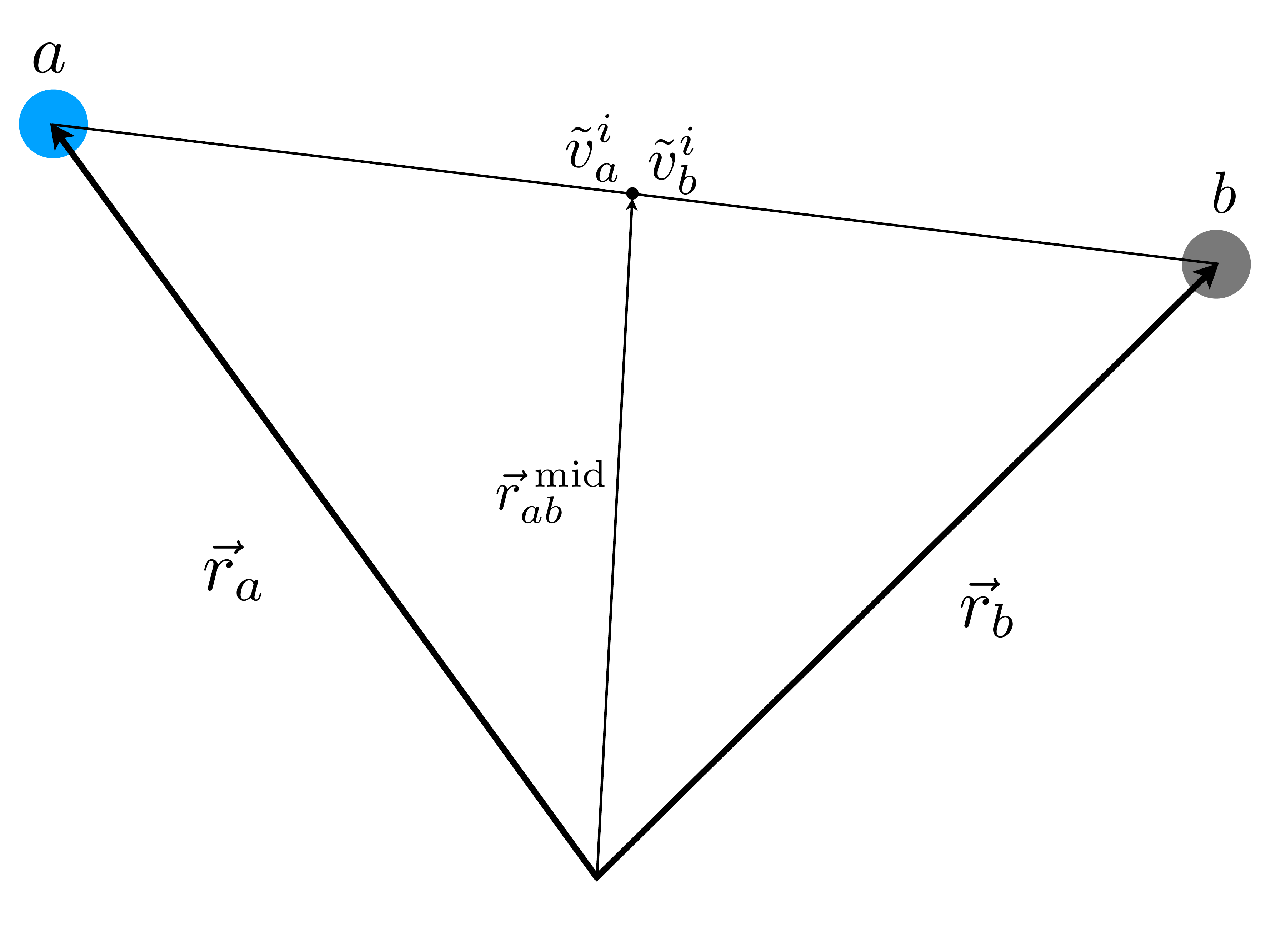}
\caption{The artificial dissipation scheme uses the difference of the  velocities  
reconstructed from particle $a$ and particle $b$
to their inter-particle midpoint at $\vec{r}_{ab}$.}
\label{fig:rec_sketch}
\end{figure}
%
\subsection{Dissipation with reconstruction and slope-limiting}
\label{sec:AV}
\subsubsection{Artificial viscosity}
A common approach to deal with shocks is to enhance the physical pressure $P$ by additional  viscous pressure 
terms $Q$ \citep{vonNeumann50}, i.e. to replace the physical pressure, wherever it occurs, by
$P + Q$. The pressures $Q$ are typically the sum two terms, one is proportional to the velocity jump 
between  computational elements (either cells or particles) while the other is proportional to the
square of the jump. In SPH, the expression for the viscous pressure at particle position $a$
is often written as \citep{monaghan83}
\be
Q_a= \rho_a \left(- \alpha c_{{\rm s},a} \mu_a + \beta \mu_a^2\right),
\label{eq:Qvis}
\ee
where the velocity jump is
\be
\mu_a= \rm{min} \left(0, \frac{\sum_\delta v_{ab}^\delta \eta_a^\delta }{\eta_a^2 + \epsilon^2}\right).
\label{eq:mu_vis}
\ee
Here, $\alpha, \beta$ and $\epsilon$ are numerical parameters (typical values are 1, 2 and $0.1$),
$c_{{\rm s}}$ is the sound speed and 
\be
\eta^\delta_a= \frac{\left(\vec{r}_a - \vec{r}_b\right)^\delta}{h_a}, \quad \eta_a^2= \sum_\delta \eta^\delta_a  \eta^\delta_a
\ee
are separations between particles, de-dimensionalised via the smoothing length $h_a$.
We follow the convention that particles are labelled usually by $a$ and 
$b$ and we use greek letters to denote summation indices (such as the $\delta$ above).\\
In SPH it is common 
practice to use $v_{ab}^\delta = v_a^\delta -  v_b^\delta$ in Eq.~(\ref{eq:mu_vis}), 
i.e. one applies the velocity difference between
the two particles. One can think of artificial dissipation as a simple way of solving an inter-particle
Riemann problem \citep{monaghan97}. Translated to the language of finite volume methods, the 
common practice of using straight forwardly
the differences between particle velocities corresponds to a zeroth-order, or constant, velocity reconstruction. 
Zeroth-order reconstruction is known to introduce excessive dissipation in finite volume methods, 
but higher order reconstructions yield successively less dissipative numerical schemes. 
This idea is translated here into an SPH context.\\
Most modern SPH implementations use time-dependent dissipation parameters to avoid excessive
and unnecessary dissipation \citep{morris97,rosswog00,cullen10,rosswog15b,wadsley17,price18a,rosswog20b}.
Recently, \cite{frontiere17}, following ideas from a Eulerian finite volume context \citep{christensen90}, 
have explored an alternative refinement where the same form of artificial pressure, Eq.~(\ref{eq:Qvis}), 
is used, but the velocity is (in their approach linearly) reconstructed to the particle midpoint. While still being a simple 
artificial viscosity scheme their approach showed excellent performance even with fixed dissipation 
parameters $\alpha$ and $\beta$.\\
Instead of using the difference of the two velocities, we quadratically reconstruct the velocities
of particle $a$ and $b$ to their midpoint at $\vec{r}^{\rm \; mid}_{ab}= 0.5(\vec{r}_a+\vec{r}_b)$, see Fig.~\ref{fig:rec_sketch}. 
The velocities reconstructed from particle $a$  to the midpoint read
\bea
\tilde{v}_a^i &=& v_a^i + \Phi_{ab} \left[ (\partial_j v^i) \delta^j  + \frac{1}{2} (\partial_l \partial_m v^i) \delta^l \delta^m \right]_a,
\eea
where  $\Phi_{ab}$ is a slope limiter, see below, the index at the square bracket indicates that the derivatives
at the position of particle $a$ are used and the increments
from point $a$ to the midpoint are $(\delta^i)_a= \frac{1}{2} (\vec{r}_b - \vec{r}_a)^i$. The reconstructed 
velocities from the $b$-side, $\tilde{v}_b^i$, are calculated correspondingly,
but now with derivatives at position $b$ and increments $\delta^i_b= -\delta^i_a$. In Eq.~(\ref{eq:mu_vis}) we use the
difference in the {\em reconstructed} velocities, i.e. 
 $\tilde{v}_{ab}^{\delta}= \tilde{v}_a^\delta - \tilde{v}_b^{\delta}$. We calculate the first derivatives
as in \cite{rosswog15b}
\be
(\partial_j v^i)_a= C_a^{jk} \sum_b{\frac{m_b}{\rho_b} \; [v^i_b - v^i_a] \; (\vec{r}_b - \vec{r}_a)^k} \; W_{ab}(h_a),
\label{eq:acc_der}
\ee
where $C$ is the correction matrix from Eq.~(\ref{eq:corr_mat}). For the second derivatives we proceed in two steps. First, we calculate
an auxiliary gradient  that does not require the density \citep{price04c,rosswog07c} and therefore can be conveniently
calculated alongside the density loop 
\be
(\partial_j v^i)_a^{\rm aux} \equiv D_a^{jk} \sum_b m_b (v^i_b - v^i_a) \partial_k W_{ab}(h_a),
\ee
where the corresponding density-independent correction matrix is given by
\be
D_a^{jk}= \left\{ \sum_b m_b (\vec{r}_b - \vec{r}_a)^j  \partial_k W_{ab}(h_a) \right\}^{-1}.
\ee
The second derivatives are then calculated by applying 
Eq.~(\ref{eq:acc_der}) to the auxiliary first derivatives
$(\partial_j v^i)^{\rm aux}$. While this procedure to calculate second derivatives
still comes at some cost, it does not require an additional loop over the neighbour particles
compared to just linear reconstruction.
\\
We use a  modification
of van Leer's slope limiter \citep{vanLeer74,frontiere17}
\be
\Phi_{ab}= \rm{max}\left[0, \rm{min}\left[ 1, \frac{4A_{ab}}{(1+A_{ab})^2}\right] \right] \begin{cases}
    1, & \text{if $\eta_{ab} > \eta_{\rm{crit}}$}.\\
    e^{-\left(\frac{\eta_{ab}-\eta_{\rm crit}}{0.2}\right)^2}, & \text{otherwise}
  \end{cases}
\ee
with
\be
A_{ab}= \frac{\sum_{\delta \gamma}(\partial_\delta v_a^\gamma) \; x_{ab}^\delta \; x_{ab}^\gamma}{\sum_{\delta \gamma}(\partial_\delta v_b^\gamma) \; x_{ab}^\delta \; x_{ab}^\gamma}
\ee  
where the $x_{ab}^\delta$ are the components of $\vec{r}_a-\vec{r}_b$ and
\be
\eta_{ab}= \rm{min}(\eta_a,\eta_b) =  \rm{min}\left(\frac{r_{ab}}{h_a},\frac{r_{ab}}{h_b}\right)  \; {\rm and} \;
\eta_{\rm crit}= \left( \frac{32\pi}{3N_{\rm nei}}\right)^{1/3},
\ee
with $r_{ab}= |\vec{r}_a-\vec{r}_b|$ and $N_{\rm nei}$ being the number of neighbours for the chosen kernel,
see Sec.~\ref{sec:kernel}.
This prescription yields excellent, oscillation-free shock results as we demonstrate below.
\subsubsection{Artificial conductivity}
\label{sec:conduct}
The analogy  with Riemann solvers \citep{monaghan97,chow97,price08a} also suggests 
to include thermal conductivity in the artificial dissipation terms. Such approaches 
have been found advantageous in certain shock problems \citep{noh87,rosswog07c},
but --as with all artificial dissipation terms-- one has to ensure that no unwanted side 
effects are introduced. We add an artificial conductivity term to our energy equation
\be
\left( \frac{du_a}{dt}\right)_C= -\alpha_u \sum_b m_b \frac{v_{\rm sig,u}^{ab}}{\rho_{ab}} \left(\tilde{u}_a - \tilde{u}_b \right)
\frac{|\vec{G}_a+\vec{G}_b|}{2},
\label{eq:conductivity}
\ee
where for the standard gradient version the average of the $\vec{G}$-terms is replaced by
$[\nabla_a W_{ab}(h_a) + \nabla_a W_{ab}(h_b)]/2$. As for the velocities in the artificial
viscosity, we use the differences in the quadratically reconstructed internal energies
at the interparticle midpoint, $\tilde{u}_a$ and $\tilde{u}_b$, which are calculated analogously to
the velocities. For the conductivity signal velocity we use
\be
v_{\rm sig,u}^{ab}= (1-iG) \; v_{\rm sig,nG}^{ab} + iG \; v_{\rm sig,G}^{ab}, 
\ee 
where $iG$ is a flag indicating whether gravity is used ($iG=1$) or not ($iG=0$) and
\be
v_{\rm sig,nG}^{ab} = \sqrt{\frac{|P_a - P_b|}{\rho_{ab}}} \quad {\rm and} \quad
v_{\rm sig,G}^{ab} = |\vec{\tilde{v}}_a - \vec{\tilde{v}}_b|.
\label{eq:vsig_u}
\ee
These expressions are very similar to those used in the \Pha code \citep{price18a}, but,
as with the artificial viscosity terms, we are using here differences in the reconstructed 
velocities and internal energies (as indicated by tildes) rather that "flat" differences and
matrix-inversion based gradient functions.\\
Throughout this study, we always use constant dissipation parameters $\alpha=1$ and $\beta= 2\alpha$. 
As we will demonstrate below, and consistent with a similar approach in a reproducing kernel  context \citep{frontiere17},
the described velocity reconstruction produces excellent results and reduces drastically unwanted 
effects of artificial viscosity. Nevertheless, one may still try to additionally steer the dissipation parameter $\alpha$
so that --if it safe to do so-- it decays towards zero.  A novel way of steering $\alpha$ via violations
of exact entropy conservation has recently been explored in a separate study \citep{rosswog20b}.
It has also been implemented in \Ma, but all test shown here use fixed parameter values to demonstrate how powerful
the artificial viscosity approach with reconstructed quantities is.\\
Concerning the conductivity, we very much stay on a conservative side with 
only small conductivity effects: we choose a  low value of $\alpha_u=0.05$ as default choice,
we use the difference of reconstructed $u$-values in Eq.~(\ref{eq:conductivity}) and we use
a switch of signal velocities, see Eq.~(\ref{eq:vsig_u}), to protect a hydrostatic equilibrium
configuration from being destroyed by conductivity. As shown below, none of the tests
is substantially impacted by conductivity, not even the Kelvin-Helmholtz tests, see 
Fig.~\ref{fig:KH_conductivity} in Sec.~\ref{sec:KH_instab}, which were a major motivation
for introducing conductivity in the first place. While on the time scales usually shown in standard
tests (and also here), conductivity effects are hardly noticeable in \Ma simulations, we do see some positive effects in the
long term evolution of Kelvin-Helmholtz instabilities, where, without conductivity, the flow
looks more granular. With our very conservative conductivity implementation we have not 
encountered any negative side effect, see for example, the test in Sec.~\ref{sec:oscil_poly}.
It is worth noting that \cite{price18a} actually use a value of unity for $\alpha_u$ and report that 
they have not encountered artefacts for  even this large value. 
\begin{figure}
\includegraphics[width=1.1\columnwidth]{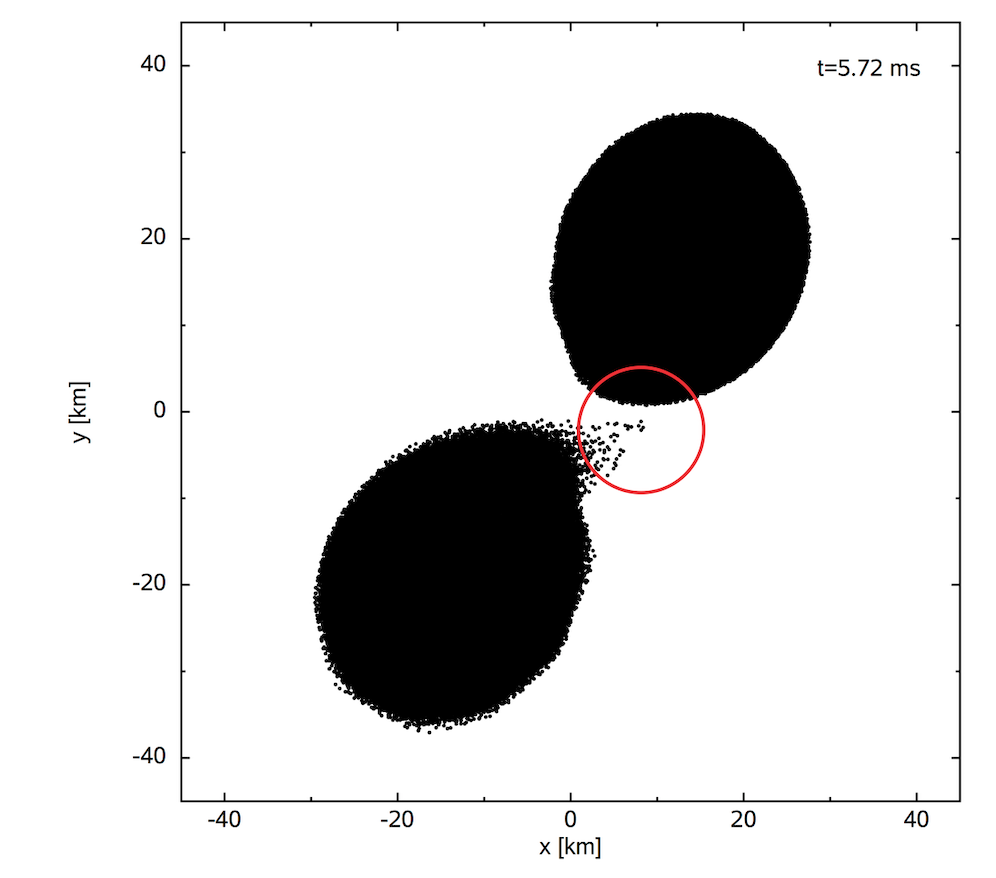}
\caption{Astrophysical example (merger of a 1.3 with a 1.4 \Msun neutron star) where an
expanding flow encounters a sharp surface with a very large SPH particle density. The smoothing
length of the leading particle is indicated by the red circle.}
\label{fig:particle_dist}
\end{figure}

\subsection{Kernel choice}
\label{sec:kernel}
SPH needs a kernel function to estimate densities and gradients, see Eqs.(\ref{eq:dens_sum}) - (\ref{eq:std_eqs}).
Cubic spline kernels \citep{schoenberg46,monaghan05} have traditionally been the standard choice in SPH.
In recent years, however, a number of alternatives have been explored, see, for example, 
\cite{cabezon08,read10,dehnen12,rosswog15b}. In particular the family of Wendland kernels \citep{wendland95}
has received a fair amount of attention, because these kernels avoid the so-called "pairing instability" \citep{schuessler81}
as pointed out by \cite{dehnen12}. While these kernels require a large neighbour number for an
accurate density and gradient estimate, see Figs.~4 and 5 in \cite{rosswog15b}, they are exceptionally good
at keeping an ordered particle distribution\footnote{For an illustration see, e.g., Fig.~2 in \cite{rosswog15c}.} 
and in suppressing sub-resolution noise. We had experimented with other high-order kernels \citep{rosswog15b}
and in static density and gradient estimation tests on fixed particle lattices they actually showed better accuracy
than the Wendland kernels.
In dynamic tests, however, the Wendland kernels delivered more ordered particle distributions and were
overall the better choice. To illustrate this, compare the high-order $W_{\rm h,9}$ and a $C^6$ Wendland kernel 
(WC6)\footnote{In the mathematical literature this kernel is referred to as  $\Phi_{3,3}$.} in Figs. 4 
and 11 in \cite{rosswog15b}.  For these reasons we 
choose the WC6  kernel as a default
\be
WC6(q)= \frac{\sigma}{h^3} \left( 1 - q \right)^8_+ \left(32 q^3 + 25 q^2 +  8 q + 1 \right),
\ee
where $\sigma= 1365/(512 \pi)$ and $q= r/(2 h)$. Note that for practical purposes/consistency
with the other kernels in our module, we have written the kernel in a form so that it vanishes
at $r/h= 2$. At every sub-step in the ODE integration we set the  smoothing lengths so that there are exactly
300 contributing neighbours within the $2 h_a$-support of each particle $a$, see Sec.~\ref{sec:adapt_h}.
 This neighbour number is motivated by the tests shown in Figs. 4a and 5
 in \citet{rosswog15b}. All tests in this paper use this combination of kernel and neighbour number.
\begin{figure*}
\centerline{
\hspace*{-0.5cm}{\includegraphics[width=0.65\textwidth,angle=0]{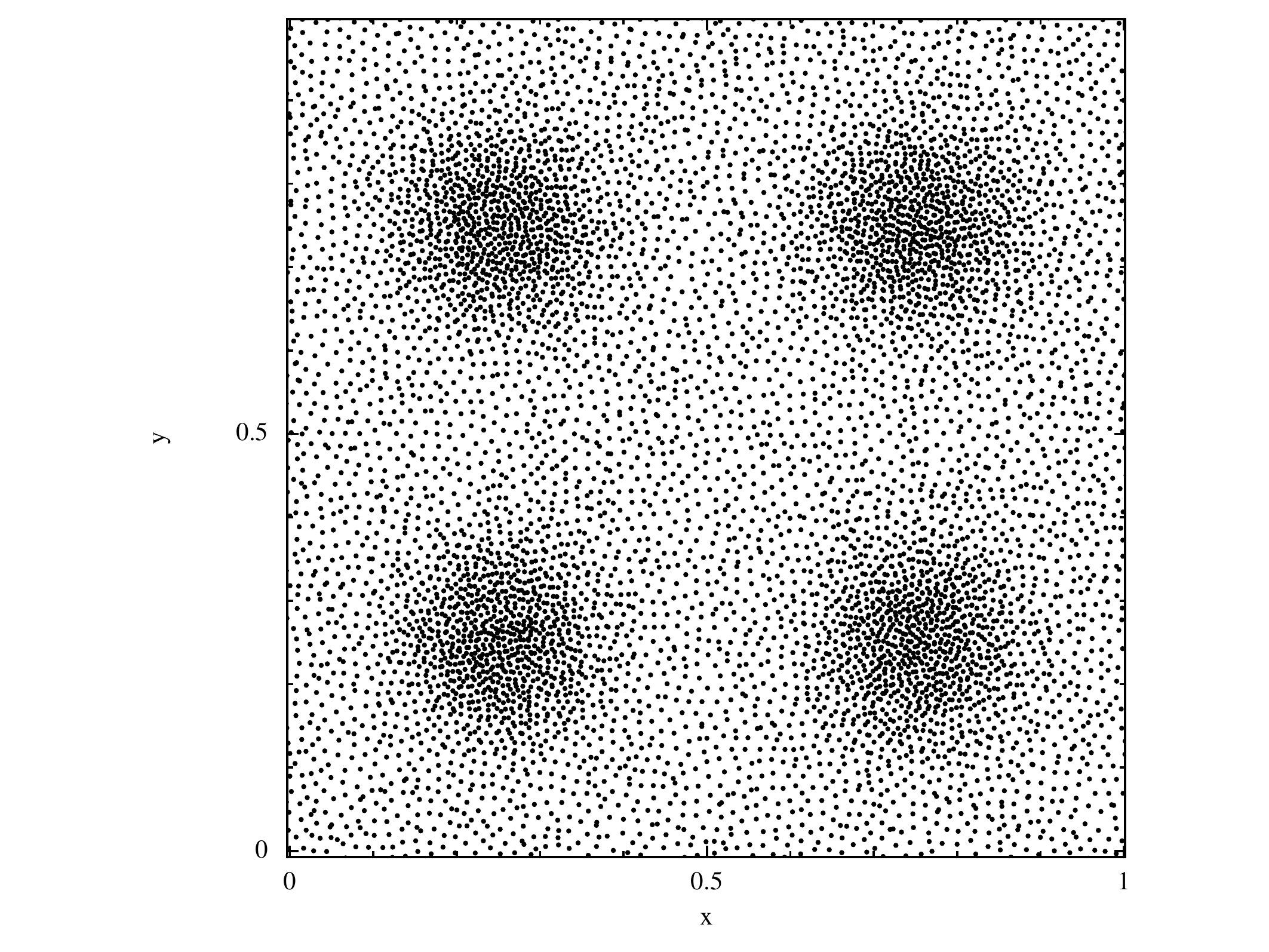}}\hspace*{-2cm}
\hspace*{-0.5cm}{\includegraphics[width=0.65\textwidth,angle=0]{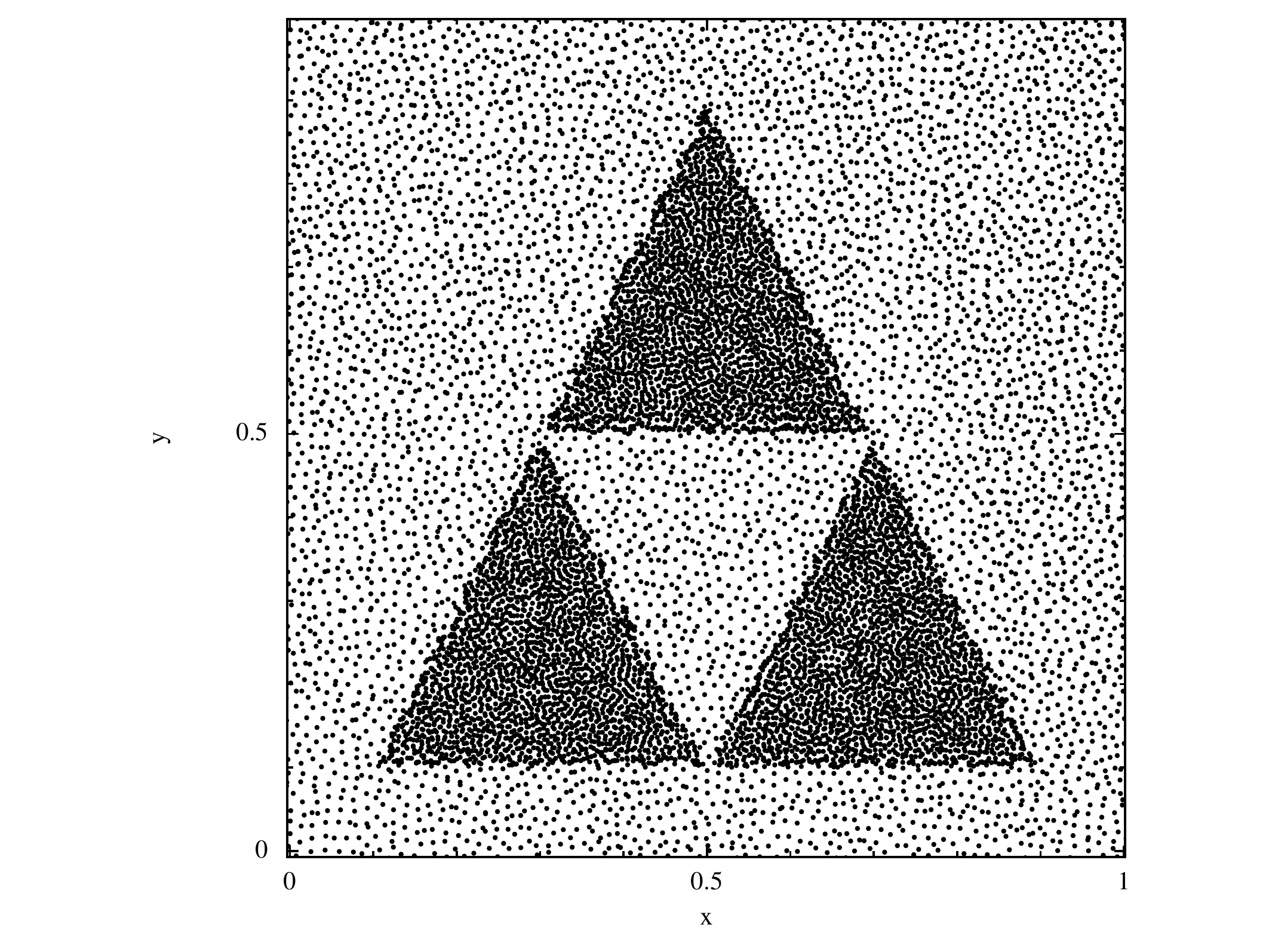}}
}
\caption{Two examples of  "glass-like", equal-mass particle distributions that have been
constructed with the "artificial pressure method" (APM). In this method, particles are assigned
an artificial pressure value based on their current density error and, driven by hydrodynamics-type
accelerations, gradients in these errors make the particles settle into a configuration that minimises this  error.
Left: superposition of four smooth
Gaussian pulses, right: three triangles with enhanced density in their interior and sharp density transitions.}
\label{fig:APM}
\end{figure*}
%
\begin{figure*}
\vspace*{-2cm}
\centerline{
\includegraphics[width=17cm,angle=0]{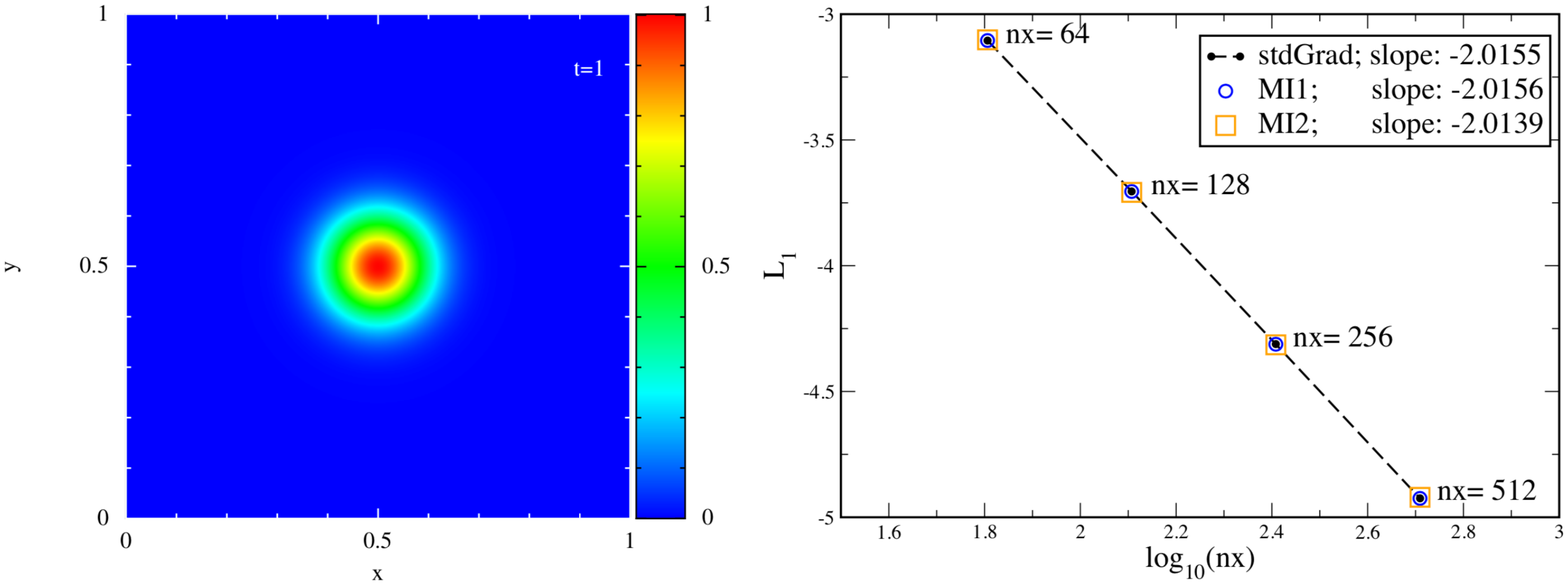}
}
\vspace*{-3cm}
\caption{Advection of a smooth Gaussian pulse. Left: density distribution, right:
$L_1$-error as function of number of points in x-direction $n_x$ for the kernel gradient (stdGrad)
and the matrix-inversion versions (MI1, MI2). All three converge very close to the expected second order.}
\label{fig:Pulse_advection}
\end{figure*}
%
%
\begin{figure}
\vspace*{-1cm}
\hspace*{-0.3cm}\includegraphics[width=1.\columnwidth,angle=0]{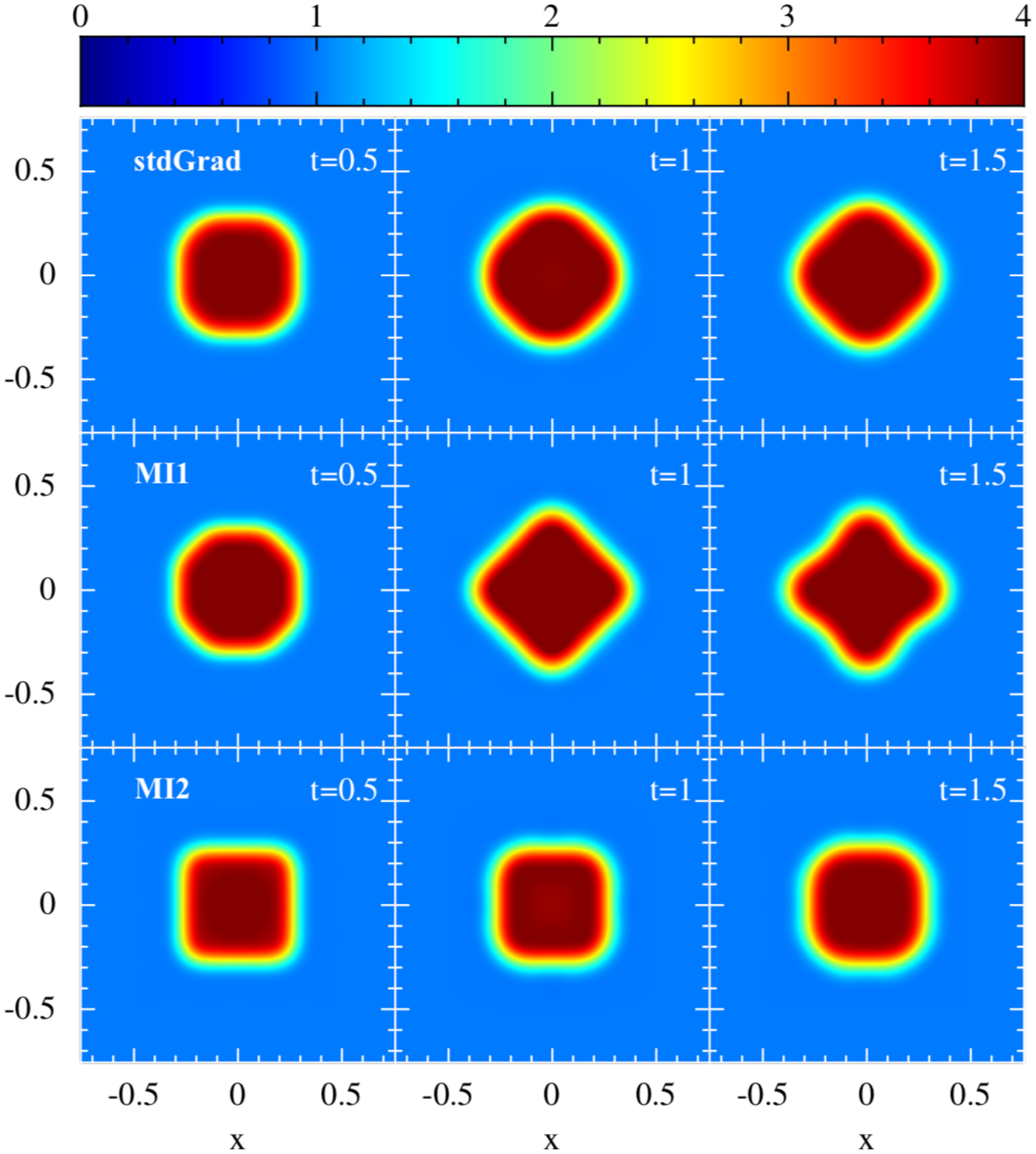}
\vspace*{-1.5cm}
\caption{Results of a surface tension test (density) similar to \citet{saitoh13}. The upper row
refers to the SPH-version with standard gradients, row two to the matrix inversion formulation 1,
the lower row to the matrix inversion formulation 2 with a different symmetrisation of the SPH equations.}
\label{fig:Surf_tens}
\end{figure}
%
\begin{figure}
\vspace*{-0.2cm}
\hspace*{-0.3cm}\includegraphics[width=1.\columnwidth,angle=0]{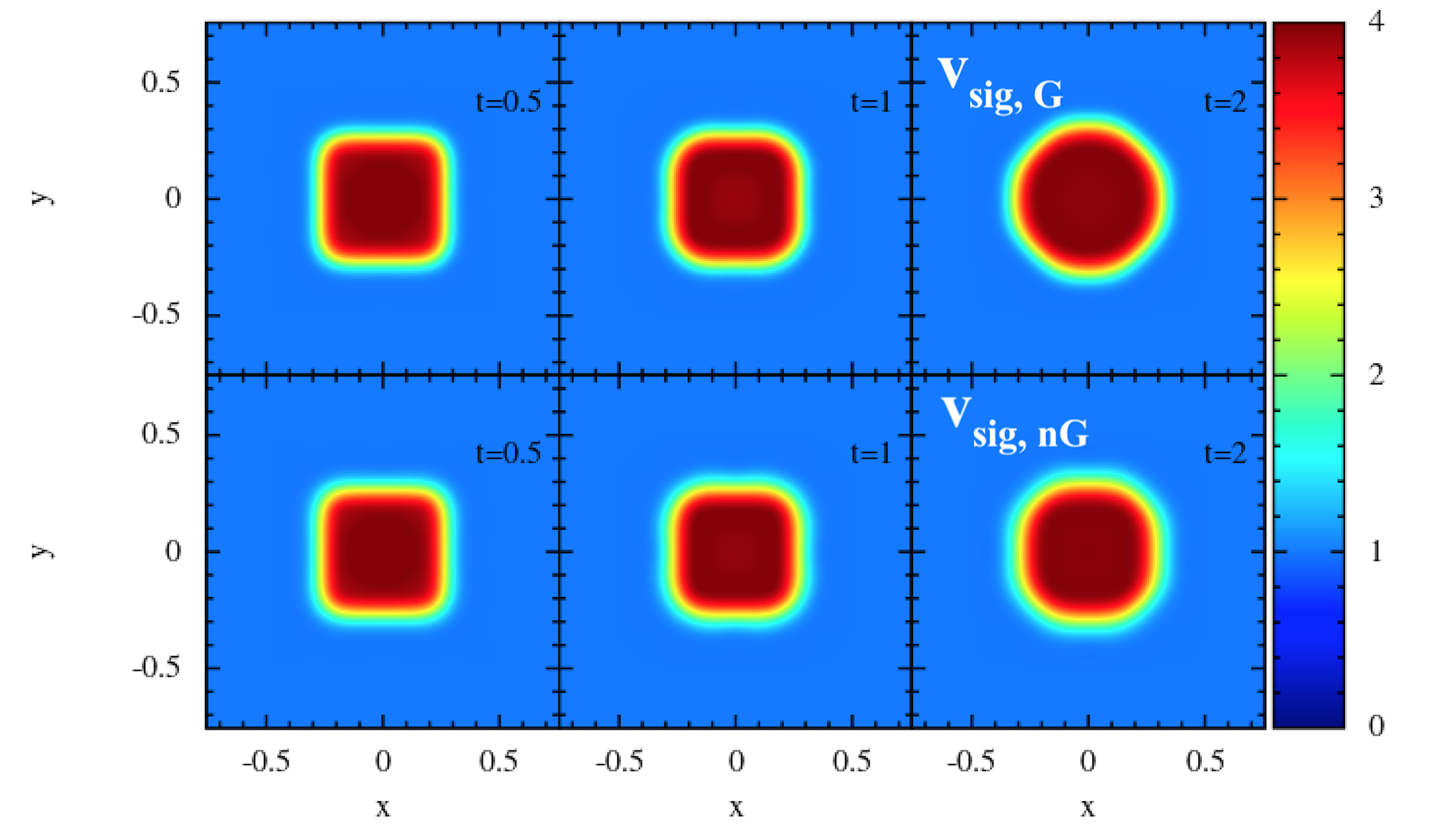}
\vspace*{-0.3cm}
\caption{Same as previous figure, but the impact of varying the conductivity signal speed is shown (MI2).
The top row uses $v_{\rm sig,G}$, the bottom row $v_{\rm sig,G}$, see Sec.\ref{sec:conduct}.}
\label{fig:Surf_tens2}
\end{figure}

%
\subsection{Neighbour search and gravitational forces}
We use a non-recursive tree that is based on "recursive coordinate bisection" (RCB)  
\citep{gafton11} to search for neighbour 
particles and to calculate gravitational forces. The main idea is to start from a
cuboid that contains all the particles and then one recursively splits the longest side of
each cuboid so that it contains (to high accuracy) the same number of particles
in each resulting daughter cell. The procedure is repeated until, on the deepest
level of the tree, each cell contains no more than $N_{ll}$ particles. Such cells
are referred to as lowest-level or ll-cells. We use $N_{ll}=12$, but as shown in 
our original paper \citep{gafton11}, the results are not very sensitive to  this choice. Our tree 
yields very simple integer relations between different nodes in
the tree, so that nodes can be addressed via simple integer arithmetics. The tree is not
walked down for each particle, but instead only for each ll-cell. This means
that the number of required tree-walks to find the neighbour particles is reduced
by a factor of $\approx N_{ll}$. As a result of a neighbour tree walk, the tree returns
a list of potential neighbour particles ("candidates"). Also for gravity, we only descend the
tree for the centre of mass of each ll-cell, the forces at the particle positions within the
ll-cell are obtained via a high-order Taylor-expansion. As shown in our original paper,
this tree-build is for $4 \times 10^6$ particles approximately 30 faster than the Press tree
\citep{benz90b}. Neighbour search and gravity are at this particle number about a factor
of 6 faster with our RCB with the discrepancy becoming increasingly larger for higher
particle numbers $N$. Our RCB-tree scales close to $O(N)$ while the Press tree
scales like most tree methods proportional to $O(N \log N)$.\\
Similar to the hydrodynamic equations, one can also derive the gravitational
accelerations consistently from a Lagrangian \citep{price07a} and the resulting 
equations contain corrective, "gravitational grad-h terms". Consistent with our treatment 
of  hydrodynamics, we also neglect the grad-h terms  here, so that the gravitational 
acceleration reads
\be
\left( \frac{d\vec{v}_a}{dt}\right)_{\rm grav} = -G \sum_b m_b \left[ \frac{\varphi'_{ab}(h_a) + 
\varphi'_{ab}(h_b)}{2}\right] \hat{e}_{ab},
\ee
where $\hat{e}_{ab}= (\vec{r}_a - \vec{r}_b)/|\vec{r}_a - \vec{r}_b|$ and the gravitational
potential that is used for monitoring the total energy
\be
\Phi_a= G \sum_b m_b \varphi_{ab}(h_a).
\ee
The gravitational smoothing kernel for the force, $\varphi'$, and for the gravitational
potential, $\varphi$, can be calculated directly from the density kernel $W$ via
\be
\varphi'(r,h) = \frac{4 \pi}{r^2} \int_0^r W(r',h) r'^2 dr'
\ee
\bea
\varphi(r,h) = 4 \pi &&\left[ -\frac{1}{r} \int_0^r W(r',h) r'^2 dr' + \int_0^r W(r',h) r'dr'  \right. \nonumber \\
                   && - \left. \int_0^{Qh} W(r',h) r' dr' \right].
\eea
For commonly used density kernels $W$ these integrals can be solved analytically. However,
to make it possible to change $W$ with minimal  modifications in the code, we calculate $\varphi'$
and $\varphi$ numerically and tabulate all kernels, so that for a different choice of $W$ automatically the
consistent kernels $\varphi'$ and $\varphi$ are available for force and potential calculation.
As stressed in our original paper,  large speed gains  have been obtained by mapping the tree
variables into a "tree-vector" in exactly the order in which they are addressed during the tree-walk.
\begin{figure*}
\hspace*{0cm}{\includegraphics[width=18cm,angle=0]{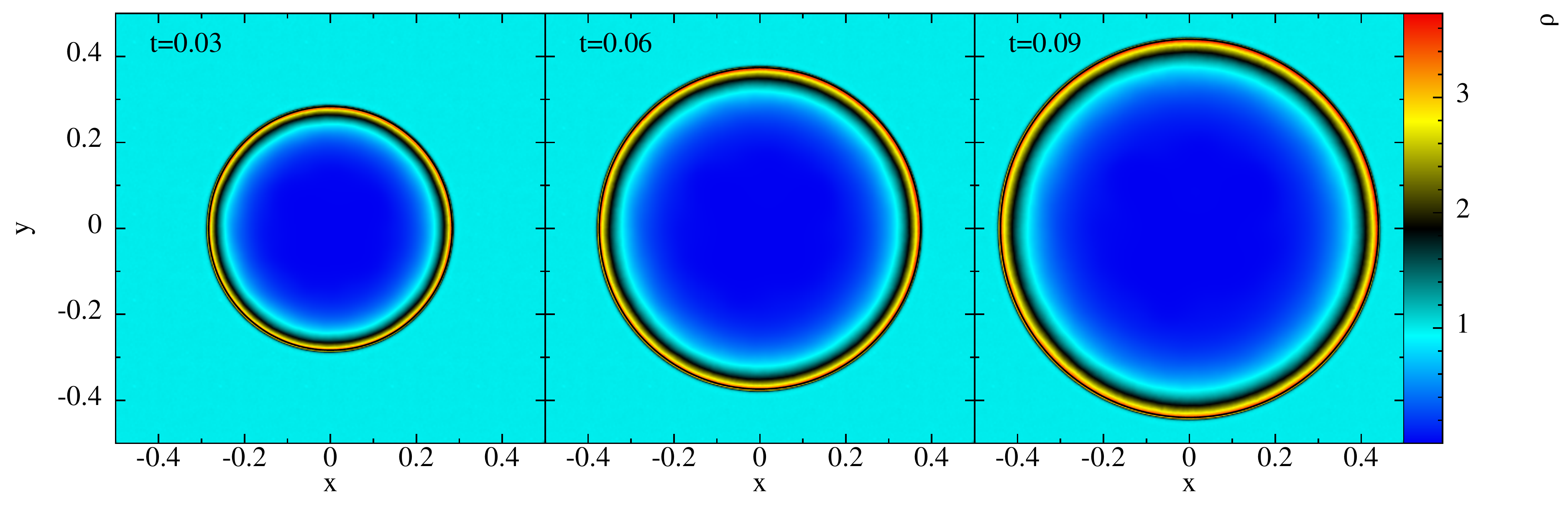}}
\caption{Time evolution of a 3D  Sedov blast ($256^3$ particles). Colour coded is the density in $xy$-plane,
              the black circle at the leading edge of the blast is the analytical shock position.
              Note the absence of any visible deviations from spherical symmetry.}
\label{fig:Sedov_256}
\end{figure*}

\begin{figure}
\hspace*{0cm}{\includegraphics[width=\columnwidth,angle=0]{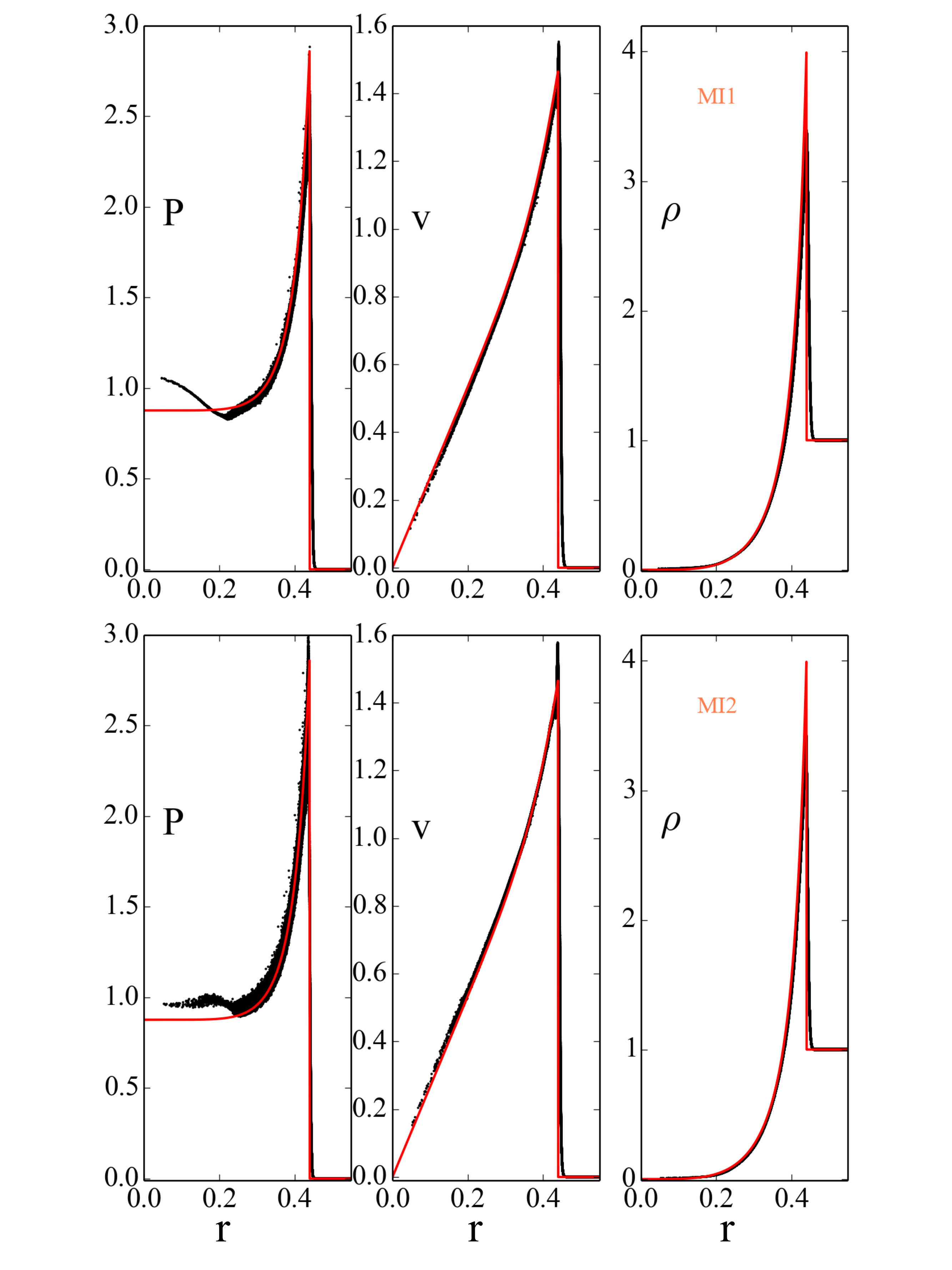}}
\caption{Sedov blast  ($200^3$ particles) for different SPH-symmetrisations (matrix inversion formulation MI1 and MI2).
The red lines denote the exact solution, the black dots are the numerical \Ma results.}
\label{fig:Sedov_F3_GDF}
\end{figure}

\subsection{Adaption of smoothing lengths}
\label{sec:adapt_h}
The so-called "grad-h terms" \citep{springel02,monaghan02} have  been introduced 
in Newtonian, special- \citep{rosswog10b} and general-relativistic SPH \citep{rosswog10a}
to  ensure exact conservation.
For pragmatic reasons, however, we set them here to unity since a consistent update of
smoothing length and density requires an iteration between the two and this comes at 
a non-negligible computational expense. More importantly for us, our
earlier experiments \citep{rosswog07c}  showed that while 
the conservation properties improved somewhat, these were rather small corrections
to an already excellent conservation. Even in a violent 
head-on collision of two stars, historically considered a "worst case scenario" for energy 
conservation in SPH \citep{hernquist93}, the relative energy conservation was better  than $\approx 10^{-3}$,
even when ignoring the grad-h terms. These results are consistent with those presented below,
see Sec.~\ref{sec:stellar_collision}.
Finally, and most relevant for our decision to {\em not} use an iteration between $\rho$ and $h$, is that
we kept running into numerical problems when an expanding flow suddenly encounters
a sharp surface with a very large  number density of SPH particles. Such an example, the merger of
two neutron stars with 1.3 and 1.4 \Msun is shown in Fig~\ref{fig:particle_dist}: the density of the
first particles that flow over the inner Lagrange point towards the heavier neutron star
drops rapidly, thereby causing a strong increase in the smoothing lengths. Once close enough, 
the particles suddenly encounter the other neutron star
with an enormous particle number density.  It is a challenge to assign a good value for the 
smoothing length $h$ of such front particles since a tiny change in $h$ can easily change the 
neighbour number by an order of magnitude.  This can lead to problems with the size of neighbour
lists or --in case counter measures are taken-- this can lead to very erratic changes of the smoothing
length of this particle and therefore to a substantial amount of numerical noise.\\
To avoid such problems, we {\em assign to each particle an exact neighbour number before its force
is calculated}. Consistent with our SPH equations, we consider as  "neighbours" of particle $a$
all those particles that are in the kernel support of $2 h_a$, where $h$ is the smoothing length.
We first build our RCB-tree with smoothing lengths that are 10\% larger than those from
the previous time step. A neighbour tree walk then returns a substantially longer candidate list for
each "lowest-level cell" 
than the desired neighbour number $n_{\rm des}$ of each particle. From this candidate list the particle 
with the $(n_{\rm des} + 1)$th largest distance, $d_a^{n_{\rm des}+1}$, to the particle of interest $a$ is selected via  a 
partitioning  algorithm \citep{press92} and this distance sets  the smoothing length: $h_a= 0.5 d_a^{n_{\rm des}+1}$
(keep in mind that all our kernels are scaled so that they have a support size of $r/h=2$).\\
Although this algorithm may seem at first sight very computationally expensive, it is actually not: 
if we calculate all the derivatives needed for an SPH simulation for the case of a star made of $5 \times 10^6$ particles, 
the assignment of the smoothing length takes only $\approx 7$ \% of the total time to calculate the derivatives
(usually dominated by self-gravity). 
Apart from being very robust in extreme situations, this
procedure has the additional advantage that  the smoothing lengths evolve very smoothly and without 
introducing unnecessary noise. 

\subsection{Time integration}
\label{sec:integ}
We perform all shown tests with the total variation diminishing (TVD) second order Runge-Kutta (RK2)
method \citep{gottlieb98}
\bea
y^\ast &=& y^n + \Delta t f(y^n)\\
y^{n+1}&=& \frac{1}{2} \left[y^n + y^\ast + \Delta t f(y^\ast) \right].
\eea
If desired, the derivatives $f(y^\ast)$ can be ``recycled'' for the next prediction step, 
so that only one derivative is effectively calculated per time step.  We have
implemented this option as a simple switch, so that we can choose whether one or
two derivatives are calculated per time step. We have explored the accuracy of this
"force-recycling" in practical tests, but have not found noticeable differences in any of them.
Nevertheless, we use the  TVD RK2 integrator with two derivative calculations  in the tests
presented below.\\
The time step is chosen as  minimum from a "force-" and "Courant-criterion"
\citep{monaghan92}
\be
\Delta t= C \; \rm{min}(\Delta t_f,\Delta t_C)
\label{eq:t_step}
\ee
where 
\bea
\Delta t_f &=&    {\rm min}_a \left( \sqrt{\frac{h_a}{|\vec{f}_a| }} \right), \\
\Delta t_C &=& {\rm min}_a \left(  \frac{h_a}{c_a + 0.6 \alpha(c_a + 2 \tilde{\mu}_a)}   \right),
\eea
 where $\tilde{\mu}_a= {\rm max}_b \left[h_a \tilde{v}_{ab}^\delta r_{ab}^\delta/(r_{ab}^2 + 0.01)\right]$ and
 for the prefactor we choose $C= 0.2$. While being very simple and efficient, 
this time integration algorithm provides excellent numerical conservation of energy 
and angular momentum, as we will show below.  \\
We restrict ourselves for now to a global time step for all particles.
Obviously, for problems with a large range of different possible time steps
among the particles, one can substantially reduce the computing time by allowing
for individual time steps. This comes, however, at a price. It makes the code
more involved  due to the time step book-keeping, it deteriorates the conservation
properties and if the  time step bins between neighbouring particles are not restricted
properly, particles can "be surprised", say, by an approaching blast wave and this  can 
lead to wrong results \citep{saitoh09}. Individual time steps make it
also more cumbersome to remove particles, say, in an accretion process. For all these
reasons we stick for now with a global time step, but if future problems will require it,
we will implement an integration scheme with individual time steps.

\subsection{Implementation}
 \Ma has been written from scratch in clean and modular Fortran 95/2003. 
We have paid particular attention to separate technical infrastructure (such as the
tree for neighbour search)  from the physical modules. 
\Ma uses exclusively double precision
and much attention has been payed to keep the "hot loops" fast. This is of particular importance for \Ma
since due to the large neighbour number  the density and hydrodynamic derivative loops become computationally
very expensive. The strategy is to analyze in which order different variables are addressed in the 
expensive loops and to map them exactly in this order in "cache arrays", very similar to how this is done
within our RCB-tree, see for example Sec. 2.2.1 in \cite{gafton11}.  To keep the code clean and (relatively)
simple, we have so far only included a global time step for all particles. At the current 
stage, \Ma is parallelised with OpenMP and is able to perform tests (global time step, 2nd order Runge-Kutta 
and 300 neighbour particles) with $\sim 10^8$ SPH particles  when no self-gravity is involved and a few $10^7$ otherwise 
(e.g. on Intel Skylake Gold compute nodes).  Further performance improvements are a subject for the future. 
%
\begin{figure}
\includegraphics[width=\columnwidth,angle=0]{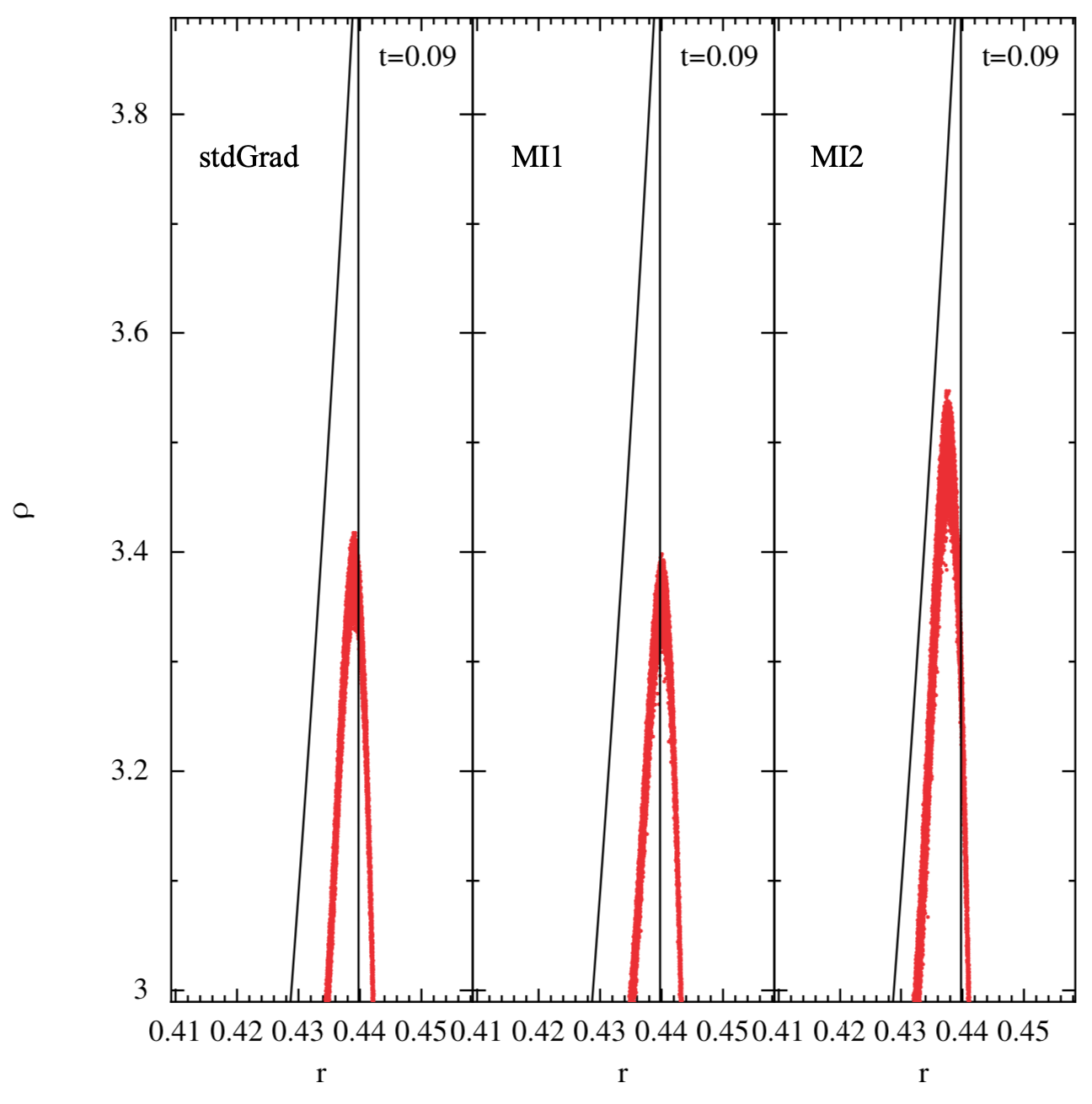}
\caption{Comparison of the density peak hight ($200^3$ particles) for the kernel- (stdGrad) and the two 
matrix-inversion formulations (MI1, MI2).}
\label{fig:Sedov_density_peak}
\end{figure}
%
\begin{figure*}
\hspace*{0cm}{\includegraphics[width=18cm,angle=0]{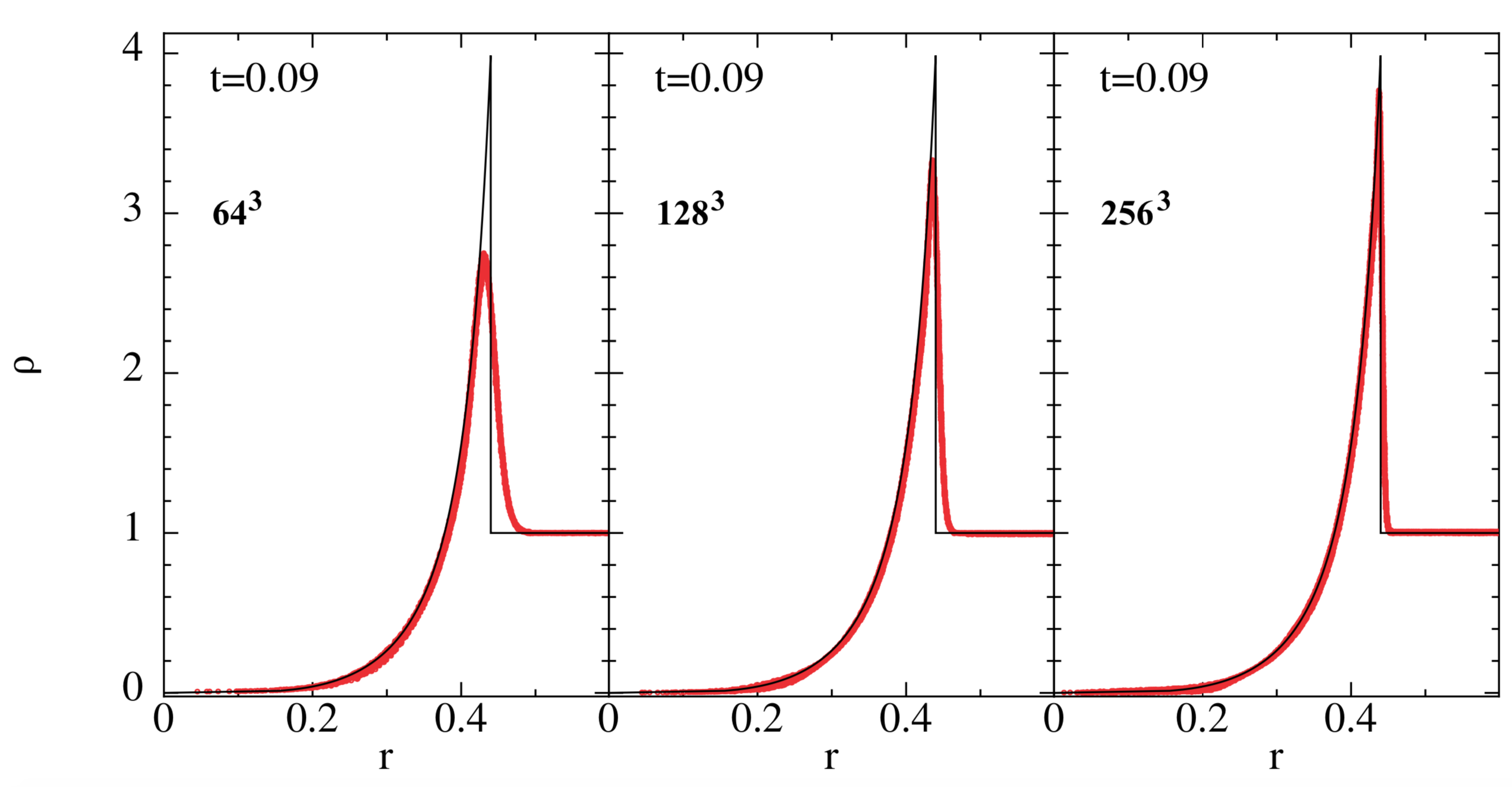}}
\caption{3D Sedov blast of increasing resolution ($64^3, 128^3, 256^3$ particles; MI2 formulation), every single particle is plotted.}
\label{fig:Sedov_64_128_256}
\end{figure*}

\begin{figure*}
\hspace*{0cm}{\includegraphics[width=0.9\textwidth,angle=0]{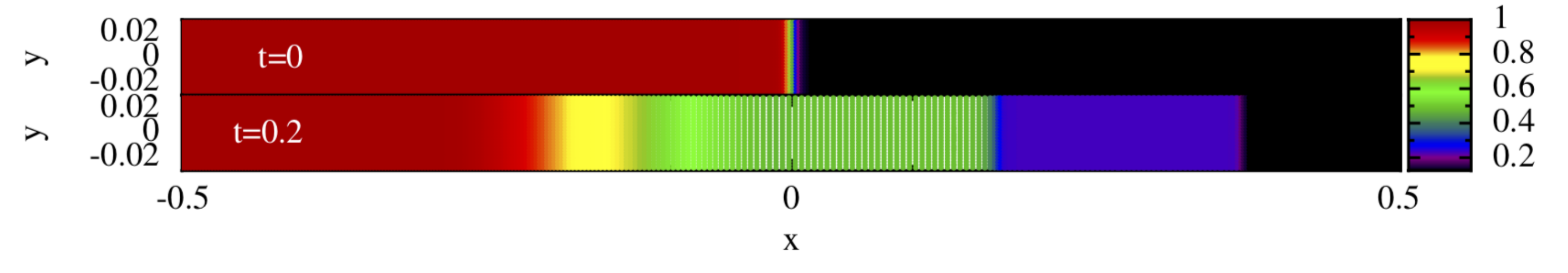}}
\hspace*{0cm}{\includegraphics[width=0.9\textwidth,angle=0]{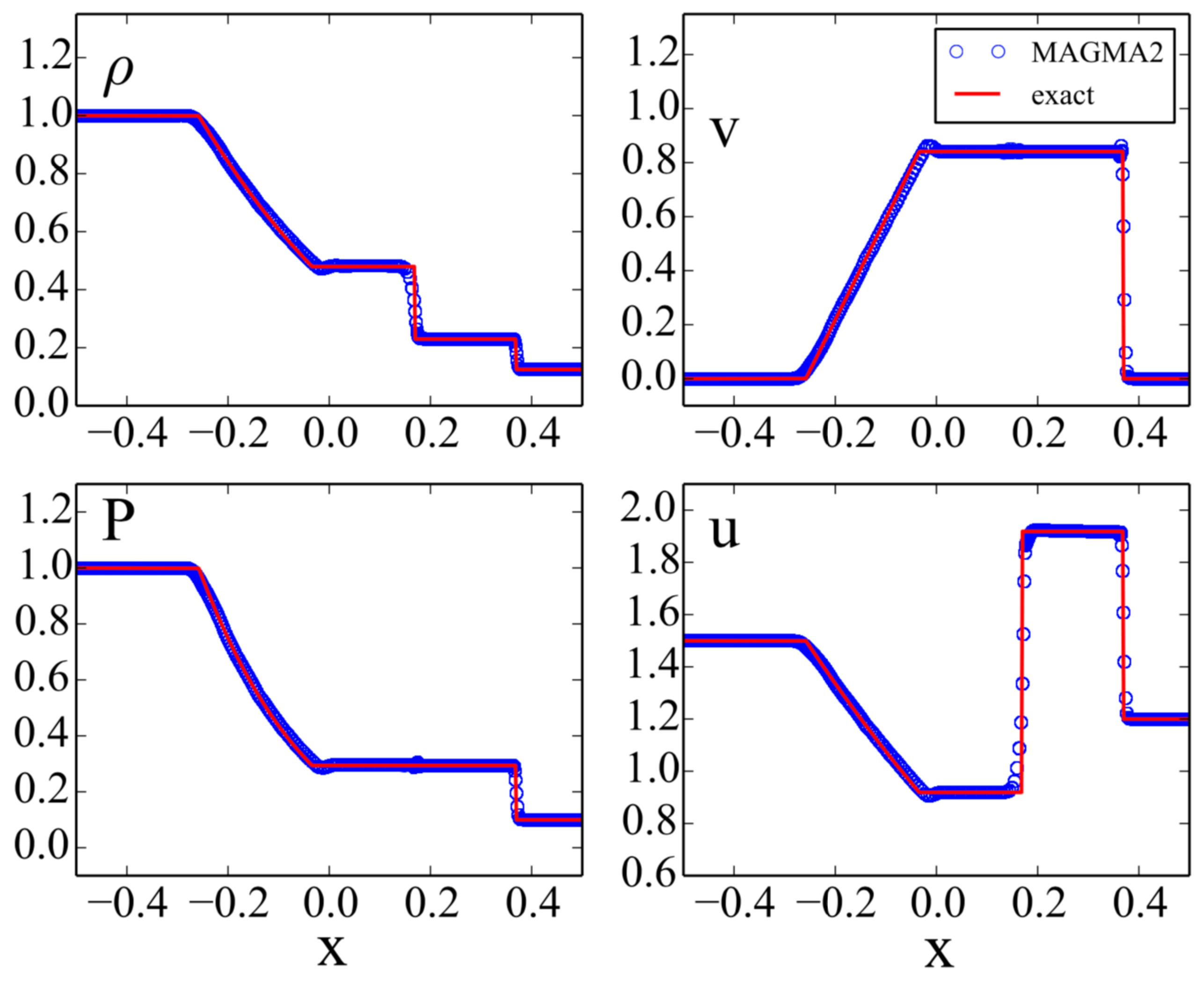}}
\caption{Sod shock test performed in 3D with $ 400 \times 24 \times 24$ particles.
The upper  two panels shows the particle distribution (density is colour-coded) at t=0 and t= 0.2.
The four panels below show (upper left to lower right) density, velocity, pressure and
internal energy of all particles at t= 0.2 together with the exact solution (red). All particles are plotted.}
\label{fig:Sod3D}
\end{figure*}
\section{Tests}
\label{sec:tests}
We begin with an advection test to measure the order of convergence
in smooth flows.
We then show a number of shocks, a Kelvin-Helmholtz and a
Rayleigh-Taylor instability test. We further show combined tests
where shocks go along with vorticity creation ("Schulz-Rinne tests") and we conclude
with astrophysical tests, a stellar collision and a tidal disruption, 
that demonstrate \ma's robustness for practical applications.\\
All of the tests shown below are performed with the full 3D code.
"2D" tests are performed by simulating a slice that is thick enough
for the central plane not to be affected by edge effects.

\begin{figure}
\vspace*{0cm}
\hspace*{-0cm}\includegraphics[width=\columnwidth,angle=0]{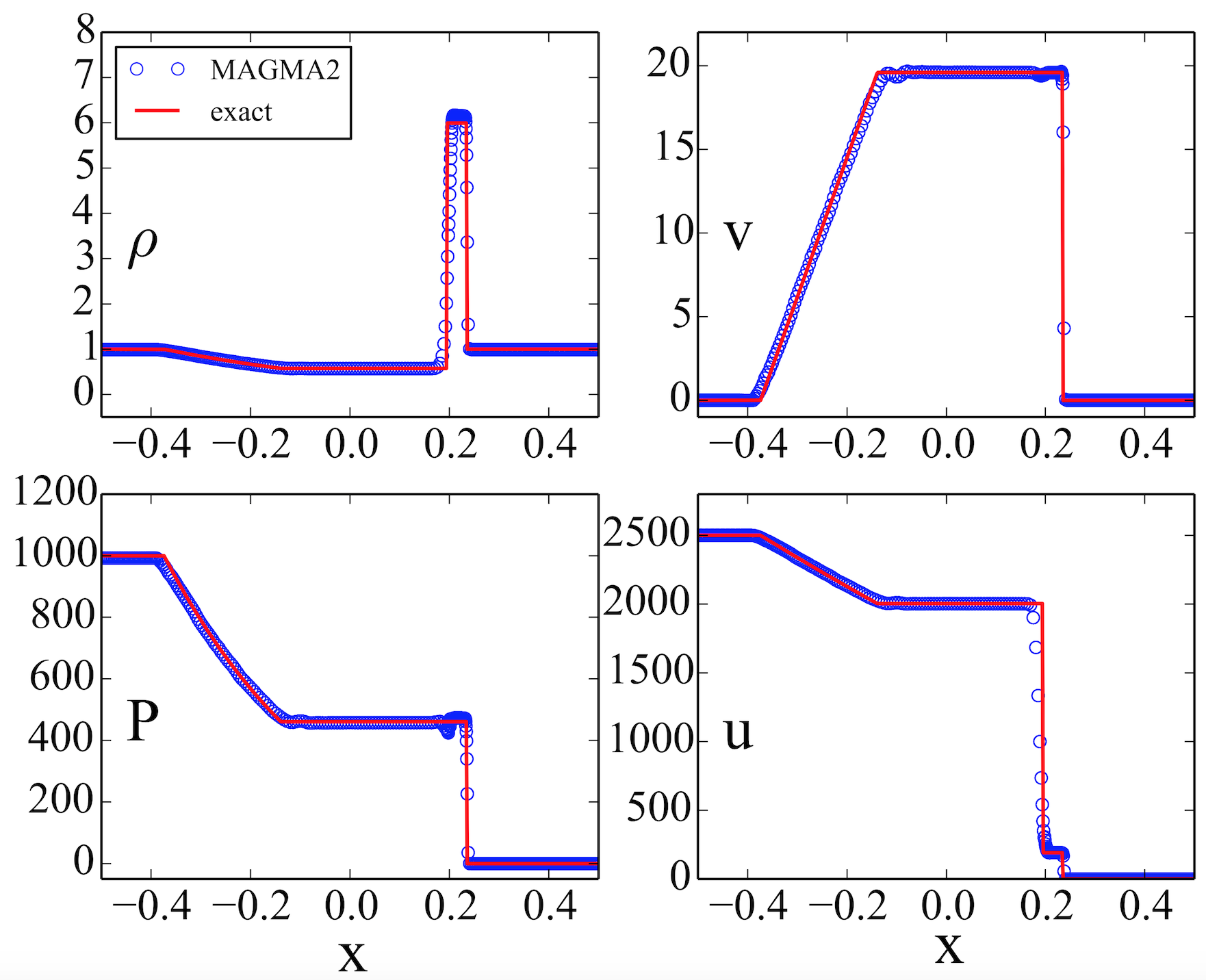}
\vspace*{-0.5cm}
\caption{3D strong blast wave test at $t=0.01$ ($800 \times 24 \times 24$ particles).} 
\label{fig:strong_blast}
\end{figure}

\subsection{Initial conditions: the Artificial Pressure Method (APM)}
\label{sec:APM}
As pointed out earlier \citep{rosswog15b} good initial conditions
are crucial for obtaining accurate results in SPH simulations, but 
unfortunately it is sometimes non-trivial to construct initial conditions 
with all the desired properties. The particles should be distributed with 
high regularity so that they guarantee a high interpolation accuracy 
(see, e.g., Sec. 3 of \cite{rosswog15b} for an accurate interpolation 
quality indicators). This is, however, not enough, since the particle 
distribution should also not contain any preferred directions, but 
simple lattices usually do. This may lead to artefacts such
as the piling up of particles in a shock ("particle ringing") 
along the grid direction, see   Fig. 17 in  \cite{rosswog15b} for an 
illustration.\\
Setting up SPH-initial conditions for a predescribed non-constant density distribution can become
challenging. The simplest approach is to start with some regular lattice, say a cubic lattice (CL) 
or (better) a  "close-packed" (CP) lattice,  and assign particle masses
so that the desired density is reproduced. For small density differences this delivers acceptable results,
but  for large density contrasts this results in particles with very different masses and this
is known to introduce unwanted noise into simulations, see e.g. \cite{lombardi99}. Setting up configurations 
with equal mass particles would be desirable, but it is substantially more challenging since the particles
need to be placed in a way so that their local number density reflects the desired mass density. 
For simple cases, stretching a uniform lattice to the desired density distribution can be used, e.g. 
\cite{rosswog09a,price18a}. For more general cases methods based on
Centroidal Voronoi Tesselations have been suggested \cite{diehl12}.\\
Here we suggest a novel approach that is very close to the spirit of SPH. The idea is to start from 
an SPH-like momentum equation that uses artificial pressures, $\tilde{P}_a$,  which  are based
on the current density error. If the density estimated by all particles is in perfect agreement
with the desired density profile, all particles have the same pressure and will therefore not feel a net force.
 If in contrast, density errors exist, then the
gradients in the artificial pressures drive the equal mass particles into positions that minimise
density errors. Rather than actually integrating this artificial equation, we use a Courant-type
time step to translate the ordinary differential equation into  a position update formula for each 
particle, $\vec{r}_a \rightarrow  \vec{r}_a + \Delta \vec{r}_a$. Once a particle has reached an 
optimal position, the pressure gradient vanishes and it stays at the reached position.\\
Assume that we start from an initial distribution of equal mass particles whose mass $m$ has been
calculated from the desired density profile $\rho^P(\vec{r})$ and the number of particles. 
As a next step we measure the densities at the particle positions, $\rho_a$, via 
Eq.~(\ref{eq:dens_sum}) and then assign to each particle an artificial pressure based on its relative density
error:
\be
\tilde{P}_a= P_0  \left(\rm{max}\left[1 + \frac{\rho_a - \rho^P(\vec{r}_a)}{\rho^P(\vec{r}_a)}, 0.1\right]\right)
\equiv P_0 \;  \Pi_a,
\ee
where $P_0$ is a so far arbitrary base pressure and the max-function avoids negative pressures
as they might otherwise occur for  bad initial guesses. This means that particles with too low (too high)
density estimates have  smaller (larger) pressures and therefore more particles will move into (out of)
this region. With these artificial pressures we construct an SPH-type momentum equation (similar to
Eq.~(\ref{eq:dvdt_IA2}))
\begin{equation}
\vec{f}^{\rm APM}_a= - m P_0 \sum_b \frac{\Pi_a+\Pi_b}{\rho_a\rho_b} \nabla_a W_{ab}(h_a),
\label{eq:f_APM}
\end{equation}
and we now choose a Courant-type time step
\begin{equation}
\Delta t^{APM}_a \propto \frac{h_a}{c_{{\rm s}, a}} \propto h_a \sqrt{\frac{\rho_a}{P_0}},
\end{equation}
where we have used the polytropic relation for the sound speed $c_s= \sqrt{\Gamma P/\rho}$.
We obtain a dimensionally correct update formula by multiplying our acceleration formula Eq.~(\ref{eq:f_APM})
by $(\Delta t^{APM}_a)^2$ so that the final position correction becomes
\begin{equation}
 \Delta \vec{r}_a = - \xi h_a^2 m \sum_b \frac{\Pi_a + \Pi_b}{\rho_b} \nabla_a W_{ab}(h_a).
 \label{eq:position_update}
 \end{equation}
Note that the base pressure $P_0$ has dropped out. The prefactor  $\xi$ is not critical for the iteration
process, but has some impact on how quickly the desired density profile is reached. After some experimenting
we settled on a value of $\xi=$ 0.5.\\ 
During the iteration process  we monitor the maximum and average density error and 
\begin{equation}
\delta PU_a \equiv 1 -  \sum_b \frac{m_b}{\rho_b} W_{ab}(h_a),
\end{equation}
to see by how much the particle distribution deviates from a perfect partition of unity at each particle position, see 
e.g. Sec. 2.1 in \cite{rosswog15c} for a discussion of interpolation quality. As an example, in our setup of
 the Kelvin-Helmholtz problem, see below,
we find after 500 iterations typically average density errors of a few times $10^{-3}$ with maximum devdeviationsations
in transition regions of very few percent.  The average $\delta PU_a$ values are at this point
below $10^{-5}$, so that we have an excellent interpolation quality.
To further improve the agreement between the set up and the desired density profile after the iteration process 
has stopped, we measure the local SPH particle number density
\begin{equation}
n_a = \sum_b W_{ab}(h_a)
\end{equation}
and assign final particle masses $m_a= \rho^P(\vec{r}_a)/n_a$, so that the setup contains
particles that are to within percent level of the same mass.\\
To illustrate the versatility of this method, we set up a non-trivial density profile with equal mass
SPH particles. As density profile we choose
\be
\rho^P(x,y)= \Delta \rho \sum_{j=1}^4 \exp\left\{ -\frac{(x-x_j)^2 + (y-y_j)^2}{\sigma^2}\right\}+ \rho_0,
\ee
where $\rho_0=1$, $\Delta \rho=5$, $\sigma= 0.1$.
We start by placing 20 layers  (in z-direction) of 100$\times$100 particles in [-0.2,1.2] $\times$ [-0.2,1.2]
and we place six layers of  "frozen" particles (which are not updated in the iteration process) as boundary
conditions in the surrounding volume. The centres of the density pulses, $(x_j,y_j)$, are offset from
the corners by 0.25.
The particles are initially distributed on a cubic lattice and then randomised
to erase the undesired lattice structure.
The particle distribution ($|z| < 0.02$) after 1000 APM-iterations is
shown in Fig.~\ref{fig:APM}, left panel. The particles have settled into a "glass-like" structure with
their number density reflecting the mass distribution.\\
As a second example, we set up sharp density profile that the particles try to approximate
(within the limits of their finite resolution and uniform mass) as well as possible. For the density
profile we choose 
\begin{equation}
  \rho^T(x,y) =
    \begin{cases}
      \rho_0 + \Delta \rho& \text{inside the outer, but not in inner triangle}\\
      \rho_0 & \text{else,}
    \end{cases}       
\end{equation}
where the outer triangle refers to the interior of the points  $(0.1,0.1), (0.1,0.9), (0.5,0.9)$ and 
the inner triangle is given by the midpoints of the outer triangle's sides and we use
$\rho_0=1$ and $\Delta \rho=5$, as before. The particles are initially placed as 
in the previous example and their distribution after 1000 APM-iterations ($|z| < 0.02$)  is
shown in Fig.~\ref{fig:APM}, right panel. Note that in all regions the particles
have arranged into uniform glasses with sharp transitions between the different regions.\\
While the method is very flexible and powerful, it  costs some computational effort: each iteration
requires a density loop with the corresponding neighbour search. We therefore stick to the
pragmatic approach that we use simpler particle setups where this delivers good results. 

\begin{figure*}
\vspace*{-0cm}
\hspace*{-0.3cm}\includegraphics[width=18cm,angle=0]{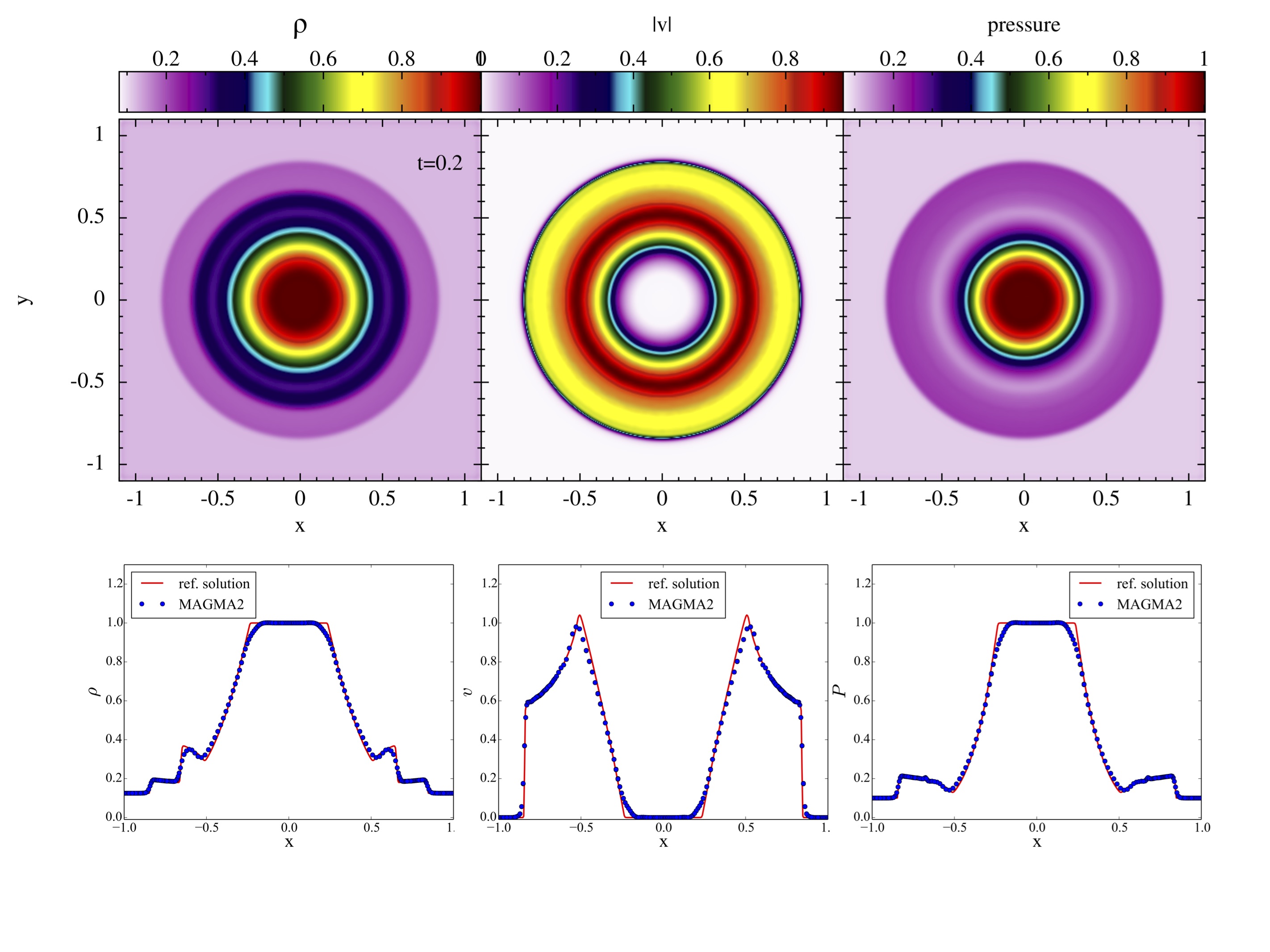}
\vspace*{-1.5cm}
\caption{Circular blast wave  problem 1 in 3D at $t=0.2$ ($200^3$ particles). Upper row: \Ma result for density, $|v|$ and pressure.
Lower row: corresponding quantities from \Ma results compared with a 
reference solution  (red line) obtained by the Eulerian  weighted average flux method  ($400^3$ grid cells)  \citep{toro99}.} 
\label{fig:Riemann1}
\end{figure*}

\subsection{Smooth advection}
\label{sec:sm_ad}
In this test we place a Gaussian density pulse of initial shape
\be
\rho^{\rm in}(x,y)= (\rho_2-\rho_1)\exp\left[ -\frac{(x-0.5)^2 + (y-0.5)^2}{\sigma^2}\right] + \rho_1,
\ee
where $\rho_1= 10^{-3}$, $\rho_2= 1$ and  $\sigma= 0.1$ in a (quasi-)2D
box with $[0,1] \times [0,1]$, see Fig.~\ref{fig:Pulse_advection}, left panel. 
The gas with polytropic index $\Gamma=5/3$ is represented by 20 CL-layers of
$n_x \times n_x$ particles. The density profile is advected with
uniform pressure $P_0=10^{-6}$ and velocity $v_x=1$ through the box with
periodic boundaries. After crossing the box once the $L_1$-error, $L_1= \sum_b^N |\rho^{\rm in}(\vec{r}_b) - \rho_b|/N$,
is measured. The $L_1$-results for different resolutions $n_x$ are plotted in the 
right panel of Fig.~\ref{fig:Pulse_advection} for all of our three variants. The measured slopes 
are always very  close to the 2nd order that is theoretically expected for our code.

\subsection{Surface tension test}
\label{sec:surface}
SPH has the peculiarity that density and internal energy can be of different smoothness: 
the density is calculated via Eq.~(\ref{eq:dens_sum}), so that even when there
is a sharp transition in the particle masses the density transition is smooth.
The internal energy equation, in contrast, is a straight-forward translation of
the first law of thermodynamics and $u$ is not necessarily smooth. If 
contact discontinuities, i.e. interfaces where both density and internal energy 
exhibit a  discontinuity, but the pressure, $P=(\Gamma-1)\rho u$, is continuous,
 are not set up carefully, spurious pressure gradients can emerge which lead to 
 unwanted "surface tension effects" \citep{springel10a,hess10}.
In the worst case, this can suppress weak instabilities \citep{agertz07}. \\
We set up a surface tension test similar to \cite{saitoh13}
and compare the performance of our three different SPH formulations. We place $40^3$ particles
on a cubic lattice with spacing $\Delta$ within $[-0.5,0.5]^3$ and assign them masses 
$m_a= \tilde{\rho}(\vec{r}_a) \Delta^3$ with a sharp transition according to
 \[
  \tilde{\rho}(x,y,z)  =  \left\{\begin{array}{lr}
       4   \quad {\rm for} -0.25 < x, y, z < 0.25 \\
       1  \quad {\rm otherwise} 
        \end{array}\right.
  \]
to test for the presence of spurious forces. Note, however, that this is
{\em not} how we would usually set up reliable simulation. We then
calculate the density according to Eq.~(\ref{eq:dens_sum}) and 
assign the internal energies according to $u_a= P_0/(\Gamma-1) \rho_a$. Following
\cite{saitoh13} we use constant pressure $P_0= 2.5$ everywhere.\\
The results for the different SPH formulations are shown in Fig.~\ref{fig:Surf_tens}. Both
the standard gradient version (top row) and the MI1 formulation show (with this somewhat pathological test setup) clear signs of
surface tension. The MI2 formulation performs by far best, but is still not entirely free
of surface tension effects (at least for this setup with a sharp mass transition). Here further progress could be obtained 
by using volume elements that are different from $V_b=m_b/\rho_b$ \citep{saitoh13,hopkins13,rosswog15b,cabezon17}.
We also show the impact of varying the conductivity signal speed, see Fig.~\ref{fig:Surf_tens2}.
Clearly, $v_{\rm sig,nG}$ shows the better result, which
argues for using the switch in Eq.~(\ref{eq:vsig_u}). We note, however, that despite small possible surface
tension effects both MI1 and MI2  formulations perform very well in the instability tests, see below.

\subsection{Shocks}
\subsubsection{3D Sedov-Taylor explosion}
The  Sedov-Taylor explosion test, a strong, initially point-like blast 
into a low density environment, has an analytic self-similarity solution \citep{sedov59,taylor50}. 
For an explosion
energy $E$ and a density of the ambient medium $\rho$, the blast wave
propagates after a time $t$ to the radius $r(t)= \beta (E t^2/\rho)^{1/5}$,
where $\beta$ depends on the  adiabatic exponent of the gas
($\approx 1.15$ in 3D for the $\Gamma=5/3$ that we are using). Directly
at the shock front, the density jumps by the strong-explosion limit factor of
$\rho_2/\rho_1= (\Gamma + 1)/(\Gamma-1)= 4$, where  the numerical value
refers to our chosen $\Gamma$. Behind the shock the
density drops quickly and finally vanishes at the centre of the explosion.\\
To set up the test numerically, we distribute a given number of SPH particles 
according to a {\em Centroidal Voronoi Tesselations} (CVT) \citep{du99} 
 in the computational volume
 [-0.5,0.5]$\times$[-0.5,0.5]$\times$[-0.5,0.5]. While this produces
 already nearly perfect initial conditions, they can be further improved
 by  additional sweeps according to Eq.~(\ref{eq:position_update}). Even if the
 differences in the particle distributions are hard to see by eye, they still
 improve the outcome of the Sedov test. We use here 500 of such sweeps.
Once the particle distribution is settled, we assign masses so that their density is $\rho=1$. 
This is done in an iterative way where we first assign a guess value
for the masses, then measure the resulting density via Eq.~(\ref{eq:dens_sum})
and subsequently correct the particle masses. The iteration is stopped once
the density agrees everywhere to better than 0.5\% with the desired value.
The energy $E= 1$ is spread across
a very small initial radius $R$ and it is entirely distributed as internal energy, the
specific internal energy $u$ of the particles outside of $R$ is entirely negligible
($10^{-10}$ of the central $u$). For the initial radius $R$ we choose twice
the interaction radius of the innermost SPH particle. 
Boundaries play no role in this test as long as the blast does not interact with them.
We therefore place "frozen" particles around the computational volume as boundary particles.\\
Fig.~\ref{fig:Sedov_256} shows the time evolution of the density for $256^3$ particles for 
our MI2-formulation (the other formulations look virtually identical). 
The overall agreement with the exact solution (shock position is indicated by a
black circle at shock front) is excellent and there are no noticeable deviations 
from spherical symmetry. 
We show in Fig.~\ref{fig:Sedov_F3_GDF} comparisons of the solutions 
(pressure, velocity and density) obtained with the MI1 and MI2 formulations,
the particle solutions are shown as black dots (downsampled by a factor of 10), the red lines indicates
the exact solutions. Note in particular the absence of density and velocity
oscillations in the wake of the shock. Such oscillations plague practically all (even modern) 
SPH-simulations of this test, see for example \cite{rosswog07c,hu14a,cabezon17,wadsley17,frontiere17}, but they
are virtually absent in our tests\footnote{The \Pha code paper \citep{price18a} does not show velocities, but some 
oscillations are visible in the densities.}. We attribute this  to our carefully constructed initial conditions
together with the use of a Wendland kernel  with 300 neighbour particles.\\
Despite this overall very good agreement,  a closer inspection
shows some interesting differences that are due to different SPH symmetrisations
(gradient accuracy plays a minor role in this test)
see Fig.~\ref{fig:Sedov_F3_GDF}. The MI2-formulation seems to be substantially more sensitive to density
variations which can be seen in a larger post-shock pressure variance, probably picking up
on small density variations that are residuals of our iterative approach to set up the initial 
conditions. More importantly, MI2 reaches a noticeably larger density peak (apart from
the symmetrisation everything else is exactly the same), see Fig.~\ref{fig:Sedov_density_peak}.
Note also that the SPH-peak for the more commonly used symmetrisations (stdGrad
and MI1) occurs at the shock-front while the MI2-case peaks between both lines indicating 
the exact solution. The MI2-symmetrisation result also captures the flat central pressure profile in a better way.\\
Obviously, the peak height could easily be raised by using
a lower order kernel with fewer neighbours, if one was willing to accept more post-shock noise
(which we are not).
For completeness we also show Sedov tests (with MI2) with increasing resolution
of $64^3$, $128^3$ and $256^3$ SPH particles within $[-0.5,0.5]^3$, see Fig.~\ref{fig:Sedov_64_128_256}.
\subsubsection{3D Sod shock}
The "Sod shock tube" \citep{sod78} is a classic code test to ensure the correctness
of a hydrodynamics code implementation. As initial conditions we use 
\be
(\rho,\vec{v},P)=
 \left\{
\begin{array}{l}
(1.000,0,0,0,1.0) \quad {\rm for \; \; x < 0.}\\
(0.125,0,0,0,0.1) \quad {\rm else.}
\end{array}
\right.
\ee
and $\Gamma=5/3$.
In most SPH papers this test is only shown
in 1D. Here we perform a quasi-1D test with the 3D code, employing $ 400 \times 24 \times 24$ particles
placed in $[-0.5,0.5] \times [-0.03,0.03]  \times [-0.03,0.03]$.  
Here even the simplest approach  with particles placed on  a uniform cubic lattice gives good results, see Fig.~\ref{fig:Sod3D}. 
The numerical results (all particles are shown) agree well with the exact solution, there is only a small velocity overshoot
at the shock front and small over-/undershoots at one of the edges of the rarefaction region ($x\approx -0.05$). The plot
shows the results of the MI1-formulation, there are no noteworthy differences between the different formulations for this test.

\subsubsection{3D strong blast wave}
A substantially more challenging version of the Sod test, the so-called  strong blast wave test, 
is set up with initial conditions
\be
(\rho,\vec{v},P)=
 \left\{
\begin{array}{l}
(1.0,0,0,0,1000.0) \quad {\rm for \; \; x < 0}\\
(1.0,0,0,0,0.1) \quad {\rm else}
\end{array}
\right.
\ee
and with $\Gamma=1.4$. Again we set up this test as quasi-1D with our 3D code in an analogous way
to the Sod test with $800 \times 24 \times 24$ particles 
between $[-0.5,0.5]$ in x-direction. The result (obtained with MI1; no noteworthy differences for MI2 and
stdGrad) at time $t=0.01$ is shown in 
Fig.~\ref{fig:strong_blast}. Overall, the numerical result is in very good agreement with the exact 
solution, but there are small overshoot in density and dips in pressure and velocity at the contact 
discontinuity.

\subsubsection{Spherical blast wave 1}
As another benchmark we use a three-dimensional shock-tube problem.
We follow \cite{toro99} in the parameter choice (apart from a shift of the origin): 
the computational domain is $[-1,1]^3$ and the initial conditions are chosen as:
\be
(\rho,\vec{v},P)=
 \left\{
\begin{array}{l}
(1.000,0,0,0,1.0) \quad {\rm for \; \; r < 0.5}\\
(0.125,0,0,0,0.1) \quad {\rm else.}
\end{array}
\right.
\ee
The solution exhibits a spherical shock wave, a spherical contact surface traveling in the same direction
and a spherical rarefaction wave traveling towards the origin.
We show the \Ma solution ($200^3$ particles, cubic lattice;  shown is MI1, other formulations 
nearly identical) at time $t=0.2$ in Fig.~\ref{fig:Riemann1}. We 
show in the upper row the density, $|v|$ and pressure at $t= 0.2$ in the XY-plane. In the lower row,
we compare the SPH-result ($|y|<0.018, |z|< 0.018$) with a reference solution obtained by the Eulerian 
weighted average flux method with $400^3$ grid cells  \citep{toro99}. 
The SPH solution is essentially oscillation-free and in very close
agreement with the reference solution. Only the sharp edges (e.g. in $|v|$ near$|x|\approx 0.3$) are somewhat
smoothed out and there is a small oscillation in the pressure at the contact discontinuity.

\subsubsection{Spherical blast wave 2}
\begin{figure*}
\vspace*{-0cm}
\hspace*{-0.3cm}\includegraphics[width=18cm,angle=0]{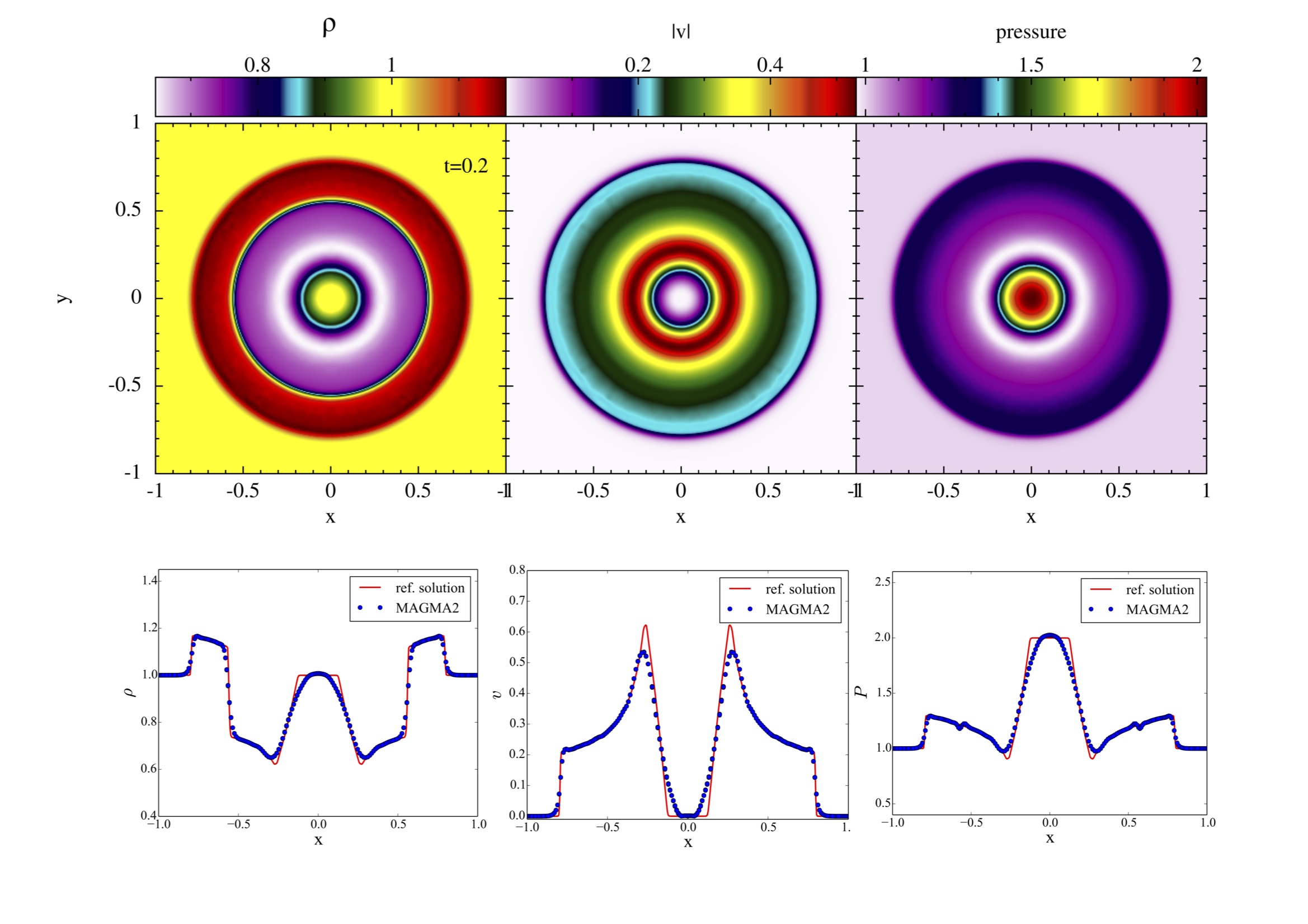}
\vspace*{-1.5cm}
\caption{Circular blast wave problem 2 in 3D at $t=0.2$ ($200^3$ particles). Upper row: \Ma result for density, $|v|$ and pressure.
Lower row: corresponding quantities from \Ma results compared with a 
reference solution  (red line) obtained by the Eulerian  weighted average flux method  ($400^3$ grid cells; \citep{toro99}).} 
\label{fig:Riemann2}
\end{figure*}

\begin{figure}
\vspace*{-0cm}
\hspace*{0cm}\includegraphics[width=\columnwidth,angle=0]{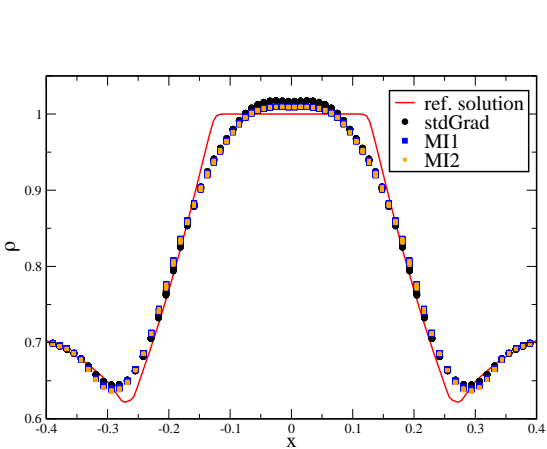}
\vspace*{-0.8cm}
\caption{Zoom into the central density region of the circular blast wave problem 2 for the three different
SPH formulations (stdGrad, MI1 and MI2; $200^3$ particles).  The reference solution  (red line) has been
obtained by the Eulerian weighted average flux method   \citep{toro99} with $400^3$ grid cells.} 
\label{fig:riem2_comp}
\end{figure}

As a second spherical blast wave problem \citep{toro99} we start from  
\be
(\rho,\vec{v},P)=
 \left\{
\begin{array}{l}
(1.0,0,0,0,2.0) \quad {\rm for \; \; r< 0.5}\\
(1.0,0,0,0,1.0) \quad {\rm else}.
\end{array}
\right.
\ee
We show the numerical solution ($200^3$ particles, cubic lattice;  MI1) at time $t=0.2$ in Fig.~\ref{fig:Riemann2}.
Again, the SPH result is in very good agreement with the (higher resolved) reference solution, only the inner
density plateau is somewhat smeared out, and there is a small wiggle at the contact discontinuity. This could be cured, 
for example, by applying a larger amount of conductivity (keep in mind we only use a small value, $\alpha_u=0.05$).
In this test all three the SPH formulations perform well, the matrix-inversion versions capture
the central density plateau better and show a smaller overshoot than the kernel-gradient version, see Fig.~\ref{fig:riem2_comp}.
Note that in both circular blast wave problems the spherical symmetry is very well preserved, despite the initial particle 
setup on a cubic lattice. %
%

\subsubsection{Noh test}
\begin{figure}
\vspace*{-0cm}
\centerline{
\includegraphics[width=\columnwidth,angle=0]{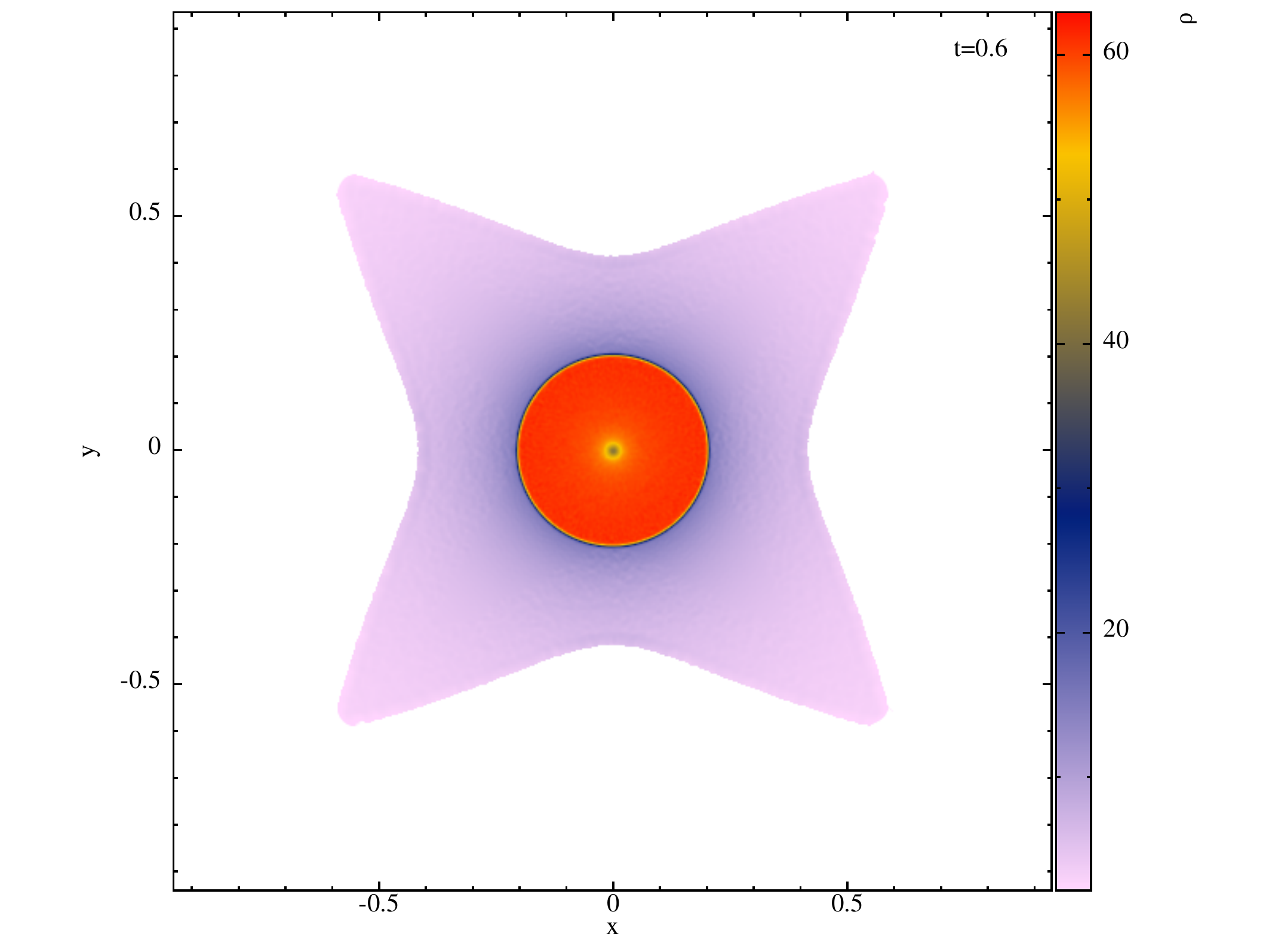}
}
\vspace*{0cm}
\caption{Result of the challenging 3D Noh implosion test (with the MI1 equation set). Shown is the  density distribution in the XY-plane 
at $t= 0.6$.} 
\label{fig:Noh1}
\end{figure}

\begin{figure}
\vspace*{-0cm}
\centerline{
\includegraphics[width=\columnwidth,angle=0]{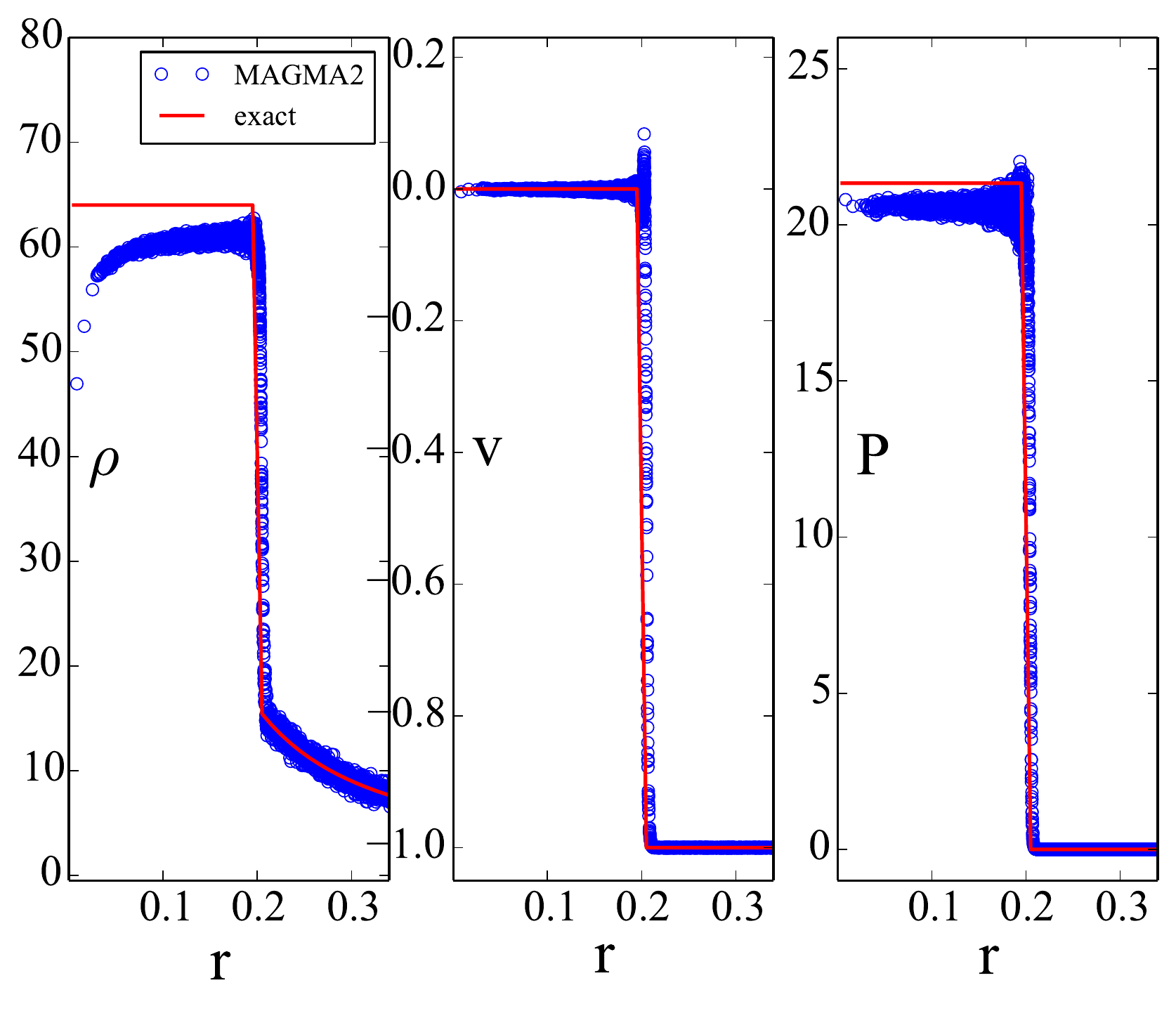}
}
\vspace*{-0cm}
\caption{Result of the challenging 3D Noh implosion test (MI1), the exact solution is shown as solid red line.} 
\label{fig:Noh2}
\end{figure}

Next we consider the very challenging 3D Noh implosion test \citep{noh87} 
which has the reputation as a "code breaker",  since a number of established 
methods are not able to handle this test without breaking or simply producing
wrong results \citep{liska03}.  The test is performed with 
a polytropic exponent of $\Gamma=5/3$ and with initial conditions $[\rho_0,v_0^i,P_0]=
[1,-\hat{r}_a^i,10^{-10}]$, where $\hat{r}_a^i$ is the $i$-component of the radial unit vector 
of particle $a$. This corresponds to the spherical inflow to a point and results in a self-similar
shock moving outwards at a velocity $v_s= 1/3$. The shocked region ($r < v_s t$) has
density $\rho_s= \rho_0 [(\Gamma+1)/(\Gamma-1)]^3$.
The pre-shock flow ($r > v_s t$) undergoes shock-less compression according to
$\rho(r,t)= \rho_0(1-v_0 t/r)^2$.\\
The test is challenging for two reasons. First, the initial conditions contain an unresolved
point of convergence that gives rise to the well-known "wall heating problem" where the thermal
energy overshoots at the origin while the density undershoots in order to strive for the correct pressure. 
This phenomenon lead to the original suggestion to apply an artificial conductivity \citep{noh87} 
to alleviate the problem. The second difficulty is that the adiabatic pre-shock compression 
should not produce entropy which is a serious challenge for most artificial viscosity schemes.\\
We setup this challenging test similar to Sedov test case: we distribute $2.3 \times 10^7$ particles according
to a Centroidal Voronoi Tessellation \citep{du99} and subsequently perform 1000 correction sweeps according to Eq.~(\ref{eq:position_update}).
We show in Fig.~\ref{fig:Noh1} the density for this test (for the MI1 equation set; the other equation sets 
give very similar results), $\rho$, $v$ and $P$ are shown in
Fig.~\ref{fig:Noh2} compared with the exact solution\footnote{http://cococubed.asu.edu}. 
At the centre the "wall-heating problem" with a 
substantial density undershoot is encountered and the pressure is about 4\% below the
theoretical value. Nevertheless, compared to most existing methods \Ma performs
rather well in this challenging test. The wall heating effect could be alleviated
by employing larger $\alpha_u$-values. The PHANTOM code \citep{price18a}, for
example, uses $\alpha_u= 1$ and the authors report to have not found serious artefacts from
this large value. 
However, given the somewhat pathological initial conditions with its unresolved point of convergence,
we are not excessively worried about the  encountered wall-heating problem (that is shared
by most other methods) and do not see this as a strong incentive  to increase the dissipation 
parameter $\alpha_u$ beyond the chosen, low value.


\subsection{Instabilities}

\subsubsection{Kelvin-Helmholtz Instability}
\label{sec:KH_instab}
\begin{figure*}
\hspace*{-0.cm}\includegraphics[width=2.4\columnwidth,angle=0]{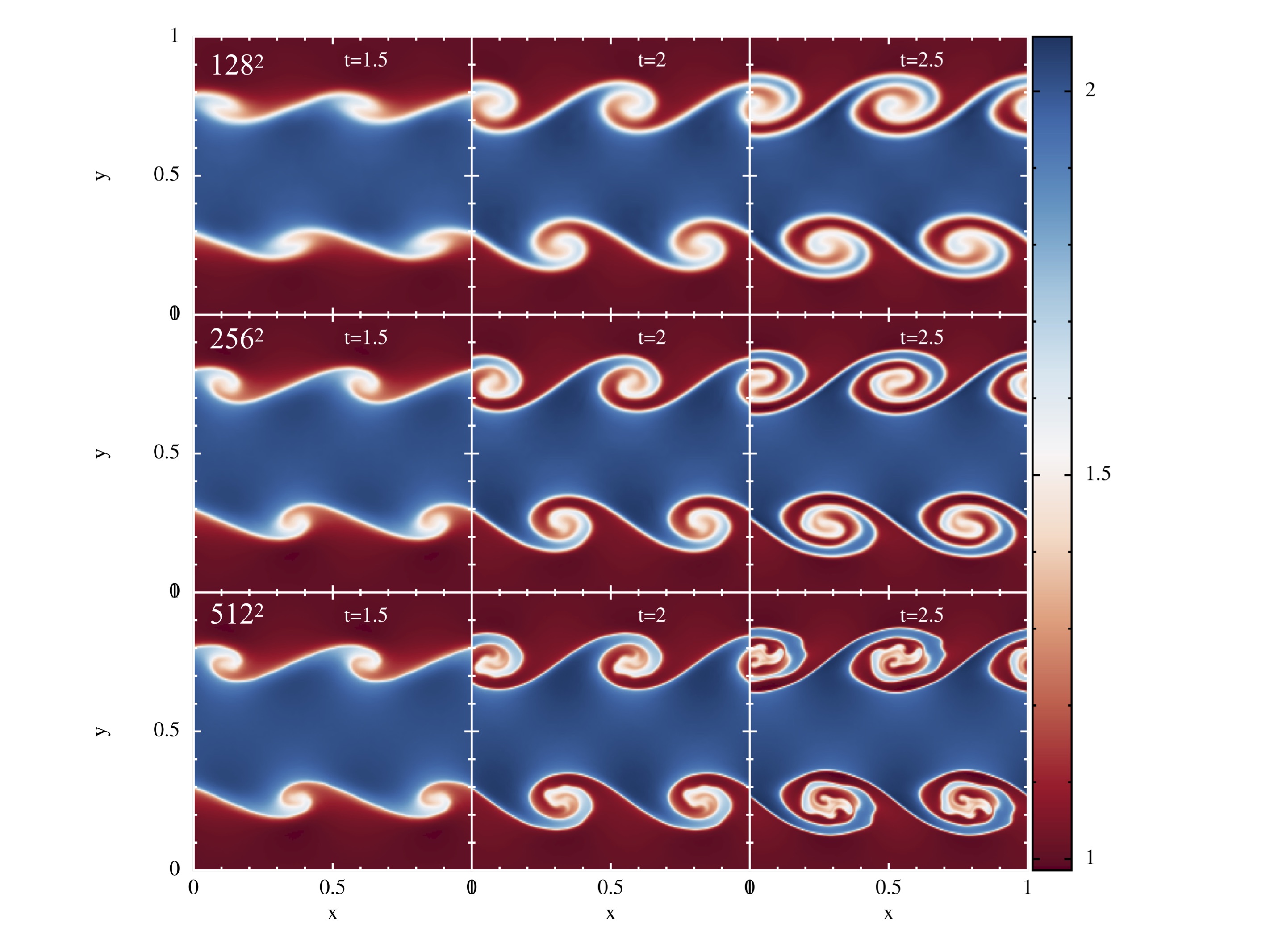}
\vspace*{-1cm}
\caption{Density evolution for weakly triggered Kelvin-Helmholtz instability at ($128 \times 128 \times 8$),  ($256 \times 256 \times 8$) and ($512 \times 512 \times 8)$
particles resolution.}
\label{fig:KH_resolution}
\end{figure*}

\begin{figure}
\hspace*{-0.4cm}\includegraphics[width=1.1\columnwidth,angle=0]{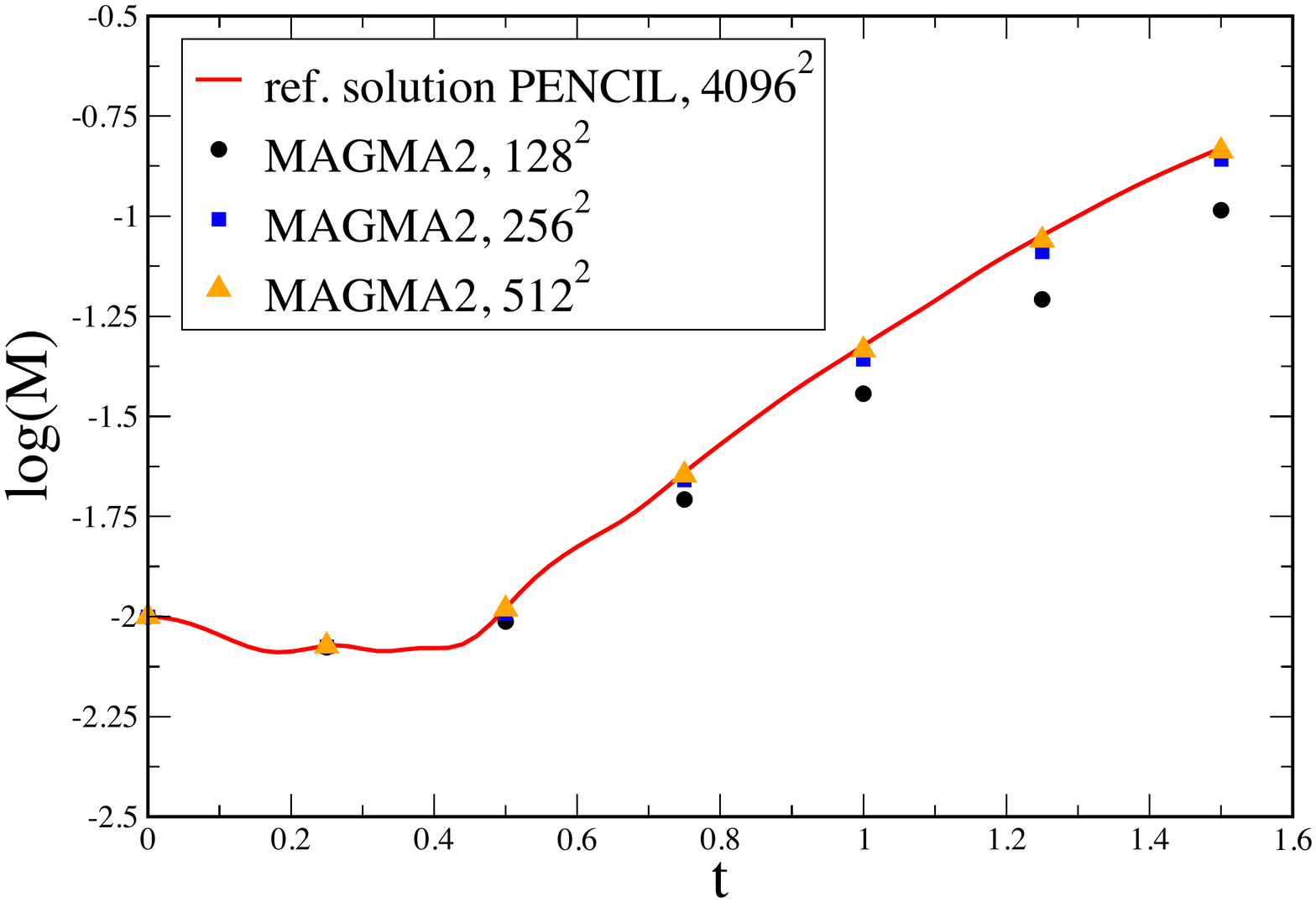}
\vspace*{-0.5cm}
\caption{Resolution dependence of the Kelvin-Helmholtz mode amplitude growth (MI1). As reference 
solution (solid, red) we use a simulation with the Pencil Code at a resolution of $4096^2$. Note that all 
cases show a growth close to the reference solution, even at the lowest resolution ($128^2$ particles). 
For comparison, traditional SPH simulations struggle with this only weakly triggered test, even at 
substantially higher resolution of $512^2$ \citep{mcnally12}.}
\label{fig:KH_mode_amplitude}
\end{figure}

\begin{figure*}
\vspace*{-0cm}
\hspace*{-0.3cm}\includegraphics[width=17cm,angle=0]{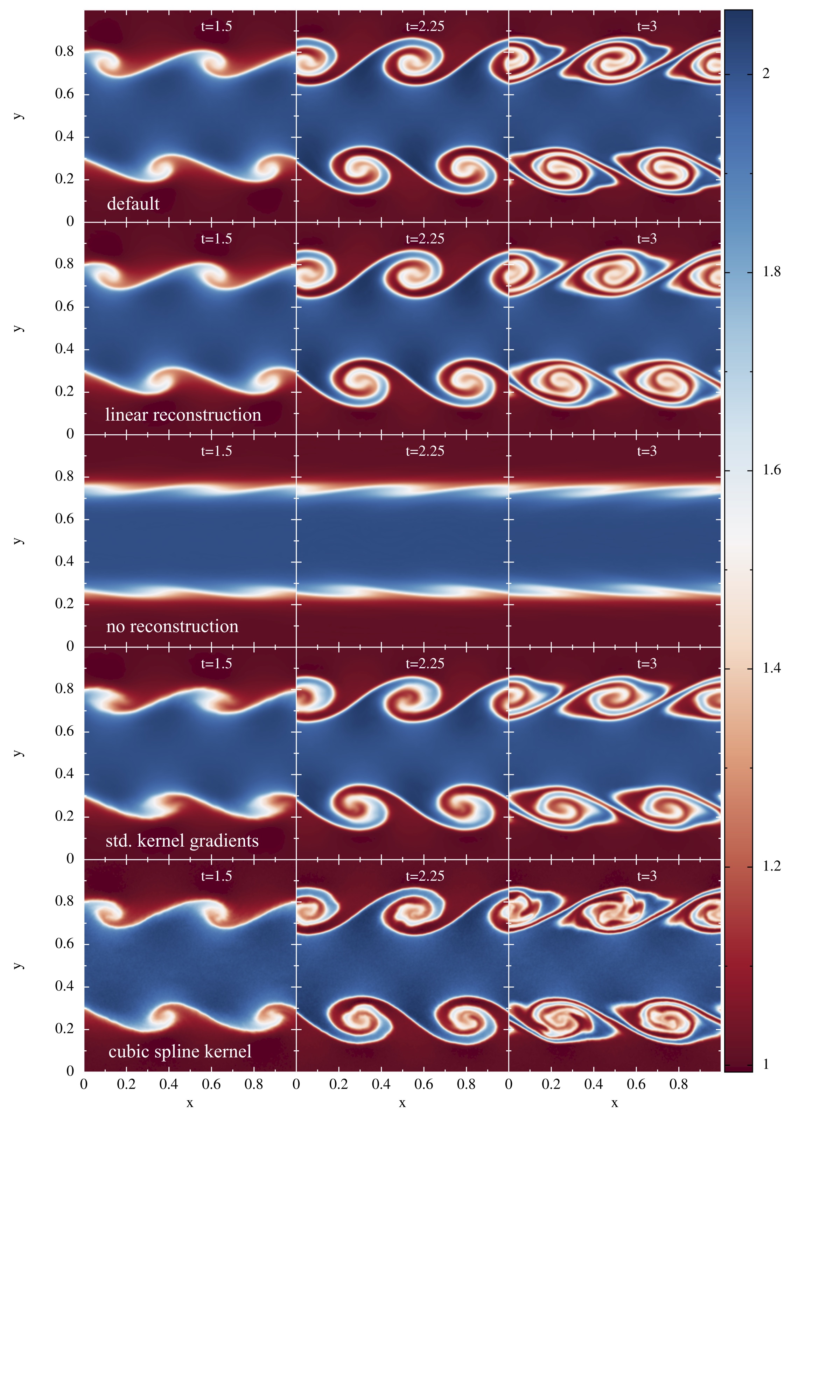}
\vspace*{-6cm}
\caption{3D Kelvin-Helmholtz test ($256 \times 256 \times 20$ particles) where we explore the impact of different methodological choices. The first
row shows our default choice with matrix-inversion gradients (MI1), artificial dissipation that uses a slope-limited, quadratic velocity reconstruction and the high-order Wendland kernel. Second row: as row 1, but only using linear reconstruction. Third row: as default, but no reconstruction, just using the velocity differences between particles; this is the conventional way of implementing artificial viscosity in SPH; Fourth row: as default, but suing kernel gradients; Fifth row: as default, but using the cubic spline kernel.} 
\label{fig:KH256_comparison}
\end{figure*}

\begin{figure}
\hspace*{-0.4cm}\includegraphics[width=1.1\columnwidth,angle=0]{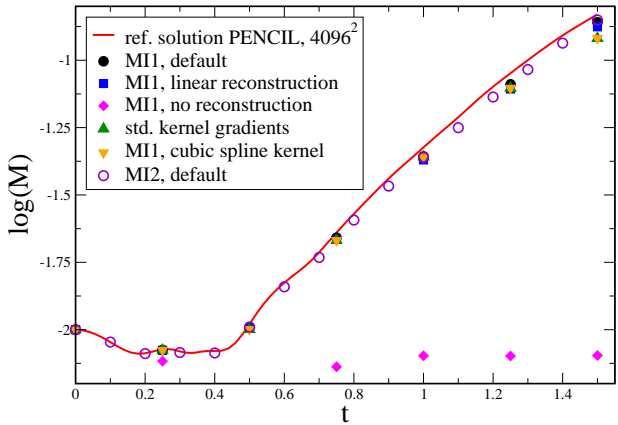}
\vspace*{-0.5cm}
\caption{Growth of the Kelvin-Helmholtz mode amplitude for the $256^2$-case where we vary
methodological choices as in Fig.~\ref{fig:KH256_comparison}.  The results for the MI1 default choices are
shown as black circles, the results  with only linear reconstruction (blue square) grow at a similar rate, but slightly slower. Both the case with
standard kernel gradients (green triangles) and the one using the cubic spline kernel (orange triangles)) 
grow somewhat slower. The largest impact in this test, however, has the velocity reconstruction: if it
is not applied and the standard SPH prescription is applied instead, the instability 
is suppressed (magenta diamonds). The symmetrisation in the matrix inversion formulation does
make a small difference in the growth rate, with MI2 growing slightly faster.}
\label{fig:KH_mode_amplitude_ingredients}
\end{figure}

\begin{figure}
\hspace*{-2.5cm}\includegraphics[width=1.7\columnwidth,angle=0]{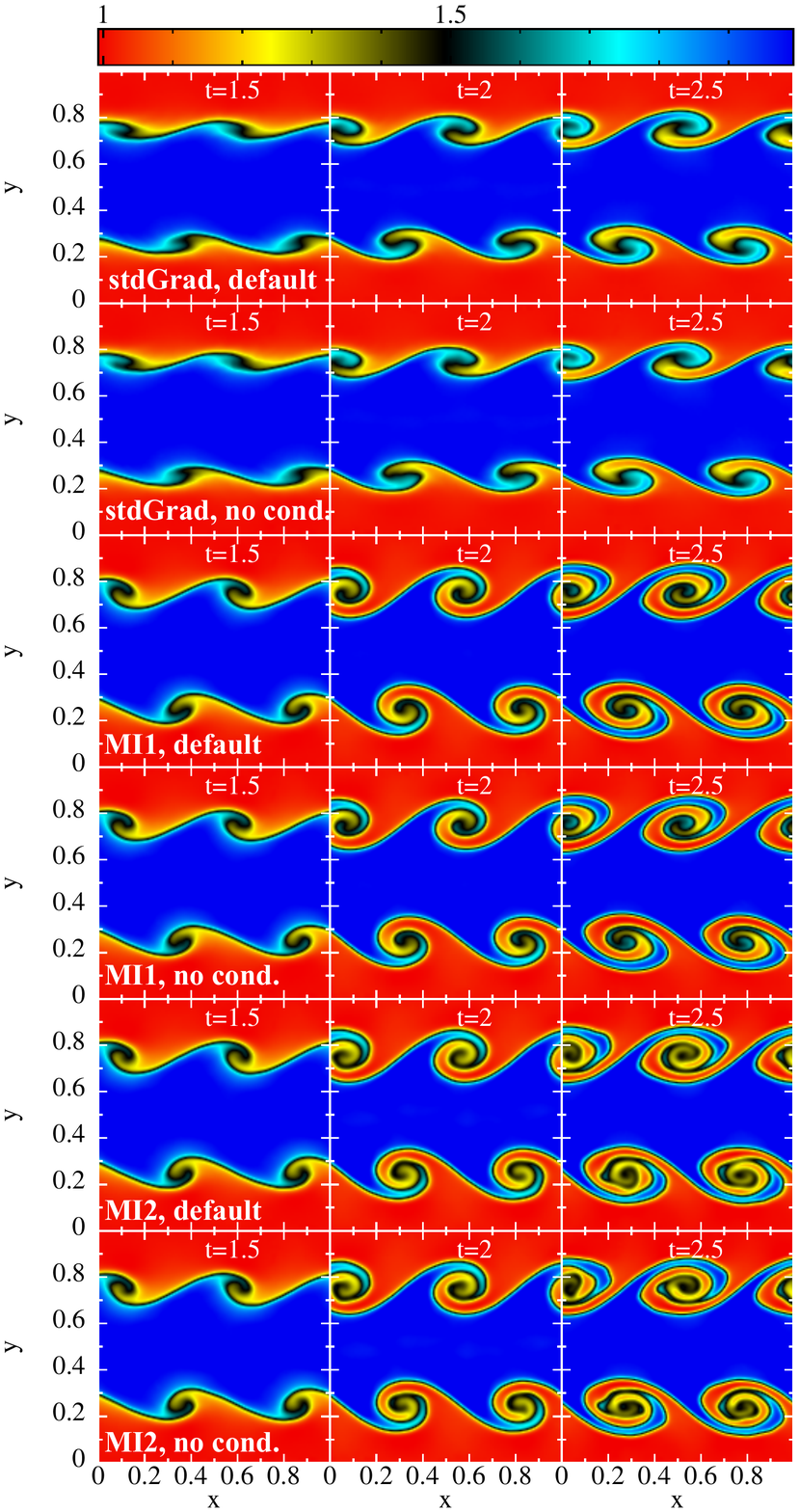}
\vspace*{-2cm}
\caption{Low resolution ("$128^2$ particles") test of the importance of thermal conductivity for our different
SPH formulations in a Kelvin-Helmholtz problem: standard kernel gradients ("stdGrad"; rows 1 and 2), matrix inversion formulation 1 ("MI1"; rows 3 and 4) and
matrix inversion formulation 2 ("MI2"; rows 5 and 6), each time once with conductivity and once without. For none of our SPH-variants does
the conductivity play a major role, results are mostly determined by artificial viscosity (velocity reconstruction or not?) and the accuracy
of the hydrodynamic gradients.
}
\label{fig:KH_conductivity}
\end{figure}

Kelvin-Helmholtz (KH) instabilities occur in shear flows with 
a perturbed interface.They play an important role in astrophysics and occur
in a broad range of environments, e.g. in mixing processes in novae \citep{casanova11}, amplification 
of magnetic fields in neutron star mergers \citep{price06,giacomazzo14,kiuchi15} or planetary atmospheres
\citep{johnson14}, to name just a few. Traditional versions of SPH, however, have been shown to struggle with weakly triggered 
KH-instabilities \citep{agertz07,mcnally12}. We focus here on a test setup in which traditional SPH has been shown
to fail, even at rather high resolution  in 2D, see \cite{mcnally12}.
We follow the latter paper in  setting up the test  with the only difference that we use our full 3D code to perform the test in quasi-2D. To this end
we set up a thin 3D slice with $N \times N \times 20$ particles (referred to as $"N^2"$), for simplicity initially placed on a cubic lattice. 
Periodic boundary conditions are obtained by placing appropriate particle copies outside of the "core" volume. 
The test is initialised as:
\be
\rho(y)=
 \left\{
\begin{array}{l}
\rho_1 - \rho_m e^{(y - 0.25)/\Delta} \quad {\rm for \; \; 0.00 \le y < 0.25}\\
\rho_2 + \rho_m e^{(0.25 - y)/\Delta} \quad {\rm for \; \; 0.25 \le y < 0.50}\\
\rho_2 + \rho_m e^{(y - 0.75)/\Delta} \quad {\rm for \; \; 0.50 \le y < 0.75}\\
\rho_1 - \rho_m e^{(0.75 - y)/\Delta} \quad {\rm for \; \; 0.75 \le y < 1.00}\\
\end{array}
\right.
\ee
where $\rho_1= 1$, $\rho_2= 2$, $\rho_m= (\rho_1 - \rho_2)/2$ and $\Delta= 0.025$.
The velocity is set up as
\be
v_x(y)=
 \left\{
\begin{array}{l}
v_1 - v_m e^{(y - 0.25)/\Delta} \quad {\rm for \; \; 0.00 \le y < 0.25}\\
v_2 + v_m e^{(0.25 - y)/\Delta} \quad {\rm for \; \; 0.25 \le y < 0.50}\\
v_2 + v_m e^{(y - 0.75)/\Delta} \quad {\rm for \; \; 0.50 \le y < 0.75}\\
v_1 - v_m e^{(0.75 - y)/\Delta} \quad {\rm for \; \; 0.75 \le y < 1.00}\\
\end{array}
\right.
\ee
with $v_1$= 0.5, $v_2= -0.5$, $v_m= (v_1-v_2)/2$ and a small velocity perturbation in
$y$-direction is introduced as $v_y=  0.01 \sin(2\pi x/\lambda)$ with the perturbation
wave length  $\lambda= 0.5$. In the linear regime, the instability grows on 
a characteristic time scale of 
\be
\tau_{\rm KH}= \frac{(\rho_1+\rho_2) \lambda}{\sqrt{\rho_1 \rho_2} |v_1-v_2|},
\ee
with $\tau_{\rm KH}\approx 1.06$ for the chosen parameters.
The test is performed with a polytropic equation of state with exponent $\Gamma=5/3$. \\
We show in Fig.~\ref{fig:KH_resolution} the evolution of the instability at times $t=1.5, 2.0$
and 2.5 for resolutions of $128^2$, $256^2$ and $512^2$ particles and the MI1 formulation.
All cases grow at very similar rates and produce the characteristic "Kelvin-Helmholtz billows",
even at the lowest resolution. For comparison, traditional SPH implementations struggle with this 
only weakly triggered instability ($v_y= 0.01$), see, for example, Fig. 9 
of \cite{mcnally12}, where even at a resolution 512$^2$ particles the instability hardly grows (for the
cubic spline kernel, label "Ne512") or much too slowly (for the quintic spline kernel, label "No512").
In Fig.~\ref{fig:KH_mode_amplitude} we show the mode growth (calculated exactly as in \cite{mcnally12})
for all our cases compared to a high-resolution reference solution ($4096^2$ cells) obtained by the
PENCIL code \citep{brandenburg02}. Even our low-resolution case with $128^2$ particles is very 
close to the reference solution, the growth rates of the higher resolution cases are hard to distinguish from the reference solution.\\
It is instructive to repeat this test (at fixed resolution of $256^2$ particles) and each time vary one of 
the choices we have made, starting the MI1 formulation with default choices as baseline.
Density snapshots of these experiments are shown in Fig.~\ref{fig:KH256_comparison},
the corresponding growth rates are shown in Fig.~\ref{fig:KH_mode_amplitude_ingredients}. As first line in 
Fig.~\ref{fig:KH256_comparison} we show the results obtained with our  MI1 default choices. For the results
in line two we used the same choices, but only linear reconstruction; line three used no reconstruction at all;
line four applied the default choices, but used kernel gradients (instead of matrix-inversion gradients) and, finally, line five
used the cubic spline kernel with 50 neighbours rather than the WC6 kernel with 300 neighbours.
The reconstruction, which substantially reduces the net dissipation, has clearly the largest impact in this test. 
The differences between the quadratic and linear reconstruction are only moderate, the interfaces between
the high and low density parts remain sharper in the former case (this is confirmed by running the simulations
for longer). Not using any reconstruction at all, i.e. applying the common SPH approach of using the velocity 
differences at the particle positions, suppresses the growth of the instability all together, see the magenta
diamonds in Fig.~\ref{fig:KH_mode_amplitude_ingredients}. Note, however, that this could also be improved
by applying time-dependent dissipation schemes \citep{morris97,rosswog00,cullen10,rosswog15b,price18a,rosswog20b}.
The version with standard gradients also shows healthy growth, albeit at a somewhat lower rate
and at t= 1.5 the edges of the high-density region are rather noisy. Although the cubic spline kernel has been
found to be inferior to higher order kernels \citep{rosswog15b,tricco19a} it delivers in this test (together with
matrix-gradients and large dissipation, but velocity reconstruction) satisfactory results. With all other
options being the same, we see a small difference in the growth rate between the two different symmetrisations
of the matrix-inversion formulations, but MI2 is even closer to the high-resolution reference solution (open circles
vs filled circles in Fig.~\ref{fig:KH_mode_amplitude_ingredients}).\\
Obviously, the growth rates are in good agreement with the reference solution, even at very low resolution
and one may wonder how important the thermal conductivity is for this result. To find out, we perform the following 
experiment. We set up $128^2$ particles with the APM\footnote{We hardly see a difference for a grid setup.}, Sec.~\ref{sec:APM}, 
and run for each simulation the test once with default parameters and once with default parameters, but without 
conductivity. As can be seen in Fig.~\ref{fig:KH_conductivity}, the impact of conductivity is rather small and it
seems that (apart from the reconstruction) the largest impact is made by the gradient accuracy: both matrix
inversion formulations (columns 3 to 6) show even at this low resolution a healthy growth, rather independent of conductivity, while the
standard gradient version struggles and only grows with some noticeable delay, both with and without conductivity (columns 1 and 2). 


\subsubsection{Rayleigh-Taylor instability}
The Rayleigh-Taylor instability is a standard probe of the subsonic growth of a small perturbation.
In its simplest form, a layer of density $\rho_t$ rests on top of a layer with density $\rho_b < \rho_t$
in a constant acceleration field, e.g. due to gravity. While the denser fluid sinks down, it develops
a characteristic, "mushroom-like" pattern.  Simulations with traditional SPH implementations 
have shown only retarded growth or even a complete suppression of the instability \citep{abel11,saitoh13}.\\
As before, we adopt  a quasi-2D setup and use the full 3D code for the evolution.
We place the particles on a CL in the XY-domain $[-0.25,0.25] \times [0,1]$ and we use
8 layers of particles in the Z-direction. Similar to \cite{frontiere17} we use 
$\rho_t=2$, $\rho_b=1$, a constant acceleration $\vec{g}= -0.5 \hat{e}_y$ and
\be
\rho(y)= \rho_b + \frac{\rho_t-\rho_b}{1+\exp[-(y-y_t)/\Delta]}
\ee
with transition width $\Delta=0.025$ and transition coordinate $y_t=0.5$. We apply a small velocity perturbation to the interface
\be
v_y(x,y)= \delta v_{y,0} [1 + \cos(8\pi x)][1 + \cos(5\pi(y-y_t))]
\ee
for $y$ in $[0.3,0.7]$ with an initial amplitude $\delta v_{y,0}=0.025$, and use a polytropic
equation of state with exponent $\Gamma=1.4$. The equilibrium pressure profile
is given by
\be
P (y) = P_0 -  g \rho(y) [y - y_t]
\ee
with $P_0= \rho_t/\Gamma$, so that the sound speed is near unity in the transition region.
To enforce boundary conditions we add 10 rows of extra particles ($y>1$ and $y<0$) that 
we "freeze" at the initial conditions and we use periodic boundary conditions elsewhere.\\
\begin{figure*}
\vspace*{0cm}
\hspace*{-0.3cm}\includegraphics[width=19cm,angle=0]{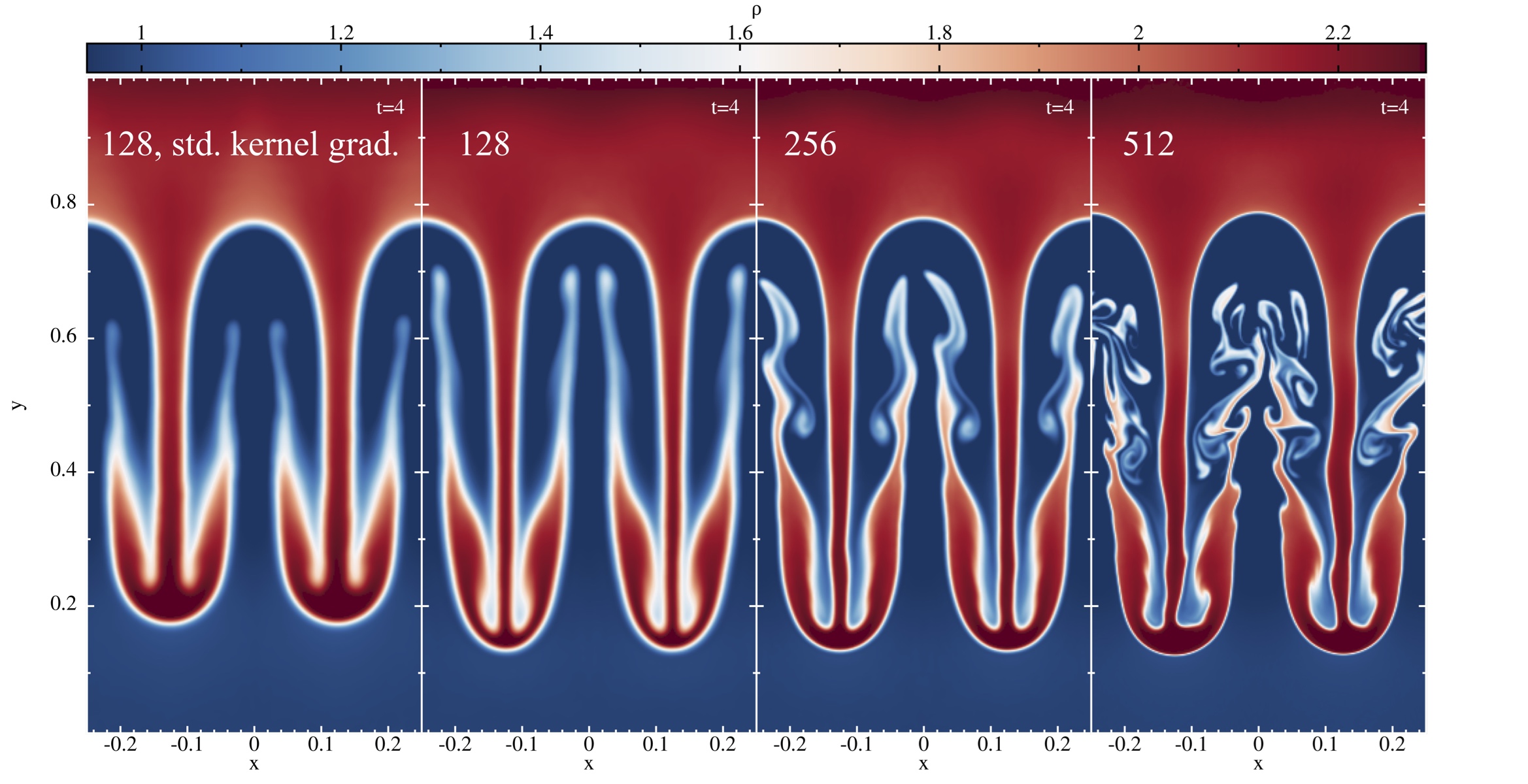}
\vspace*{-0.5cm}
\caption{The first Rayleigh-Taylor test (first panel) uses  kernel-gradients as is common practice in SPH ("$128 \times 256$" particles),
all other tests use our default choices of matrix-inversion gradients ("$128 \times 256$", "$256 \times 512$" and "$512 \times 1026$"particles).
Panels 2-4 show the result of the MI1 formulation.}
\label{fig:RaTa}
\end{figure*}
We show in Fig.~\ref{fig:RaTa}  the results of several  simulations at a time of $t= 4$. The first panel shows the result
for a simulation that uses the standard SPH approach with kernel gradients and a XY-resolution of 128$\times$256 particles which 
overall performs reasonable well. The second to fourth panel shows the results for everything else being the same (MI1 formulation), but using matrix-inversion
kernels instead. This version delivers finer resolved/less diffusive density structures and a larger plunge depth. With increasing resolution
the density transitions become sharper and more substructure appears, but the plunge depth remains the same.\\
A comparison between the MI1 and MI2 formulation, see Fig.~\ref{fig:RaTa_comp}, demonstrates that the latter 
shows a larger amount of mixing, consistent with the results from Sec.~\ref{sec:surface}.
\begin{figure}
\vspace*{0cm}
\hspace*{-0.2cm}\includegraphics[width=1.1\columnwidth,angle=0]{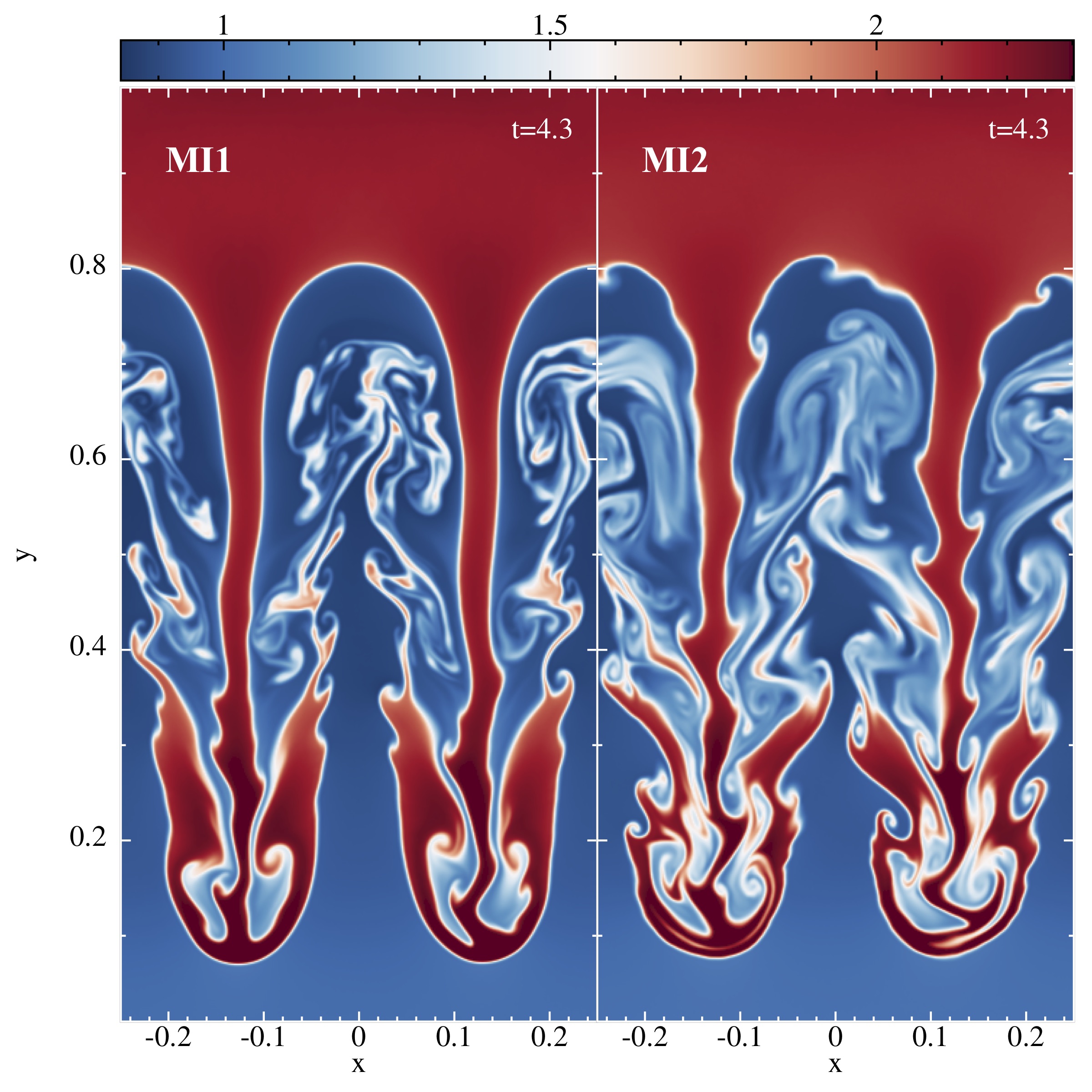}
\caption{Comparison of a Rayleigh-Taylor test (density, 512$^2$) at $t= 4.3$ between the MI1 and MI2 formulation.}
\label{fig:RaTa_comp}
\end{figure}

\subsection{Complex shocks with vorticity creation}

\begin{figure*}
\centerline{
\includegraphics[width=17cm,angle=0]{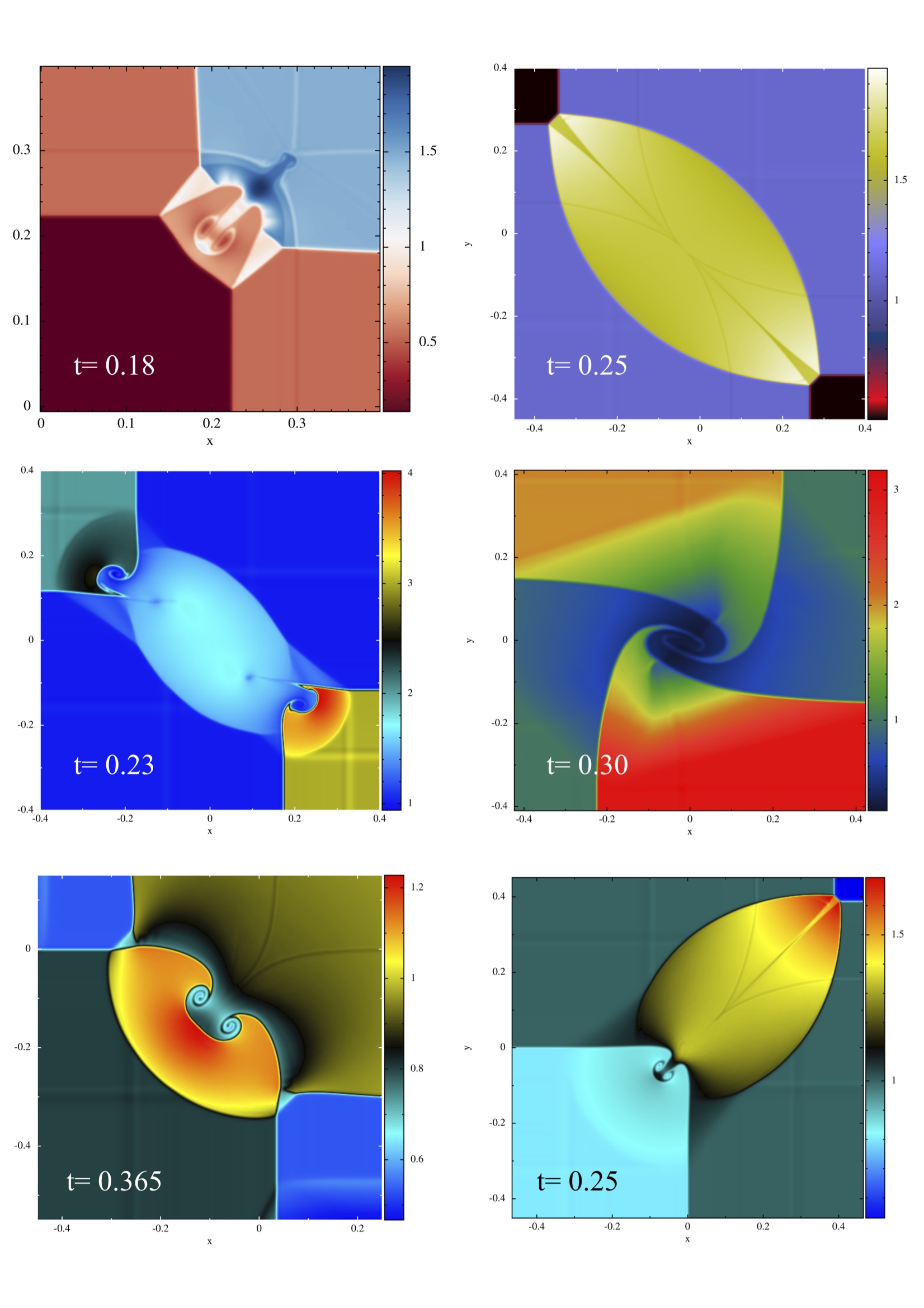}
}
\vspace*{-1cm}
\caption{\Ma solution (density) for the challenging Schulz-Rinne-type shock tests, see Tab.~1 
for the initial conditions. For these tests no exact solutions are known, results need to be compared to
other numerical schemes, see e.g. Lax and Liu (1998) or Liska and Wendroff (2003).}
\label{fig:SchulzRinne}
\end{figure*}

A set of challenging 2D benchmark tests has been suggested by \cite{schulzrinne93a}.
They are constructed in such a way  that four constant states meet at one corner and 
the initial values are chosen so that one elementary wave, either a shock,
a rarefaction or a contact discontinuity appears at each interface.  
During the subsequent evolution complex wave patterns 
emerge for which no exact solutions are known. These tests are considered
challenging benchmarks for multi-dimensional hydrodynamics codes 
\citep{schulzrinne93a,lax98,kurganov02,liska03}. Such tests are rarely shown for SPH codes,
in fact, we are only aware of the work by \cite{puri14} who show results for one such shock test
in a study of Godunov SPH with approximate Riemann solvers.\\
Here we investigate six such configurations. 
Since our code is intrinsically 3D, we simulate, as before, a slice thick enough so that the midplane is
unaffected by edge effects (we use 10 particle layers in Z-direction).  
We use 660 x 660  particles in the XY-plane arranged on a hexagonal lattice between  $[x_c-0.5,x_c+0.5] \times [y_c-0.5,y_c+0.5]$, 
$(x_c,y_c)$ being the contact point of the quadrants, and we use
a polytropic exponent $\Gamma=1.4$ in all of the tests. We refer to these Schulz-Rinne type 
problems as SR1 - SR6 and give their initial parameters for each quadrant in Tab. \ref{tab:SchulzRinne}. 
These test problems correspond to configuration 3, 4, 5, 6, 11 and 12 in the labelling convention
of \cite{kurganov02}. Our results (MI1) are shown  in Fig.~\ref{fig:SchulzRinne}.\\
\begin{table}
\caption{Initial data for the Schulz-Rinne-type 2D Riemann problems}
\begin{tabular}{ l c | c | c | c | c | c | c |}
\hline
 & & SR1; contact point: $(0.3,0.3)$ & &\\
  \hline	
  \hline		
  variable & NW & NE & SW & SE \\ \hline
  $\rho$ &  0.5323 &  1.5000  & 0.1380 &  0.5323 \\
  $v_x$ &  1.2060 &  0.0000   & 1.2060 & 0.0000  \\
  $v_y$ &   0.0000 & 0.0000   & 1.2060 & 1.2060 \\
  $P$    &   0.3000 & 1.5000   & 0.0290 & 0.3000    \\
  \hline  
  & & SR2; contact point: $(-0.15,-0.15)$ & &\\
  \hline	
  \hline		
  variable & NW       & NE        & SW      &SE \\ \hline
  $\rho$   &  0.5065 &  1.1000 & 1.1000  &  0.5065 \\
  $v_x$   &  0.8939 &  0.0000 & 0.8939   & 0.0000 \\
  $v_y$   &  0.0000 &  0.0000  & 0.8939  &  0.8939\\
  $P$      &  0.3500  & 1.1000  & 1.1000  & 0.3500    \\

  \hline  
  & & SR3; contact point: $(0.0,0.0)$ & &\\
  \hline	
  \hline		
  variable & NW & NE & SW & SE \\ \hline
  $\rho$ &  2.0000 &  1.0000 & 1.0000  &  3.0000 \\
  $v_x$ &  -0.7500 & -0.7500 & 0.7500 & 0.7500  \\
  $v_y$ &   0.5000 & -0.5000   & 0.5000   &  -0.5000\\
  $P$    &    1.0000  & 1.0000   & 1.0000   & 1.0000    \\
  \hline  
 & & SR4; contact point: $(0.0,0.0)$ & &\\
  \hline	
  \hline		
  variable & NW & NE & SW & SE \\ \hline
  $\rho$ &  2.0000 &  1.0000 & 1.0000  &  3.0000 \\
  $v_x$ &  0.7500 &   0.7500 & -0.7500 & -0.7500  \\
  $v_y$ &   0.5000 & -0.5000   & 0.5000   &  -0.5000\\
  $P$    &    1.0000 & 1.0000   & 1.0000   & 1.0000  \\
  \hline  
 & & SR5; contact point: $(0.0,0.0)$ & &\\
  \hline	
  \hline		
  variable & NW & NE & SW & SE \\ \hline
  $\rho$ &  0.5313 &  1.0000 & 0.8000  &  0.5313 \\
  $v_x$ &   0.8276 &  0.1000 & 0.1000 & 0.1000  \\
  $v_y$ &   0.0000 & 0.0000  & 0.0000  &  0.7276\\
  $P$    &   0.4000  & 1.0000  & 0.4000 & 0.4000    \\
  \hline  
   & & SR6; contact point: $(0.0,0.0)$ & &\\
  \hline	
  variable & NW & NE & SW & SE \\ \hline
  $\rho$ &  1.0000 &  0.5313 & 0.8000  &  1.000 \\
  $v_x$ &   0.7276 &  0.0000 & 0.0000 & 0.0000  \\
  $v_y$ &   0.0000 & 0.0000  & 0.0000  &  0.7262\\
  $P$    &   1.0000  & 0.4000  & 1.0000 & 1.0000    \\
  \hline  
  
\end{tabular}
\label{tab:SchulzRinne}
\end{table}
\noindent
\noindent Test SR1 is the only one of the Schulz-Rinne tests that has to our knowledge been
tackled with (Godunov-)SPH \citep{puri14}. Their  results show that existing SPH
implementations struggle to resolve the mushroom-like structure along the 
diagonal (roughly at $x\approx y \approx 0.2$ in our figure, upper left panel) and 
depending on the chosen approximate Riemann solver serious artefacts appear.
Our results, in contrast, show crisp transitions between the different regions, well developed
"mushrooms" and they look overall similar to those found with established Eulerian methods, 
see e.g. \cite{liska03}, their Fig.4.1.
\noindent In test SR2 straight 1D shocks separating constant states and two
curved shocks bordering a lens-shaped high-density/-pressure region occur. The
result should be  symmetric with respect to the lens axis and the \Ma results do
not show any noticeable deviation from perfect symmetry.
Test SR3 yields a lense-shaped central region with two vortex structures occurring
at the upper left and lower right part of the shown domain. Again, our result at $t=0.23$ 
closely resembles those in the literature, e.g. \cite{lax98}, their Fig.5.
\noindent
This is also true for the remaining tests, SR 4 can be compared, e.g. with Fig. 4.1
in \cite{liska03}, SR 5 with Fig. 11 in \cite{lax98} and SR6 with Fig. 4.4 in \cite{liska03}.
Note that all tests show a high (though not perfect) degree of symmetry which --with 
freely moving particles-- is not actively enforced. Overall, the tests are in very
good agreement with the Eulerian results found in the literature \citep{lax98,liska03} 
and they look crisp and noise free. Note in particular the appearance of mushroom-like
structures (in panels 1, 5 and 6) which are usually considered a challenge for
SPH-methods.  Some of the weak straight lines in the panels,
however, are considered spurious. But these artefacts are shared by the majority of 
methods found in the literature.
\begin{figure}
\hspace*{-0.2cm}\includegraphics[width=1.15\columnwidth,angle=0]{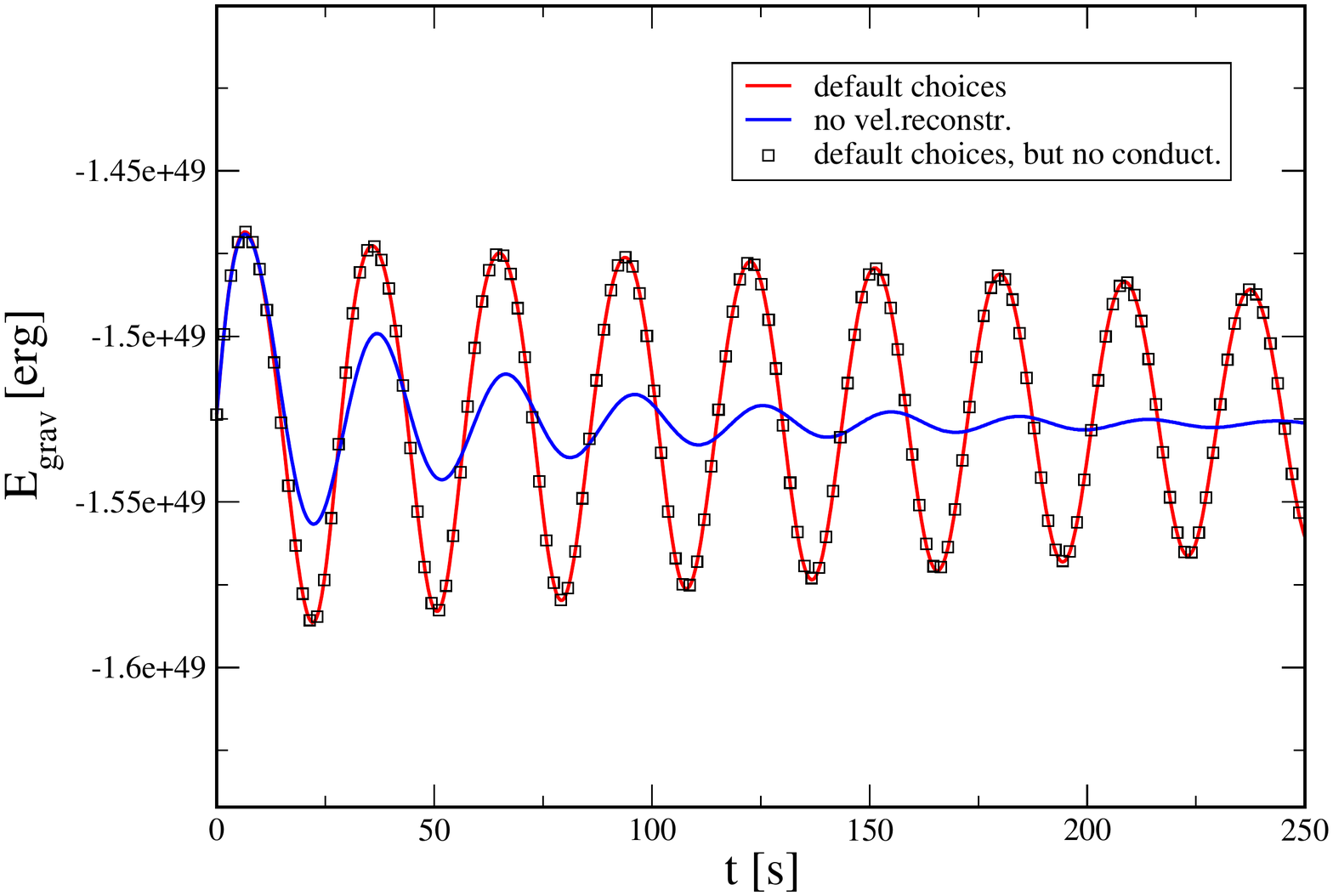}
\vspace*{-0.5cm}
\caption{Gravitational energy of an oscillating polytropic white dwarf star with 10K particles. The blue
line shows artificial viscosity applied as in standard SPH (without velocity reconstruction), the red line
is our default choice for methods and parameters and the black squares are our default choices (only every fifth point shown), but without
artificial conductivity, i.e. $\alpha_u= 0$.}
\label{fig:poly_oscillation_v1}
\end{figure}

\begin{figure*}
\includegraphics[width=2.2\columnwidth,angle=0]{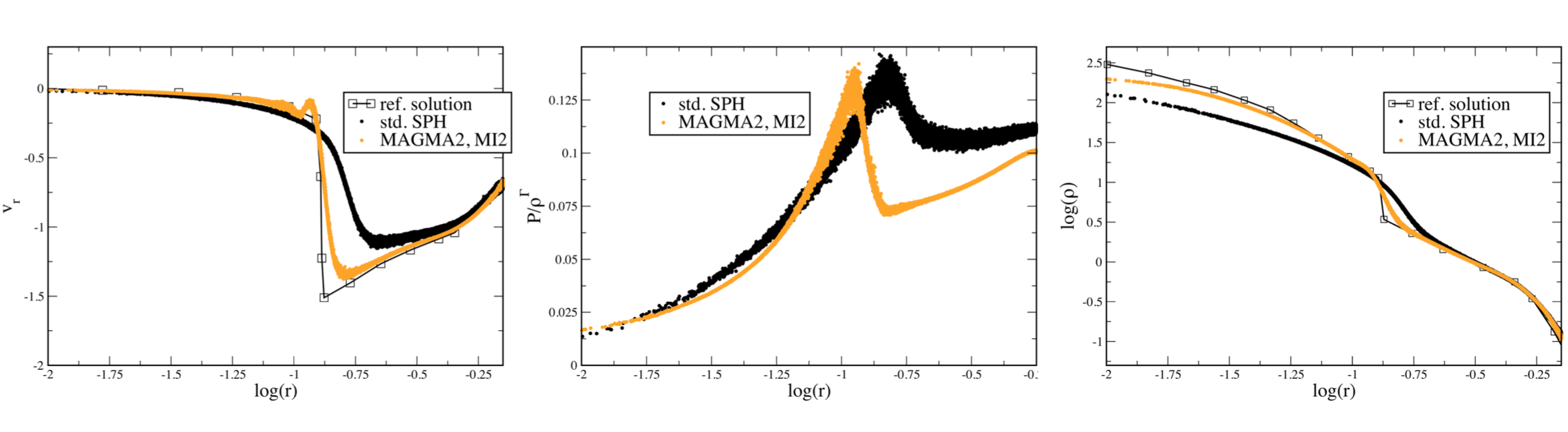}
\vspace*{0cm}
\caption{Results of Evrard's isothermal cloud collapse \citep{evrard88} for a resolution of $10^5$ particles. 
Shown are the results of "standard SPH" (no reconstruction, cubic spline kernel, kernel gradients; black)
and one of \ma's formulations (MI2; orange) for velocity (left), entropy variable $P/\rho^\Gamma$ (middle) 
and density (right) together with a 1D reference solution \citep{steinmetz93}.}
\label{fig:Evrard_100K}
\end{figure*}

\begin{figure*}
\includegraphics[width=2\columnwidth,angle=0]{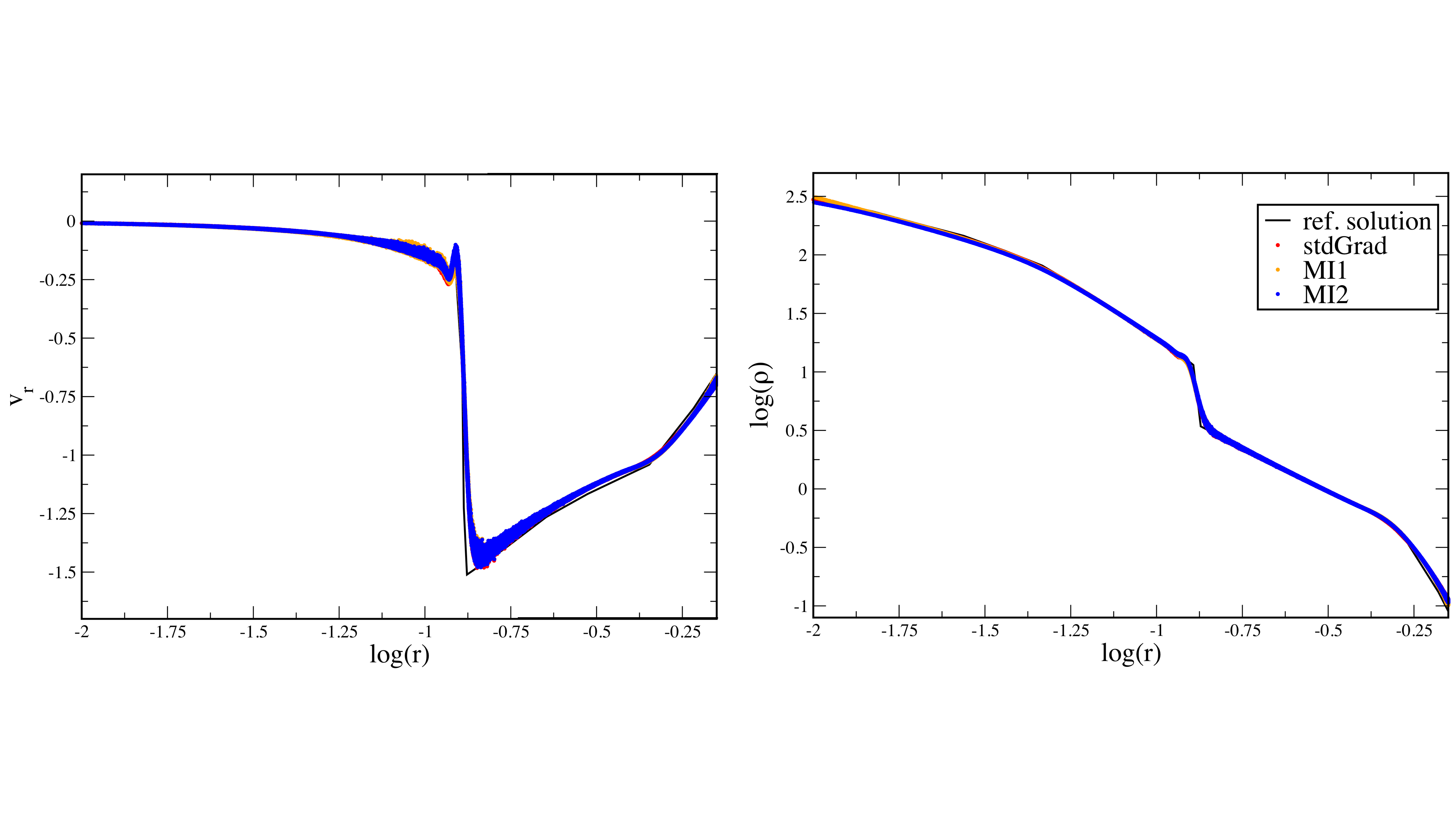}
\vspace*{-1.5cm}
\caption{Results of Evrard's isothermal cloud collapse \citep{evrard88}. Shown are the results of all three SPH formulations (stdGrad, MI1, MI2;
$10^6$ particles, all are plotted) together with a reference solution \citep{steinmetz93}. All our variants yield very similar results in this test.}
\label{fig:Evrard}
\end{figure*}

\subsection{Astrophysical applications}
In this last section we show tests that are close to astrophysical applications. The purpose 
of these tests is to demonstrate the performance of the dissipation scheme, 
measure the numerical conservation in a relevant example and to show robustness and
geometric flexibility. We did not find noteworthy differences between different SPH-formulations
in these tests and, unless explicitly stated otherwise, we show the MI1-results.

\subsubsection{Oscillating White Dwarf}
\label{sec:oscil_poly}
As another experiment, we take a relaxed $\Gamma=5/3$-polytropic star 
that represents a model for a 0.3 \Msun WD. We use only 10K SPH particles
and provide them with a radial velocity $\vec{v}_a= v_0 \vec{r}_a$, where we 
choose $v_0= 0.02$.  As before, all tests use constant 
dissipation parameters of $\alpha=1$.  To avoid dissipation from other sources, 
we run these tests with low tolerance for the tree accuracy ($\Theta=0.1$) and 
a time integration prefactor $C=0.1$.
The evolution of the oscillations in the gravitational 
energy are shown Fig.~\ref{fig:poly_oscillation_v1}. As blue line we show the standard SPH-approach,
i.e. without velocity reconstruction, the red line shows our default choice of methods and 
parameters and the black, open circles show the result for the default choices, but with
$\alpha_u=0$. Consistent with the experiments in the Kelvin-Helmholtz instability test,
see Fig.~\ref{fig:KH256_comparison}, we find a massive suppression of unwanted
dissipation when velocity reconstruction is employed. As intended, artificial conductivity
does not switch on in this test problem, the results with $\alpha_u=0$
lie nearly exactly on top of the $\alpha_u=0.05$ (default).

\subsubsection{Collapse of isothermal sphere}
The collapse of an initially isothermal cloud is another frequently performed 
complex code test \citep{evrard88,hernquist89,steinmetz93,dave97,springel01a,wadsley04,cabezon17,price18a}
that tests for the coupling of gravity and hydrodynamics. The test starts
with a gas cloud at rest that collapses under its own gravitational, then forms
a shock, bounces back with a shock wave moving outward until the system
settles into a virial equilibrium. This benchmark tests the transformation between
different forms of energy: initially mostly gravitational, then kinetic and finally
thermal.\\
We prepare the setup according to the parameters of Evrard (1988) with
\be
\rho(r)= \frac{M(R)}{2 \pi R^2 r},
\ee
where the initial cloud radius is $R=1$ and the mass $M=1$. Similar to other tests, we set up $10^6$
SPH particles according to a centroidal Voronoi tessellation \citep{du99} and subsequently 
perform 5000 sweeps according to Eq.~(\ref{eq:position_update}) to further improve the initial particle
distribution. We then assign masses so that the initial density profile is reproduced, set
the internal energy to $u= 0.05 G M/R$ and use a polytropic exponent of $\Gamma=5/3$.\\
As a first step, we compare the \Ma result with "standard SPH choices", specifically a) a cubic 
spline kernel with 50 neighbour particles, b) constant dissipation ($\alpha=1, \beta=2$) without reconstruction
and c) standard kernel gradients to calculate derivatives. Note that this is different from what
we had abbreviated before as "stdGrad": the latter uses kernel gradients, but all the other benefits of \ma.
The results of the MI2 formulation (with all default choices) and the "standard SPH choices"
for a low resolution case with $10^5$ SPH particles is shown in Fig.~\ref{fig:Evrard_100K} 
 at t= 0.77.
Compared to the reference solution \citep{steinmetz93}, our low-resolution MI2 result
shows a velocity overshoot at the shock (left panel), but otherwise agrees well. The "standard SPH choice"
version, in contrast, shows a fair amount of spurious entropy production (middle panel), so that the shock 
is broadly smeared out (left and right panel) and actually sitting at too large a radius.\\
Thus, the \Ma results are a major improvement over traditional SPH choices. Among our different
SPH variants we only find very minor differences in this test. We show a case with $10^6$
SPH particles, prepared as before,  in Fig.~\ref{fig:Evrard} together 
with the reference solution \citep{steinmetz93}. Note that all particles are plotted in the figure. 
All three formulations are nearly perfectly spherically symmetric and agree very well with the 
reference solution. Only at the shock front there is some velocity overshoot. 

\subsubsection{Collision between two main sequence stars}
\label{sec:stellar_collision}
\begin{figure*}
\centerline{
\includegraphics[width=19cm,angle=0]{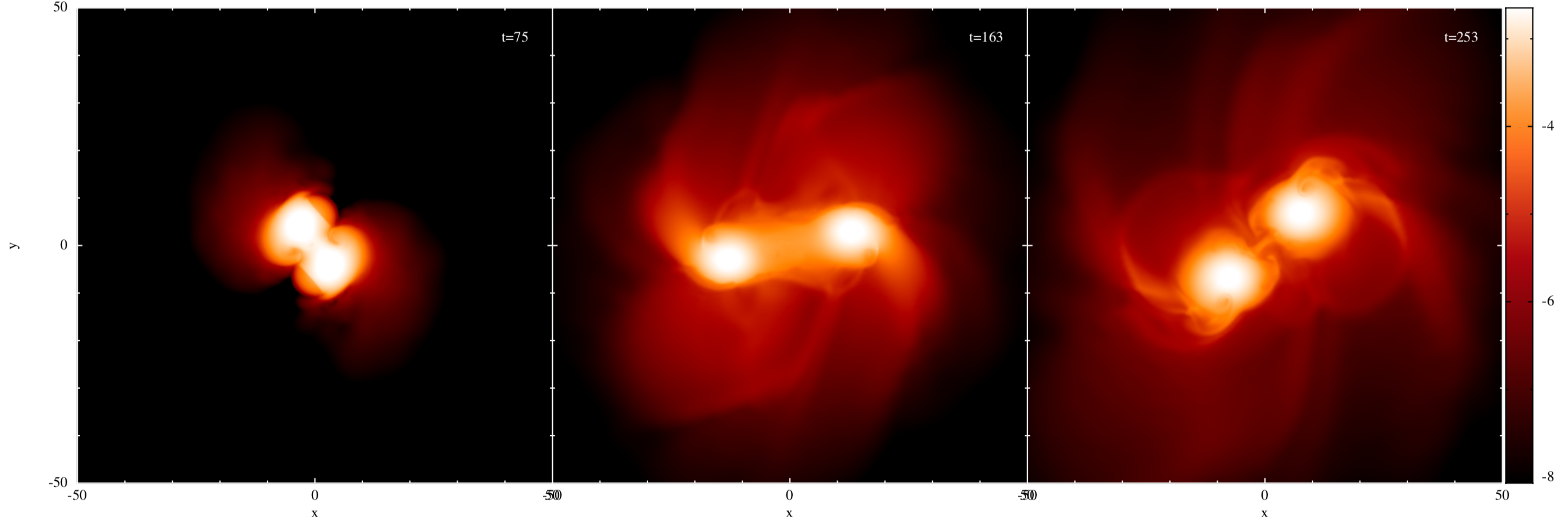}
}
\centerline{
\includegraphics[width=19cm,angle=0]{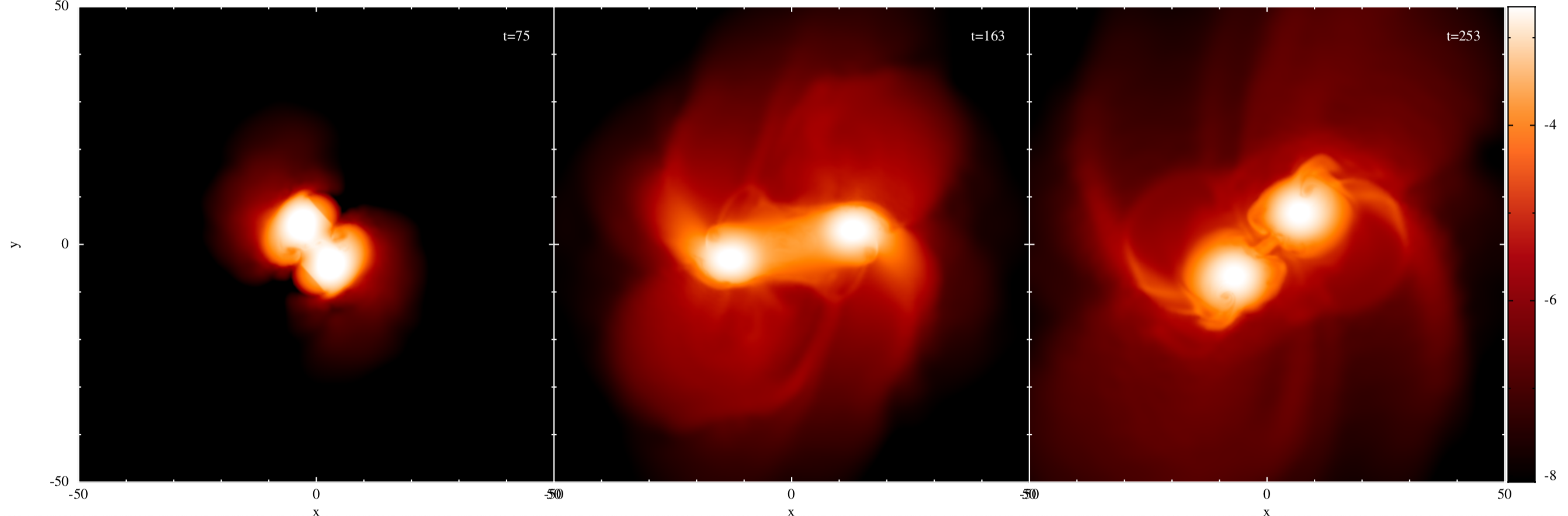}
}\vspace*{0cm}
\caption{Off-center collision between two main sequence stars with an impact strength of $\beta=3$, 
colour coded is density (in code units: 1989.1 g/cm$^3$ for density, time unit is 86.8 
seconds). The upper row shows the results from a simulation that uses standard
kernel gradients (stdGrad) in the evolution equation while the lower row shows the simulation that calculates gradients
via a matrix-inversion technique (MI1). For this test both methods are in excellent agreement.}
\label{fig:MSMS_collision}
\end{figure*}

\begin{figure}
\centerline{
\includegraphics[width=10cm,angle=0]{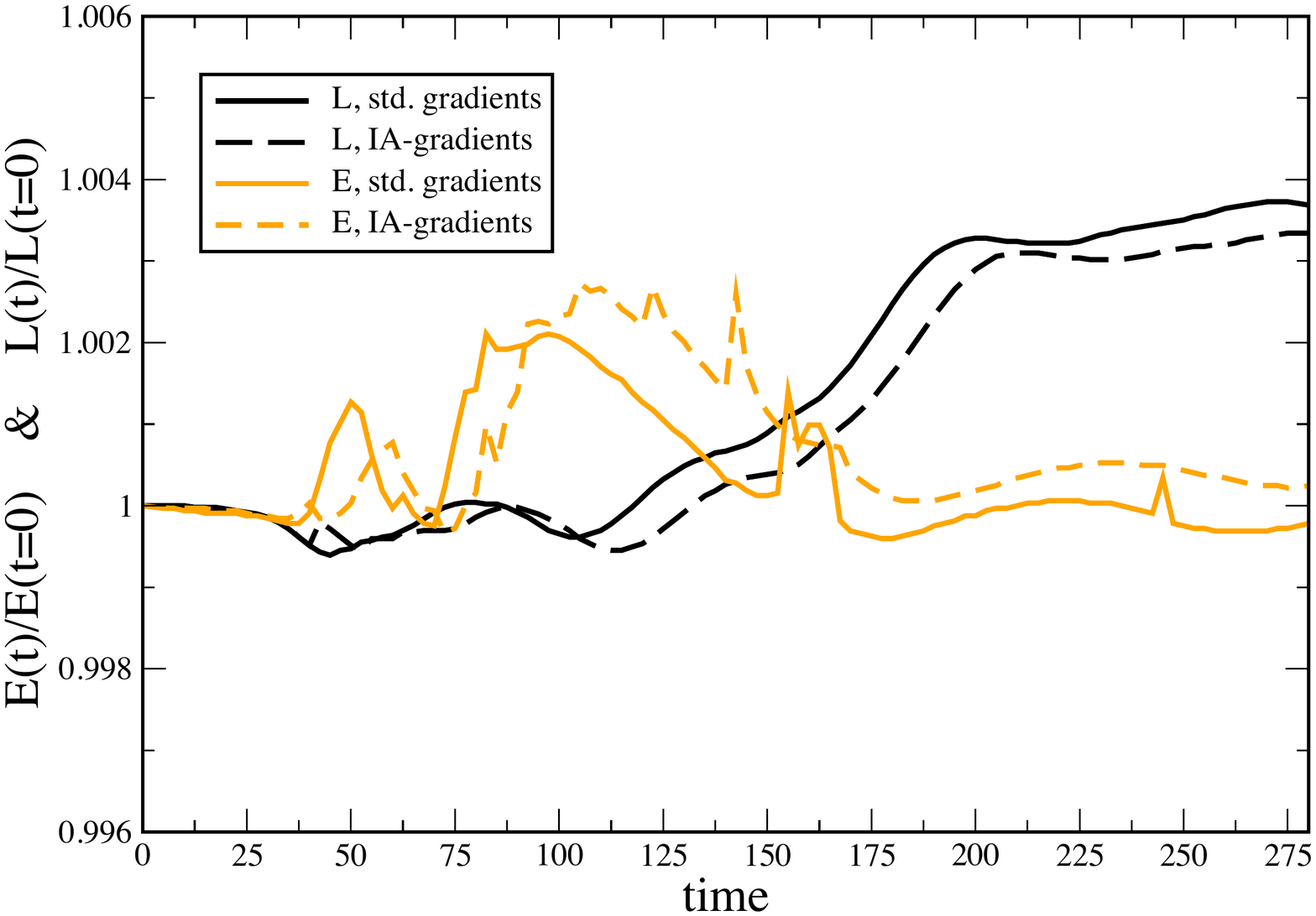}
}
\caption{Conservation of total energy and angular momentum in the stellar collisions shown
in Fig.~\ref{fig:MSMS_collision}. Both are shown for both simulations (with standard and IA-gradients)
and are normalised to their initial values. Time is given in code units of 86.8 seconds.}
\label{fig:conservation_MSMS}
\end{figure}

In this test  we simulate a collision between two main sequence 
stars. The aim is, on the one hand, to demonstrate the robustness and usefulness for the simulation 
of violent astrophysical events and, on the other hand, to measure how accurately 
matrix-inversion  formulations numerically conserve physically conserved quantities for 
typical simulation parameters (such as the tree opening criterion
and time integration prefactor.)
For the simulation we choose two identical stars, each with a mass  $M_\ast= 1 M_\odot$  and a radius
$R_\ast= R_\odot$, modelled as a $\Gamma=5/3$ polytrope with $2 \times 10^6$ equal mass SPH particles. 
The stars approach each other on a parabolic orbit with an impact strength
\be
\beta\equiv \frac{2 R_\ast}{R_{\rm peri}}= 3.
\ee
We perform this simulation twice, once with the evolution equation set that uses kernel
gradients (stdGrad; first row in Fig.~\ref{fig:MSMS_collision}) and once with the equation 
set that uses matrix-inversion based gradients (MI1; second row).
During the first collision the stars become heavily shocked (see panels 1 and 4), vorticity is created and the 
stellar cores are substantially spun up (panels 2 and 5). During the subsequent evolution the stars
fall back towards each other, thereby creating further shocks and vorticity 
 (panels 3 and 6). For this test we find very similar results.
We have performed these simulations with parameters that we would use for a practical simulation:
$C=0.2$ in Eq.~(\ref{eq:t_step}) and a (rather tolerant) tree opening criterion 
$\Theta= 0.9$, see \cite{gafton11} for a detailed description of the
used recursive coordinate bisection (RCB) tree. The latter guarantees a fast
evaluation of the gravitational forces, though at the price of sacrificing some accuracy. Despite 
this seemingly tolerant opening criterion, the conservation of both energy and angular momentum
for both approaches are better than 0.4\%, see Fig.~\ref{fig:conservation_MSMS}, and could be
easily further improved by choosing a stricter force criterion.

\subsubsection{Tidal disruption of a white dwarf star}
\begin{figure*}
\vspace*{0cm}
\includegraphics[width=16cm,angle=0]{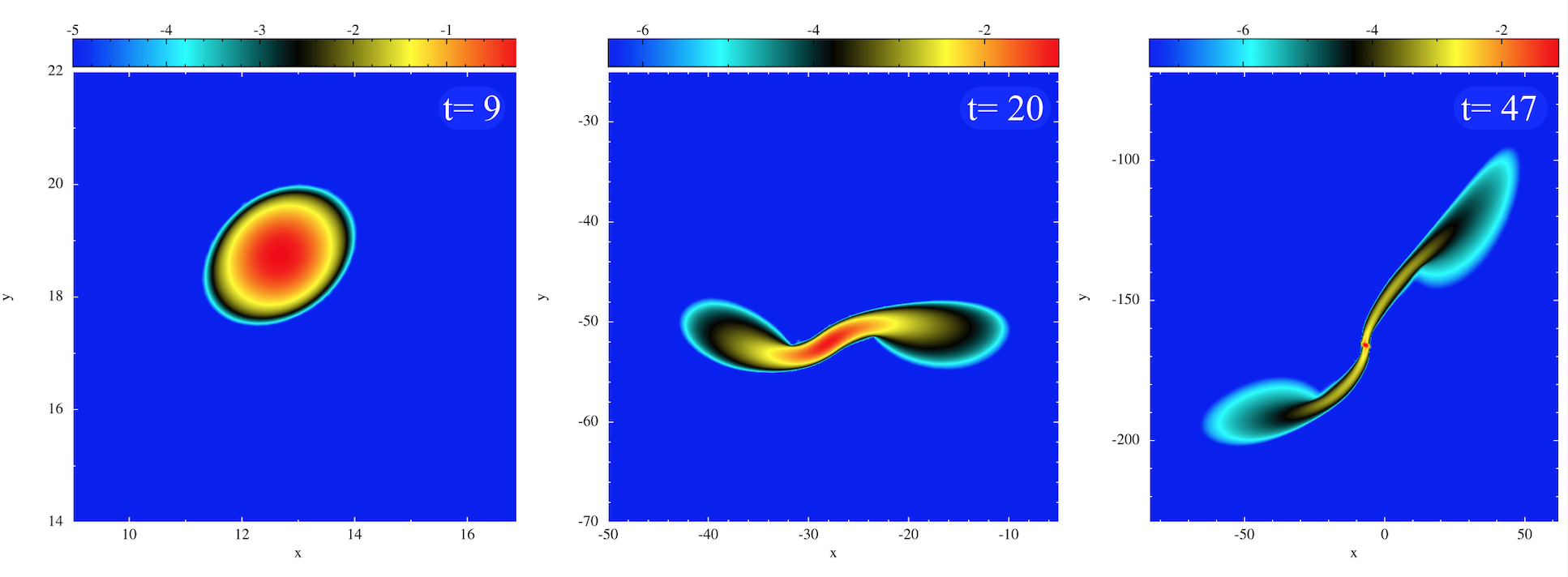}
\vspace*{0cm}
\caption{3D tidal disruption of a 0.5 \Msun white dwarf star (modelled with $5 \times 10^6$ SPH particles) by a 1000 \Msun black hole located at the coordinate origin. 
This is a weak encounter (penetration factor $\beta=0.9$) where the white dwarf actually passes 
outside the tidal radius and becomes "nearly disrupted". After the passage, however, the core contracts again due to self-gravity. 
Colour-coded is the mass density in the orbital plane.}
\label{fig:WD_TDE}
\end{figure*}

As another astrophysical test case  we show  a tidal disruption of a 0.5 \Msun white dwarf star by
a 1000 \Msun black hole. Such encounters can lead to a tidal ignition and explosion 
of the white dwarf \citep{luminet89a,rosswog08a,rosswog09a} provided that the black
hole is of "intermediate" mass (below $\approx 10^5$ \msun). We show a weak encounter with a "penetration factor" 
$\beta= R_{\rm t}/R_{\rm p}= 0.9$, where $R_{\rm t}= R_{\rm wd} (M_{\rm bh}/M_{\rm wd})^{1/3}$
is the tidal radius inside of which a star is disrupted by the black hole's tidal forces. The quantity
$R_{\rm p}$ is the distance of closest approach ("pericentre distance"). We chose a value of
$\beta < 1$, since this results in a partial disruption where the star is "nearly disrupted",
but while receding from the black hole its self-gravity overcomes the tidal pull again 
and a part of the tidal debris  re-collapses into a self-gravitating core. 
Due to the highly elongated geometry and the small self-gravitating core such partial disruptions
pose particular computational challenges.\\
We model the initial white dwarf as a polytrope with exponent $\Gamma=5/3$
with $5 \times 10^6$ equal-mass SPH-particles. Since for the chosen parameters
the pericenter distance is $R_{\rm p} > 90 G M_{\rm bh}/c^2$, we can treat 
the black hole to excellent accuracy as a Newtonian point mass\footnote{For relativistic encounters
one can use MAGMA2 with either the accurate pseudo-potential of \cite{tejeda13a}, or, even better,
within the approach suggested in \cite{tejeda17a}.}. Initially the 
white dwarf is placed at a distance $r_0= 6 R_t$ which guarantees that the tidal
acceleration is only a tiny perturbation ($\ll1\%$) compared to self-gravity.
In Fig.~\ref{fig:WD_TDE},
left panel, we show the density in the orbital plane at $t= 9$ code units (1 code unit= 2.745 s), when
the star is approaching the BH, at $t=20$, after the star has just passed it, and at $t= 47$ when the 
central region has re-contracted into a self-gravitating core.
\begin{figure*}
\vspace*{0cm}
\hspace*{-0.cm}\includegraphics[width=2.1\columnwidth,angle=0]{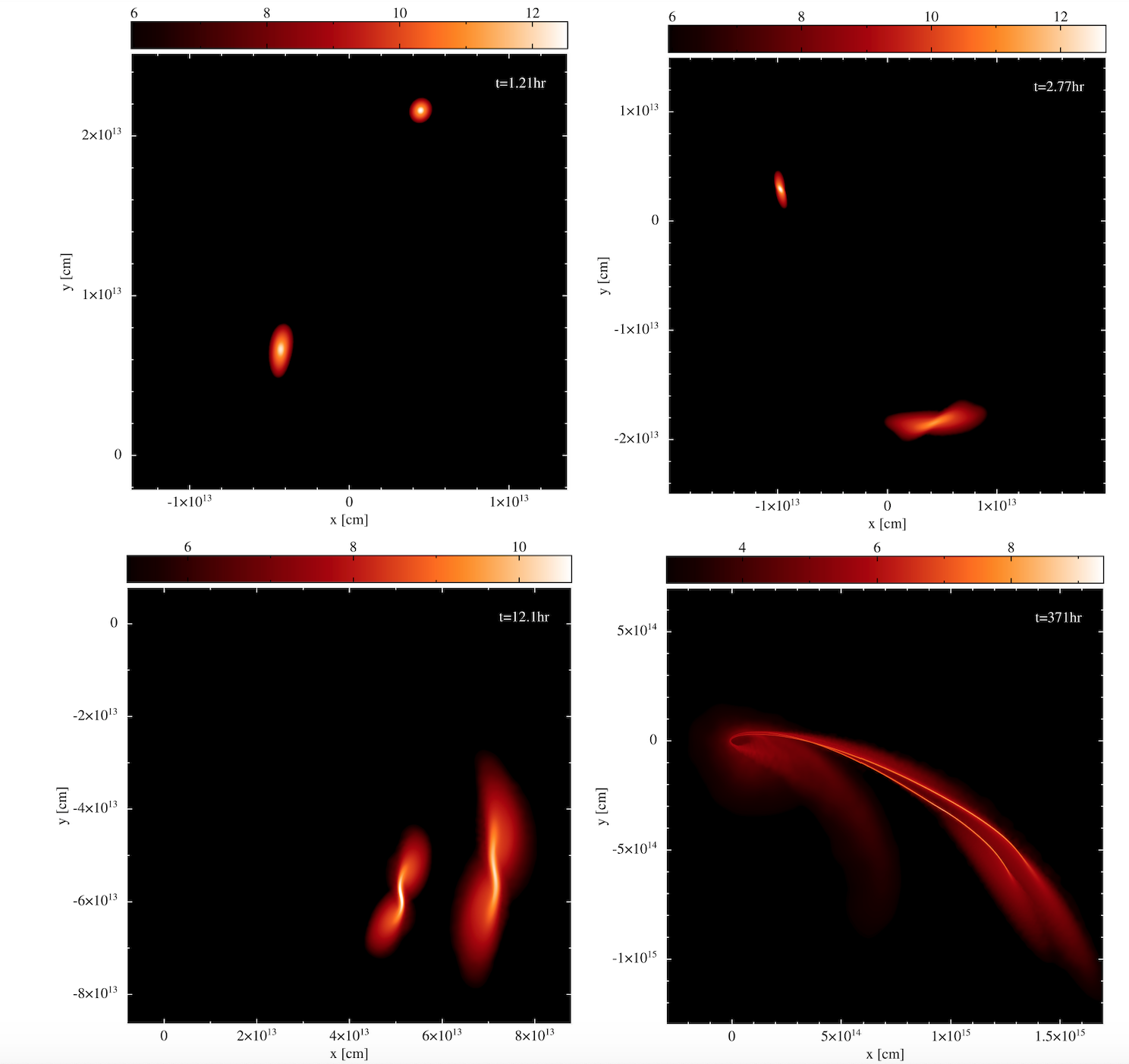}
\vspace*{0cm}
\caption{Tidal disruption of a binary system consisting of two massive stars (67.01 and 36.8 \msun) by a supermassive black hole ($10^6$ \msun).}
\label{fig:DTDE}
\end{figure*}

\subsubsection{Double TDE}
As a last astrophysical test we show the disruption of a stellar binary system of two Main Sequence
stars by a supermassive black hole. Here,  the main challenge comes from the widely varying 
geometry and the involved scales. \cite{mandel15} studied tidal interactions of stellar binary 
systems with massive black holes and found that in a substantial
fraction of cases both stars become disrupted. According to their estimate, close to 10\% of
all stellar tidal disruptions may be double disruption events.\\
We simulate here the disruption of a binary consisting of two massive
stars of 67.01 and 36.8 \Msun by a $M_{\rm bh}= 10^6$ \Msun black hole (initial conditions kindly provided
by Ilya Mandel). Fig.~\ref{fig:DTDE} shows four snapshots of this disruption (colour coded is column density). 
The first shows the stage ($t= 1.21$ hr) when the stars are approaching the black hole and the leading, more
massive star is about to be disrupted. The second snapshot ($t= 2.77$ hr) shows the leading star being disrupted while
the companion is approximately at pericentre. Snapshot three shows both disrupted stars receding
from the black hole while in the last snapshot debris is fed in two narrow streams to the hole and an
accretion disk is being assembled.

\section{Summary}
\label{sec:summary}
In this paper we have presented the new Lagrangian hydrodynamics code \ma,
which benefits from a number enhancements compared to traditional SPH codes. 
\begin{itemize}
\item \Ma uses consistently high-order kernels which substantially
reduce noise, but this comes at the price of large 
 neighbour numbers. Our default choice is a Wendland
$C^6$ kernel together with 300 neighbours  in the kernel support 
of each particle.
\item To produce entropy in shocks, our code employs artificial viscosity, but enhanced by 
techniques that are borrowed from finite volume methods. Instead of employing the velocity 
difference between two particles in the artificial viscosity terms (which is the common SPH practice),
 we use the difference of the {\em slope-limited, quadratically reconstructed velocities} at the inter-particle midpoint.
All tests shown in this paper are performed with constant dissipation parameters ($\alpha=1$
and $\beta=2$) and even with such large parameters we find excellent results in benchmark
tests. We have also implemented a new way to steer time-dependent dissipation by monitoring
for each particle how well entropy is conserved. This allows to identify 
"troubled particles" that need their dissipation increased. This approach is discussed in detail in a separate publication  \citep{rosswog20b}.
\item Apart from a conventional SPH formulation ("stdGrad") that calculates derivatives via kernel gradients, 
\Ma also offers two additional SPH formulations that use much more accurate gradient estimates 
that are based on matrix inversion techniques ("MI1" and "MI2"), see Sec.~\ref{sec:ideal_hydro}.
These two formulations only differ in the way the SPH equations are symmetrised.
All three SPH versions are implemented with the
above described kernels and artificial dissipation techniques. 
\item Self-gravity and neighbour search are  implemented via a fast tree that tessellates space 
by means of a recursive coordinate bisection and that is described in detail in \cite{gafton11}.
\item In Sec.~\ref{sec:APM} we suggest a new way to set up SPH initial conditions.  
SPH is known to perform best when equal-mass particles are used, but setting up geometrically
complicated initial conditions with equal-mass particles is  non-trivial.  To address
this problem we have introduced the {\em Artificial Pressure Method (APM)},  which is very much 
in the spirit of SPH. It starts from an initial distribution of equal mass SPH particles and compares the
currently measured density with a desired density profile. Based on the  local density errors
it calculates an artificial pressure force that steers the SPH particles  into positions 
where the deviations from the theoretical density profile are minimal.
\end{itemize}
We have scrutinised \Ma in a large number of benchmark tests including smooth
advection, a variety of shock and instability tests, vorticity creating Schulz-Rinne shocks (which
are rarely shown in SPH publications) and a number of more astrophysical tests that demonstrate
its robustness, versatility and excellent conservation properties. We find very good results in these
benchmarks, also in tests that are traditionally considered a challenge for SPH codes. \\
As expected, \Ma is second order accurate in smooth flows, see Sec.~\ref{sec:sm_ad} and it  yields 
good results in shocks. The techniques
borrowed from finite volume approaches (slope-limited reconstruction) in the artificial dissipation are
a major improvement compared to the standard approach. For example, with reconstructed velocities,
but large and constant artificial viscosity values \Ma performs excellently  in a Kelvin-Helmholtz test where 
SPH-approaches without such a reconstruction fail completely. Even the low resolution cases grow with rates
very close to the (much higher resolved) reference solution. The effect of the quadratic reconstruction 
(as compared to a linear one) is small, though welcome. But it may be a valid choice to restrict oneself 
to just linear reconstruction and to avoid the need of calculating second derivatives.\\
We have also introduced a set of matrix inversion SPH equations (MI2) with a rarely used 
symmetrisation in the particle indices. This symmetrisation (though different gradients, 
dissipation strategy, kernels etc.) has been successfully  used the \Ga code \citep{wadsley17}. 
While delivering in most tests very similar results to the other matrix inversion formulation (MI1), MI2 
has some distinct advantages:
\begin{itemize}
\item it is substantially more sensitive to density variations and --with everything else being the same--
achieves substantially better results in the Sedov explosion test than the other two formulations.
\item in addition, it  substantially reduces surface tension effects and therefore also has an advantage
in instability tests compared to the other two SPH formulations.
\end{itemize}
The only slight disadvantage that we have noticed is that it is less robust against non-ideal
particle setups. For example, in shock tests with particles placed on a cubic lattice it leads easier
to particle "ringing effects". However, such problems can be easily cured by a more sophisticated
particle setup such as via the above described APM.\\
 We have further performed a number of the challenging, vorticity-creating shocks suggested
by \cite{schulzrinne93a}. Also here, \Ma yields crisp results that are comparable with those
from established Eulerian methods.
We have also performed a number of more astrophysical simulations (stellar collision and tidal
disruptions by black holes) to demonstrate the robustness of \Ma and its accurate numerical 
conservation.\\
In none of the comparisons did we find any disadvantage of the matrix-inversion gradient prescription.
But in many tests they showed clearly superior performance. Since the computationally expensive
ingredients (long neighbour loops, matrix inversions, self-gravity) are shared by all three
SPH formulations, we do not find substantial differences in their run times.   As practically 
demonstrated in the stellar collision example, the matrix inversion formulations 
perform equally well in terms of numerical conservation as standard kernel gradients.\\
In its current version, \Ma has implemented only self-gravitating gas dynamics with polytropic
equations of state, but this framework will be enriched in the near future by more physics.  

\section*{Acknowledgements}
It is a great pleasure to thank Ilya Mandel for sharing the initial conditions of the double-TDE
and Stephen Justham for providing stellar profiles of the corresponding masses.
Thank you also to Davide Gizzi and Christoffer Lundman for their careful reading of an
earlier draft.
Some of the figures of this article were produced with the visualization software 
SPLASH \citep{price07a}. 
This work has been supported by the Swedish Research 
Council (VR) under grant number 2016- 03657\_3, by 
the Swedish National Space Board under grant number 
Dnr. 107/16, by the research environment grant 
"Gravitational Radiation and Electromagnetic Astrophysical
Transients (GREAT)" funded by the Swedish Research 
council (VR) under Dnr 2016-06012 and by the Knut and Alice
Wallenberg Foundation under Dnr. KAW 2019.0112.
We gratefully 
acknowledge support from COST Action CA16104 
"Gravitational waves, black holes and fundamental physics" (GWverse) and from COST Action CA16214
"The multi-messenger physics and astrophysics of neutron stars" (PHAROS).
The simulations for this paper were performed on the facilities of the North-German Supercomputing Alliance (HLRN)
in both G\"ottingen and Berlin, and on the SNIC resources Tetralith and Beskow.

\bibliographystyle{mn2e}
\bibliography{astro_SKR.bib}
\end{document}